\documentclass{iopart}

\usepackage{graphicx}
\usepackage{psfrag}
\usepackage{epstopdf}
\usepackage{color}
\usepackage{mathrsfs}
\usepackage{comment}
\usepackage{multirow}
\usepackage{hyperref}
\usepackage{bm}

\expandafter\let\csname equation*\endcsname\relax
\expandafter\let\csname endequation*\endcsname\relax
\usepackage{amsmath,amssymb,amsthm,wasysym}

\DeclareSymbolFontAlphabet{\mathrsfs}{rsfs}
\DeclareMathAlphabet{\mathcal}{OMS}{cmsy}{m}{n}

\def\lm{{\ell m}}
\def\p{\partial}
\def\non{\nonumber}                  

\def\ha{{\hat{a}}}
\def\H{\hat{H}}
\def\NPDelta{\underline{\Delta}} 
\def\lm{{\ell m}}

\def\hr{{\hat{r}}}
\def\hOmg{{\hat{\Omega}}}

\def\ii{{\rm i}}
\def\l{{\ell }}

\def\a{\alpha}
\def\b{\beta}

\def\del{\triangle}

\def\A{\hat{A}}

\newcommand{\scri}{\mathrsfs{I}}
\newcommand{\be}{\begin{equation}}
\newcommand{\ee}{\end{equation}}

\usepackage{float}
\definecolor{gray}{rgb}{0.5,0.5,0.5}
\definecolor{cyan}{rgb}{0,0.9,0.9}
\definecolor{orange}{rgb}{0.9,0.5,0}
\definecolor{magenta}{rgb}{1,0,1}
\definecolor{purple}{rgb}{0.8,0.4,0.8}
\definecolor{darkgreen}{rgb}{0,.6,0}
\definecolor{turquoise}{rgb}{0.25,0.88,0.82}

\begin{document}

\title
[GW generation, particles on Kerr]
{A new gravitational wave generation algorithm for particle
  perturbations of the Kerr spacetime}

\author{Enno Harms${}^1$, Sebastiano Bernuzzi${}^1$, Alessandro Nagar${}^2$, and An{\i}l Zengino\u{g}lu${}^3$}
\address{${}^1$Theoretical Physics Institute, University of Jena,
  07743 Jena, Germany}
\address{${}^2$Institut des Hautes Etudes Scientifiques, 91440
  Bures-sur-Yvette, France}
\address{${}^3$Theoretical Astrophysics, California Institute of Technology, Pasadena, CA 91125, USA} 

\begin{abstract}
  We present a new approach to solve the 2+1 Teukolsky equation for
  gravitational perturbations of a Kerr black hole.
  Our approach relies on a new horizon penetrating, hyperboloidal
  foliation of Kerr spacetime  and spatial compactification.
  In particular, we present a framework for waveform
  generation from point-particle perturbations.
  Extensive tests of a  time domain implementation in the code
  {\it Teukode} are presented.   
  The code can efficiently deliver waveforms at future null infinity. 
  The accuracy and convergence of the waveforms' phase and 
  amplitude is demonstrated.  

  As a first application of the method, we compute the gravitational 
  waveforms from inspiraling and coalescing black-hole binaries in 
  the large-mass-ratio limit. The smaller mass black hole is modeled 
  as a point particle whose dynamics is driven by an effective-one-body-resummed  
  analytical radiation reaction force. 
  We compare the analytical, mechanical
  angular momentum loss 
  (computed using two different prescriptions) to the gravitational wave angular 
  momentum flux. We find that higher-order post-Newtonian corrections are needed to 
  improve the consistency for rapidly spinning binaries. 
  We characterize the multipolar waveform as a function of the black-hole spin. Close
  to merger, the subdominant multipolar amplitudes (notably the $m=0$
  ones) are enhanced 
  for retrograde orbits with respect to prograde ones. We argue that this effect mirrors
  nonnegligible deviations from circularity of the dynamics during the late-plunge and
  merger phase. 
  For the first time, we compute the gravitational wave energy flux flowing
  into the black hole during the inspiral using a time-domain formalism proposed 
  by Poisson.
  
  Finally, a self-consistent, iterative method to compute the
  gravitational wave fluxes at leading-order in the mass of the particle is developed.
  The method can be used alternatively to the analytical radiation reaction 
  in cases the analytical
  information is poor or not sufficient. Specifically, we apply it to
  compute dynamics and waveforms for a rapidly rotating black hole with
  dimensionless spin parameter $\ha=+0.9$.
  For this case, the simulation with the consistent flux differs from
  the one with the analytical flux by $\sim35$ gravitational wave cycles
  over a total of about $250$ cycles.
  In this simulation the horizon absorption accounts for about
  $+5$ gravitational wave cycles. 
\end{abstract}

\maketitle

\section{Introduction}
\label{sec:intro}

In this work we develop a new, accurate and efficient time domain wave
generation algorithm for perturbations of the gravitational field
around the Kerr spacetime. We focus on point-particle
perturbations, and implement the algorithm to calculate gravitational
waveforms generated by a particle in equatorial motion. 
The method is then applied to modeling the late-inspiral--merger
waveforms from large-mass-ratio black hole binaries, using an
effective-one-body (EOB) approach for the particle
trajectory. We report new results on the multipolar structure of
the waveforms, EOB radiation reaction, and horizon-absorbed fluxes.

Perturbations of a field with spin $s$ (integer or semi-integer) on
a Kerr background are typically described by a master equation
derived by Teukolsky~\cite{Teukolsky:1972my,Teukolsky:1973ha}.
The Teukolsky equation (TE) is separable in the frequency domain, therefore
solutions to this equation have been historically obtained first in the
frequency domain  (see e.g.~\cite{Press:1973zz,Mano:1996vt,Fujita:2004rb}). 
Time domain methods are appealing, however, 
for the ease of treating non-circular orbits. 
The first numerical computation of gravitational perturbations of Kerr spacetime
in the time domain is by Krivan et al.~in 1997~\cite{Krivan:1997hc}. 
They use Boyer--Lindquist coordinates 
and solve the TE as a fully first-order system in 2+1
dimensions after a decomposition into azimuthal angular modes. The system is
discretized using a second-order convergent Lax--Wendroff scheme which
has favorable dissipative properties for numerical stability. The
computational boundaries are placed at finite radii and close to the
horizon using a tortoise transformation of the radial
coordinate. The errors from the boundaries are mitigated by using a large computational
domain. Applications and improvements 
of this scheme can be found, for example, in~\cite{LopezAleman:2003ik,Khanna:2003qv,PazosAvalos:2004rp,
  Sundararajan:2007jg, Sundararajan:2008zm, Sundararajan:2010sr,
  Barausse:2011vx,Taracchini:2013wfa}. 

Even though the 2+1 TE is a linear equation, its time-domain integration 
with $s\neq0$ is challenging and nontrivial due to stiff terms,
exponentially growing continuum modes, and boundary
treatment~\cite{Krivan:1997hc,Harms:2013ib}.   
Until recently, the method of~\cite{Krivan:1997hc}  was the only
successful approach to gravitational ($|s|=2$) perturbations. However, developments
in computational and geometric methods suggest that this problem
should be revisited.
On the geometric side, the accuracy~\footnote{
  We distinguish between accuracy and efficiency, referring,
  respectively, to systematic and truncation errors.
  When the only source of error is discretization, 
  an efficient code is also accurate because it allows 
  the use of more computational resources thereby reducing the
  truncation error. In general, however, numerical calculations of gravitational 
  waveforms include \emph{systematic} errors due to inaccurate numerical 
  boundary treatment or extrapolation from finite radii. 
  Our claim of accuracy refers to the removal of these latter sources of error.}
of the numerical approach can be significantly improved by the 
use of horizon-penetrating coordinates and hyperboloidal 
compactification~\cite{Zenginoglu:2007jw}.
The idea is to use a spacelike foliation that
penetrates the horizon in a regular way and asymptotically approaches
null infinity so that one can compute both the ingoing and the outgoing
radiation. This technique removes 
the two largest systematic uncertainties of Krivan's method (and of any
other based only on Boyer--Lindquist coordinates),
namely 
(i)~the inner and outer numerical boundary errors, and
(ii)~finite-radius-extraction (and/or extrapolation) errors.
On the computational side, our approach is efficient because the computational
domain is much smaller than the one needed in standard
Boyer--Lindquist tortoise coordinates. The efficiency
of the method can be further enhanced by employing high order numerical
discretization techniques (spectral methods or finite differencing).  
The accuracy and efficiency resulting from the application of the
above techniques have been demonstrated in a calculation of gravitational tail decay rates 
in Schwarzschild spacetime \cite{Zenginoglu:2008uc},
and for generic spin fields in Kerr spacetime, including the extreme case~\cite{Harms:2013ib}. 

In this paper we develop the work
presented in~\cite{Harms:2013ib} in two directions. First, we
introduce a new foliation of Kerr spacetime which leads to more efficient numerical computations
than in~\cite{Harms:2013ib}. Second, we extend the approach to
point particle perturbations and test applications relevant for
the binary black-hole problem.

Historically, particle perturbations of black hole spacetimes played a crucial
role in understanding the general relativistic two-body problem and
the related problem of modeling the emission of gravitational waves
(GWs).
Perturbation theory gave access to the first waveforms from the
strong-field/fast-motion
regime (e.g.~\cite{Davis:1971gg,Davis:1972ud,Detweiler:1978ge,Detweiler:1979xr,Sasaki:1981sx,Poisson:1995vs,Lousto:1996sx,Martel:2001yf,Martel:2003jj,Barack:2005nr,Sopuerta:2005gz}),
and has been used to study the radiation reaction 
problem including black hole
absorption (e.g.~\cite{Poisson:1993vp,Cutler:1993vq,Apostolatos:1993nu,Poisson:1993zr,Poisson:1994yf}).   
Perturbative methods interface with post-Newtonian (PN) calculations in
the test-mass limit~\cite{Sasaki:2003xr}, and have led to a very
high-order PN expansion of the circular GW
flux~\cite{Fujita:2012cm,Shah:2014tka}. Also, wave generation algorithms and 
the numerical evolution of linear perturbations sourced by particles are
key steps for the self-force
problem~\cite{Mino:1996nk,Quinn:1996am,Barack:2009ux,Poisson:2011nh}.

Most important for this work, linear point-particle
perturbations provide a natural tool to model black-hole binaries
in the large-mass-ratio limit. Various strategies have been proposed to model 
{\it adiabatic} inspirals of large-mass-ratio
binaries~\cite{Hughes:1999bq,Hughes:2001jr,Glampedakis:2002cb,Hughes:2005qb,Fujita:2009us,Sundararajan:2007jg,Sundararajan:2008zm,Sundararajan:2008bw}.   
More recently, time domain solutions of the Regge-Wheeler-Zerilli
equations (RWZE)~\cite{Regge:1957td,Zerilli:1970se} with a particle source 
term have been used as a building block of a hybrid method that models 
{\it non-adiabatic} inspirals and the transition
inspiral-plunge-ringdown, shortly {\it
  insplunge}~\cite{Nagar:2006xv,Damour:2007xr}.  
The method combines the wave generation algorithm with an
analytical, effective-one-body prescription for the particle
dynamics. 
The conservative Hamiltonian equations of motion for the particle are
augmented by a radiation-reaction term, $\hat{\cal F}$ linear in the symmetric
mass-ratio, i.e.~of order $\cal{O}(\nu)$. In this paper, following standard
practice in Numerical Relativity, we define the mass ratio as a quantity larger
than $1$, that is $q=M/\mu$, where $\mu$ is the mass of the smaller black hole
and $M$ the one of the larger black hole. In our discussion, we will also 
use the quantity $\nu\equiv 1/q=\mu/M$, that we address, with a slight abuse
of language, ``symmetric'' mass ratio~\footnote{The actual symmetric mass ratio of
a system of two masses $m_1$ and $m_2$ is defined as $\nu= m_1 m_2/(m_1+m_2)^2$.
In the large-mass-ratio limit $q=m_1/m_2\gg 1$ 
this corresponds to $m_1m_2/(m_1+m_2)\approx m_2 = \mu$ and 
$(m_1+m_2)\approx m_1= M$ so that $\nu\to \mu/M$.}. 
In the following, we will often set $M=1$, which allows to identify $\nu$
with $\mu$. The radiation reaction is built from the
factorized and resummed PN waveform for circular orbits of~\cite{Damour:2008gu}.
The hybrid RWZE/EOB method proved to be a valuable ``binary black-hole
laboratory'' and led to (i)~a detailed analysis of multipolar
merger/ringdown waveforms~\cite{Bernuzzi:2010ty}; (ii)~an improved 
analytical description of horizon-absorbed fluxes~\cite{Bernuzzi:2012ku};
(iii)~an improved EOB waveform~\cite{Bernuzzi:2010xj,Damour:2012ky}; and (iv)~an
accurate calculation of the gravitational recoil
(kick)~\cite{Bernuzzi:2010ty,Bernuzzi:2011aj}. Remarkably, the
perturbative kick calculation can be analytically extrapolated to
finite mass ratios and provides quantitative answers also for that
case~\cite{Nagar:2013sga}.   
Most of the results mentioned above rely on an accurate time-domain
RWZE solver that employs the hyperboloidal layer method to include
null infinity in the computational
domain~\cite{Bernuzzi:2011aj,Zenginoglu:2010cq,Zenginoglu:2009ey}. 
The hybrid, perturbative method has been applied to the Kerr
case in~\cite{Barausse:2011kb,Taracchini:2013wfa,Taracchini:2014zpa}, where the TE 
time domain solver of~\cite{Krivan:1997hc,Sundararajan:2007jg} is employed.

In the remainder of this paper, we extend the work
in~\cite{Bernuzzi:2010ty,Bernuzzi:2010xj,Bernuzzi:2011aj} to include
the spin of the massive black hole. We characterize the multipolar
waveform at null infinity, quantify finite radius extraction and
extrapolation uncertainties, and investigate the performance of 
analytical EOB prescriptions for the radiation reaction during the insplunge as
a function of the black hole spin. 
For the first time, we compute the gravitational wave energy flux
into the black hole during the insplunge using a time-domain formalism, proposed 
by Poisson~\cite{Poisson:2004cw}.
Finally, we present a numerical iterative method to compute  
gravitational wave fluxes at scri and at the horizon that are consistent to first 
order in the symmetric mass ratio. The method is applied to a rapidly spinning
case for which the analytical EOB prescription is inaccurate.
We discuss significant differences between the use of the consistent radiation 
reaction and the EOB one as well as the effect of including horizon-absorbed fluxes.

This paper is organized as follows. 
In Sec.~\ref{sec:coords} we introduce a new, horizon-penetrating,
hyperboloidal foliation of Kerr building
on~\cite{Bernuzzi:2011aj,Zenginoglu:2010cq,Zenginoglu:2009ey}, 
and compare the new coordinates to those in~\cite{Racz:2011qu,Harms:2013ib}.
In Sec.~\ref{sec:TE} the inhomogeneous 2+1 TE is rewritten using these new
coordinates. The derivation of the point particle source term is
presented in detail. 
In Sec.~\ref{sec:num} we summarize the numerical method employed in
the {\it Teukode}. In particular, we discuss the advantages of the new
coordinates for the numerics and the strategies to
represent the particle as a Dirac-distribution. 
In Sec.~\ref{sec:dyn_RR} we describe the EOB dynamics used for the
nonadiabatic insplunge experiments and compare it with the one
of~\cite{Barausse:2011kb,Taracchini:2013rva,Taracchini:2014zpa}.
In Sec.~\ref{sec:test} we assess the implementation by considering simple
geodetic motion (circular orbits and radial plunge), insplunge
waveforms for the nonrotating background, and self-convergence tests for waveforms. 
In Sec.~\ref{sec:inspl} we present the main results on large-mass-ratio
insplunge waveforms. First, we study  the consistency between two
different analytical prescriptions for the EOB fluxes and the
one numerically computed by solving the 2+1 TE for the given dynamics.  
Second, we present new multipolar merger waveforms at
scri for various values of the background spin up  to $|\ha|=0.9999$,
and discuss the structure of amplitudes in relation to the particle
dynamics. Third, we quantify finite radius extraction errors and discuss the
performance of the extrapolation procedure commonly employed in
numerical relativity when applied to our setup,
e.g.~\cite{Hinder:2013oqa}.  
In Sec.~\ref{sec:horabs} we present the calculation of the
energy absorbed by the black hole during the insplunge.  
In Sec.~\ref{sec:sf_flux} we describe the self-consistent numerical method for the GW
fluxes. A complete calculation for a case study of
a rapidly rotating hole is presented. 
Conclusions are given in Sec.~\ref{sec:conc}. 
In \ref{app:DataTables} we collect in tables quantitative information 
on the used dynamics and the produced waveforms.
\ref{app:ham} reviews the Hamiltonian formulation of the particle dynamics. 
\ref{app:nocirc} presents a leading-order analysis 
of next-to-quasi-circular effects in the quadrupolar waveform.

Geometric units ($c=G=1$) are employed throughout the paper.

\section{Horizon-penetrating, hyperboloidal foliations of Kerr spacetime}
\label{sec:coords}

\begin{figure}[t]
  \centering
   \includegraphics[width=0.49\textwidth]{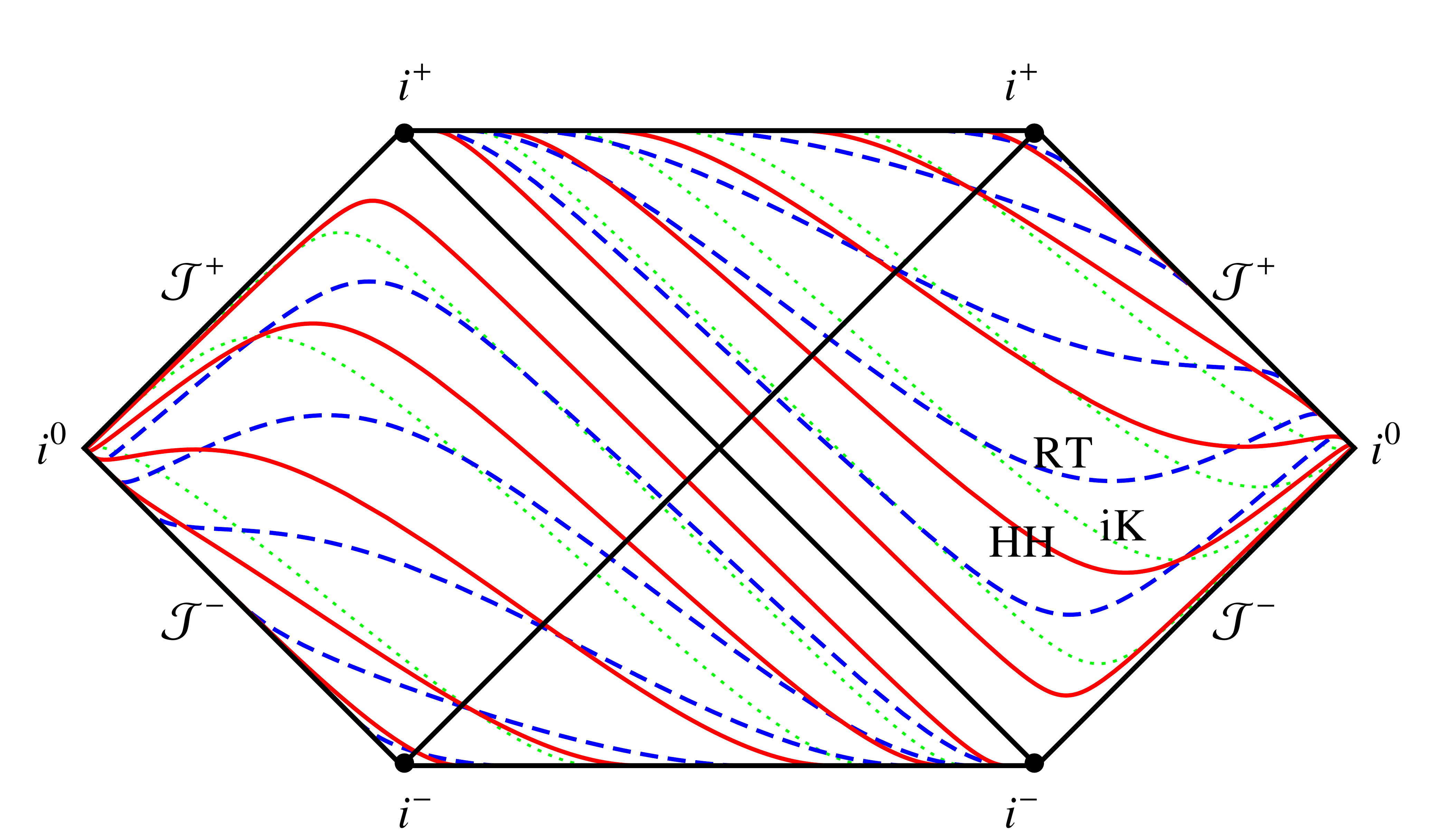}
  \caption{Foliations of Schwarzschild spacetime ($a=0$): ingoing Kerr (iK; green dotted), Racz--T\'oth
    (RT; blue dashed), and horizon-penetrating-hyperboloidal (HH; red solid) coordinates with $S=10$.}
  \label{fig:confdiag}
\end{figure}

It has long been argued that hyperboloidal foliations should have favorable properties for numerical
calculations of outgoing gravitational radiation \cite{Eardley:1978tr}. Until recently, however, it was 
not clear how to use hyperboloidal surfaces in black hole perturbation theory. An important step was
presented by Moncrief in a talk \cite{Moncrief:2000}, wherein he showed how to fix the coordinate location 
of null infinity (scri-fixing) and gave an example in Minkowski spacetime (see also~\cite{Husa:2002zc}). 
This construction was later used in numerical studies in Minkowski spacetime \cite{Fodor:2003yg}.
For black hole spacetimes there were no similarly convenient coordinates available.

A general framework for the construction of explicit, hyperboloidal coordinates with scri-fixing 
for stationary, weakly asymptotically flat spacetimes (including black hole spacetimes) was presented in
\cite{Zenginoglu:2007jw}. The idea is to introduce a new time coordinate with a height function 
that satisfies certain asymptotic properties, and to introduce a compactifying coordinate
in the radial direction along with an \emph{explicit} conformal factor.  The transformation from standard
coordinates $\{t,r\}$ to hyperboloidal coordinates $\{\bar{t},\bar{r}\}$ can be written as
\be \label{eq:general} t = \bar{t} + h(\bar{r}), \qquad r = \frac{\bar{r}}{\Omega(\bar{r})} \,,\ee
where $h(\bar{r})$ is the height function and the \emph{explicitly prescribed} $\Omega(\bar{r})$ 
acts both as a radial compactification and a conformal factor. 
Asymptotic conditions derived in \cite{Zenginoglu:2007jw} make sure that the resulting metric is regular. 
This method has been successfully demonstrated in various examples and today there are many 
choices available for $h(\bar{r})$ and $\Omega(\bar{r})$ in Minkowski, Schwarzschild, Reissner--Nordstr\"om, and Kerr spacetimes. 
How these functions are chosen beyond their asymptotic behavior is important 
for the efficiency of the related numerical computation (see Sec.~\ref{sec:num:coords}).

In the following, we discuss two such choices for the Kerr family. 
The first one has been constructed by Racz and T\'oth in \cite{Racz:2011qu} 
(``RT'' coordinates hereafter) and has been used in a numerical
study of tail decay rates in Kerr spacetime. The second one follows the general ideas of 
\cite{Bernuzzi:2011aj} (``HH" for horizon-penetrating, hyperboloidal) and has been used in numerical
studies of quasinormal mode behavior in nearly extremal Kerr spacetimes in \cite{Yang:2013uba}. 
Both coordinate systems start from the ingoing Kerr metric, which we  review now.

Consider the Kerr metric in Boyer--Lindquist (BL) coordinates $\{t,r,\theta,\phi\}$ 
\begin{align} 
\label{bl_metric}
g_{\rm{BL}} &= -\left(1-\frac{2Mr}{\Sigma}\right) dt^2 - \frac{4 a
  M r}{\Sigma}\sin^2\theta \,dt\,d\phi +
\frac{\Sigma}{\triangle}\, dr^2 \\
&+\Sigma
\,d\theta^2+\left(r^2+a^2+\frac{2Ma^2r\sin^2\theta}{\Sigma}\right)\,\sin^2\theta\,d\phi^2\ ,
\non 
\end{align}
where the two parameters are the mass of the Kerr spacetime $M$ and its angular momentum
parameter $a$, so that its angular momentum is given as $j=a M$. 
We further have $\Sigma \equiv r^2 + a^2 \cos^2\theta$ and $\Delta \equiv r^2 +
a^2 - 2Mr=(r-r_+)(r-r_-)$. The ingoing Kerr (iK) coordinates $\{\tilde{t},r,\theta,\varphi\}$ 
are obtained through the following transformation
\be
 d \tilde{t} = dt + \frac{2M r}{\Delta} dr, \qquad d\varphi =
d\phi + \frac{a}{\Delta} dr \ ,
\ee 
i.e.
\be
\label{eq:iK} 
\tilde{t} =t-r + \int \frac{a^2+r^2}{\Delta}\,dr, \qquad 
\varphi = \phi + a \int \frac{dr}{\Delta} \ .
\ee
The resulting Kerr metric reads
\begin{align}
\label{eq:ds2_iK} g_{\rm{iK}} &= -\left(1-\frac{2Mr}{\Sigma}\right)
d\tilde{t}^2 - \frac{4 a M r}{\Sigma} \sin^2\theta
\,d\tilde{t}\,d\varphi - 2a \sin^2\theta
\left(1+\frac{2Mr}{\Sigma}\right) \,dr d\varphi+ \\ 
&+\frac{4Mr}{\Sigma}\,d\tilde{t} dr +
\left(1+\frac{2Mr}{\Sigma}\right)dr^2 + \Sigma d\theta^2 +
\left(r^2+a^2+ \frac{2Ma^2r \sin^2\theta}{\Sigma} \right)\sin^2\theta
d\varphi^2 . \non
\end{align}
This representation of the Kerr metric is the starting point for both the RT and HH coordinates,  
which we discuss next.

\subsection{RT coordinates}
\label{sec:coords:RT}

The RT coordinates presented in~\cite{Racz:2011qu} are very similar to Moncrief's scri-fixing 
coordinates in Minkowski spacetime~\cite{Moncrief:2000,Husa:2002zc, Fodor:2003yg} but with 
an additional, necessary term to satisfy the asymptotic properties presented in~\cite{Zenginoglu:2007jw}. 
Denoting the time and space RT coordinates by $\{T,R\}$, the transformation from the ingoing
Kerr coordinates reads
\be
\label{eq:RT} 
\tilde{t} = T - 4 M
\ln\left(1-R^2\right)+\frac{1+R^2}{1-R^2}\ ,
\quad
r = \frac{2R}{1-R^2} \ .
\ee
So the height function and the conformal factor are given by
\be\label{eq:RThom} h(R) = \frac{1+R^2}{2 \Omega} - 4 M \ln 2\Omega \ ,   
\qquad \Omega(R) = \frac{1-R^2}{2}\,. \ee
Note that the height function blows up at infinity, where the conformal factor vanishes, in a suitable
way. The first term in the height function is the same term as in Minkowski spacetime; the second
term is needed due to the presence of the black hole.
The resulting hyperboloidal foliation of Kerr spacetime is horizon penetrating and smoothly reaches scri at
$R=1$. The event horizon $R_+$ in the new radial coordinate $R$ is located at  
\be
R_+ = \frac{2 \sqrt{2 M \sqrt{M^2-a^2}-a^2+2 M^2+1}-2}{2
   \left(\sqrt{M^2-a^2}+M\right)} \ .
\ee
Figure~\ref{fig:confdiag} shows the time surfaces of the RT coordinates in a 
conformal diagram for the non-rotating $a=0$ case (blue lines). We mention that RT coordinates can be
modified to allow a prescribed scri position~\cite{Harms:2013ib}.

\subsection{HH coordinates}
\label{sec:coords:HH}

We present a coordinate system which includes a free parameter and is more efficient for numerical
calculations as we argue in Sec.~\ref{sec:num:coords}. 

An intuitive way to construct hyperboloidal coordinates is to demand invariance of the coordinate
expression for outgoing characteristics in spatially compactifying coordinates~\cite{Bernuzzi:2011aj}.
This requirement is beneficial for numerical purposes because, for a prescribed choice of spatial
compactification, the outgoing characteristic speeds do not impose strong restrictions on the 
allowed time step due to the Courant-Friedrich-Lewy (CFL) limit. We are mainly interested in the
asymptotic behavior of outgoing null surfaces and therefore ignore the angular dependence
by setting $a=0$. 

The outgoing radial characteristics for vanishing specific angular momentum in the ingoing Kerr
coordinates are given by $u= \tilde{t} - \left(r + 4M \ln(r-2M)\right)$.
For a further simplification, we drop the subtraction term in the logarithm that is only relevant near
the horizon. Denoting the time and space HH coordinates by $\{\tau,\rho\}$, our requirement becomes
\be
\label{eq:construction_condition}
\tilde{t} - \left(r + 4M \ln r\right) \stackrel{!}{=} \tau -
\left(\rho + 4M \ln\rho\right) \ .
\ee 
For any choice of spatial compactification through $\rho$, the above requirement determines the
foliation. We choose the simplest conformal factor with a variable scri location $S$ and set
\be \label{eq:HH}
h(\rho) = \frac{\rho}{\Omega} - \rho - 4 M \ln \Omega, \qquad 
\Omega(\rho) = 1-\frac{\rho}{S}\,.
\ee
Note that the asymptotic behavior of these functions near scri is similar to the one given
for the RT coordinates in \eqref{eq:RThom}. 
The event horizon $\rho_+$ in the new radial coordinate $\rho$ is located at
\be
\rho_+ = \frac{ a^2 S + M S^2 + \sqrt{M^2 S^4 - a^2 S^4} }
	      { a^2 + 2 M S + S^2 }\,.
\ee
We will indicate with HH$_S$ these coordinates
with a specific choice of $S$, e.g.~HH$_{10}$ refers to $S=10$.
The foliation HH$_{10}$ is shown in a conformal diagram in Fig.~\ref{fig:confdiag} (red lines) 
for $a=0$.  It is qualitatively similar to the RT coordinates~\cite{Racz:2011qu} with
the main differences that the location of null infinity can be chosen freely and the outgoing radial
characteristic speeds are similar to those of the ingoing Kerr coordinates.

\subsection{Advanced and retarded time coordinates}
\label{sec:coords:u}

Gravitational waves propagate along null geodesics. The structure of null geodesics 
in Kerr spacetime is rather complicated, but their approximations by Schwarzschild null geodesics
is sufficient for our purposes. The retarded and advanced time coordinates in Schwarzschild spacetime,
$u$ and $v$, are defined by $u=t-r_*$ and $v=t+r_*$, where $r_*=r+2M\ln(r/2M-1)$ is the Schwarzschild
tortoise coordinate. We use these coordinates in Kerr spacetime to connect the particle's dynamics with 
the measured gravitational radiation. This approximation agrees with the general practice in numerical
relativity, where null geodesics in a binary black-hole spacetime are approximated by their
Schwarzschild counterparts for extrapolating gravitational waveforms \cite{Hinder:2013oqa}.

Here, we give the retarded and advanced time coordinates in the horizon-penetrating, hyperboloidal
coordinates used in the simulations. We get for the retarded coordinate
\be
\label{eq:uHH}
u(\tau,\rho) = \tau -\rho - 4M\ln\left(\frac{S\rho+2M\rho-2MS}{S}\right) +
2M\ln 2M  \ ,
\ee
and for the advanced coordinate
\be
\label{eq:uHH_adv}
v(\tau,\rho) = \tau + \rho\, \frac{S+\rho}{S-\rho} - 4 M \ln\left(
   \frac{S-\rho}{S} \right)  - 2 M \ln(2 M) \ .
\ee
The constant term $2M\ln 2M$ comes from different conventions in the tortoise coordinate.
All plots in this paper employ the above coordinates for visualizing waves and fluxes at scri
or the horizon. This allows a direct comparison of dynamical quantities
(e.g. the particle's orbital frequency) with the measured wave signal because 
the BL-$t$ used for the particle's dynamics can 
be identified with the retarded time at scri and the advanced time at the horizon.

\section{Inhomogeneous 2+1 Teukolsky equation with a particle source}
\label{sec:TE}
In this section we briefly outline the main steps in the derivation of the 2+1 Teukolsky equation 
in horizon-penetrating, hyperboloidal coordinates. The calculation of the stress-energy tensor with a
particle source term is presented in detail. Formulas for gravitational wave fluxes at scri and 
the horizon in terms of 2+1 fields are collected.

\subsection{The Teukolsky equation}

The Teukolsky equation describes the evolution of perturbations of certain Weyl tensor components 
on a Kerr background~\cite{Teukolsky:1973ha}. The equation is derived using the Newman--Penrose 
formalism~\cite{Newman:1962} which relies on the choice of a tetrad and a coordinate system.
In the original calculation Teukolsky used the Kinnersley tetrad in BL coordinates which reads
\be
\label{eq:Kinnersley}
\ell^\mu  =\frac{(r^2 + a^2, \Delta, 0,  a)}{\Delta}  \, ,
\quad 
n^\mu = \frac{(r^2 + a^2, -\Delta, 0,  a)}{2 \Sigma} \, ,
\quad 
m^\mu =  \frac{( \ii a \sin\theta, 0, 1,
  \ii\csc\theta)}{\sqrt{2} (r + \ii a \cos\theta)} \ .
\ee
In practice the equivalent, rationalized versions of $m^\mu$ and its complex conjugate $m^{\mu *}$
are preferable, e.g.
\be
  m^{\mu} = \dfrac{ (r-\ii a \cos\theta) }{\sqrt{2} \Sigma}   \left(
    \ii a\sin\theta, 0, 1, \ii\csc\theta \right) \ .
\ee

One straightforward method to obtain the Teukolsky equation in horizon-penetrating, hyperboloidal
coordinates is to substitute the transformation formulas \eqref{eq:general} directly into the equation 
as given by Teukolsky~\cite{Teukolsky:1973ha}. Subsequently, the unknown variable is rescaled 
for regularity of the coefficients at the horizon and at null infinity~\cite{Zenginoglu:2011jz,Harms:2013ib}.
The field $\psi$ with spin weight $s$ behaves as $\Delta^s$ towards the horizon 
and as $r^{2s+1}$ towards null infinity. Therefore, the rescaling goes as
\be
\label{eq:rescale}
\psi \mapsto \Delta^{-s} r^{-1} \psi \, .
\ee
This is the approach followed in \cite{Harms:2013ib} for the transformation of the homogeneous
Teukolsky equation into RT coordinates.

For this paper we found that deriving the Teukolsky equation from scratch using the new coordinates,
while equivalent to the above transformation, yields better results in the source term 
because the cancellations for regularity at the horizon are implied automatically.
To this end, we first perform a null rotation that corresponds to the rescaling given above. 
For example, for regularity at the horizon a null rotation with $\Lambda=\Delta$ can be performed 
as in~\cite{Campanelli:2000nc}. The rotation gives the tetrad fields
\be
\label{eq:Campanelli}
\ell^\mu =  (\Delta + 4 M r, \Delta, 0, 2 a) \, ,
\quad 
n^\mu = \frac{(1,-1,0,0)}{2 \Sigma} \ .
\ee
Note that, under a $\Lambda$ rotation,  $\ell^\mu \rightarrow \Lambda \ell^\mu $, 
$n^\mu \rightarrow \Lambda^{-1} n^\mu $ and $m^\mu$ stays unchanged. 
With $\Lambda=\Delta$ and  $s=-2$, the Weyl scalar 
$\psi_4 = - n^\alpha m^{*\beta} n^{\gamma} m^{*\delta} C_{\alpha\beta\gamma\delta}$ 
transforms as $\psi_{4} \to \Delta^{-2} \psi_{4}$. For general $s$, the null rotation with 
$\Lambda=\Delta$ corresponds to a rescaling with $\Delta^{-s}$. To obtain regularity at scri 
(in the homogeneous part of the equation)
a subsequent rescaling by $r^{-1}$ or an additional null rotation must be performed. 
Then the tetrad is transformed to horizon-penetrating, hyperboloidal coordinates and 
used to calculate the coefficients and the source term for the Teukolsky equation.
A more general approach can also be taken by incorporating free coordinate functions 
in the tetrad before the derivation of the TE \cite{Zenginoglu:2008uc}. 

Finally the Teukolsky equation is transformed in 2+1 form separating each Fourier
$m$-mode in the azimuthal direction. The resulting equation has the general form
\begin{align}
 \label{eq:TE} 
 &  C_{\tau\tau} \partial_{\tau\tau}\psi +  C_{\tau\rho} \partial_{\tau\rho} \psi + 
 C_{\rho\rho}  \partial_{\rho\rho}\psi + C_{\theta\theta}  \partial_{\theta\theta}\psi + 
 C_{\tau}  \partial_\tau \psi + C_{\theta} \partial_{\theta}\psi  + C_{\rho}  \partial_\rho\psi  + 
 C_{0} \psi = S_s \ , 
\end{align}
with coefficients $C(\rho,\theta; m,s)$ depending on the background
coordinates, the spin weight $s$, and the azimuthal mode-index $m$.
The index $m$ in the variable $\psi_m$ has been suppressed for brevity.  
The general formulas for the coefficients under time transformation, 
spatial compactification, and rescaling can be found in \cite{Zenginoglu:2009hd}.

\subsection{Particle source term}
\label{sec:src:tmunu}

The calculation of the stress-energy tensor for a particle source is well documented for the 
frequency-domain Teukolsky equation (see \cite{Sasaki:2003xr}). 
For time domain applications, however, we could not find a complete description. 
Therefore we present the calculation in detail for a Hamiltonian formulation 
of the particle dynamics (see~\ref{app:ham}). 

The source term $S_s$ depends on the spin weight $s$, the background metric, 
and the stress-energy tensor, $T_{\mu\nu}$, of the matter perturbation. 
The general form of $S_s$ is given in \cite{Teukolsky:1973ha}. 
Here we discuss gravitational perturbations ($s=\pm2$). In BL coordinates, 
\begin{align}
\label{eq:S-2}
S_{-2} &= 8 \pi \Sigma (r- \ii a\cos\theta)^4 T_4 \ , \\
\label{eq:S+2}
S_{+2} &= 8 \pi \Sigma T_0 \ ,
\end{align}
where $T_4$ and $T_0$ are expressions involving contractions of the stress-energy tensor 
with tetrad vector fields and their first and second partial derivatives with respect
to the background coordinates. Specifically, the contractions involved are
\be
T_{mm*} \equiv T_{\mu\nu} m^{\mu} m^{*\nu }\, , \quad
T_{nm*}\equiv T_{\mu\nu} n^{\mu} m^{*\nu }\, ,  \quad
T_{nn}\equiv T_{\mu\nu} n^{\mu} n^{\nu}\, .
\ee
For example, the term $T_4$ is given by
\begin{align}    
\label{eq:T4}
  T_4 &=(\NPDelta + 3\gamma -\gamma^* + 4\mu + \mu^*)(\delta^*-2\tau^*+2\alpha)T_{n m^*} \\
  &- (\NPDelta + 3\gamma -\gamma^* + 4\mu + \mu^*)(\NPDelta + 2\gamma-2\gamma^*+\mu^*)T_{m^* m^*}\non\\
    &+(\delta^*-\tau^*+\beta^*+3\alpha+4\pi)(\NPDelta + 2\gamma + 2\mu^*)T_{nm^*}\non\\
  &-(\delta^*-\tau^*+\beta^*+3\alpha+4\pi)(\delta^*-\tau^*+2\beta^*+2\alpha)T_{nn}\non
  \ ,
\end{align}
where $\NPDelta,\delta,\gamma,\mu,\tau,\alpha,\beta$ are the
complex Newmann-Penrose operators, e.g.~\cite{Teukolsky:1973ha}. 
In Eq.~\eqref{eq:T4} $\NPDelta=n^\a\p_\a$ (the underbar is
introduced to distinguish it from our previous definition of $\Delta$) and
$\delta^*= m^{*\a}\p_\a$ are differential operators that depend on the
coordinates. They do not commute, that is, $[\NPDelta,\delta^*]\neq 0$.  
For example, using the Kinnersley tetrad in BL coordinates they read
\begin{align}
\NPDelta & = 
\dfrac{1}{2\Sigma}
\left( 
\left(r^2+ a^2\right)\p_t - 
\Delta\p_r + 
a \ \p_\phi\right)\\
\delta^*& =
\frac{1}{\sqrt{2}\Sigma}\left( (a^2 \cos\theta \sin\theta - \ii \ r \ a \sin\theta) \p_t 
+(r+\ii \ a \cos\theta) \p_\theta + \left(a \frac{\cos\theta}{\sin\theta}- \ii \frac{r}{\sin\theta} \right)\p_\phi 
\right) \, .
\end{align}
The related 2+1 decomposed operators are obtained by the substitution $\partial_\phi \rightarrow \ii \ m $.

The stress-energy tensor for a point particle with mass $\mu$ can be written as 
\be
\label{eq:Tmunu:general}
T_{\mu\nu} = \mu \int \dfrac{d\lambda}{\sqrt{-g}} \ u_\mu
\,u_\nu\ \delta^4(x^\a - X^\a(\lambda)) 
\ee 
with $\lambda$ the proper 
time, $X^\a(\lambda)$ the particle's worldline and
$u^\a(\lambda)=dX^\a/d\lambda$ the $4$-velocity.
Assuming the particle's motion is described by Hamiltonian dynamics
(\ref{app:ham}), we write the 4-velocity in terms of the reduced
momenta, $u_\a=(-\H,p_i)$, and the coordinates as $x^\a=(t,q^i)$.
Here, $t\equiv x^0$ is a generic time coordinate, not necessarily 
the BL time. Replacing the affine parameter $\lambda(t)$ with $t$
and integrating Eq.~\eqref{eq:Tmunu:general} one gets 
\be
\label{eq:Tmunu} 
T_{\mu\nu} = \dfrac{\mu}{ \sqrt{-g}} \frac{d\lambda}{dt}
p_\a p_\b \delta^3(x^i - q^i(t)) \ .
\ee 
For the calculation of the 2+1 source term the equation above is
mode decomposed. In BL coordinates, 
we write $T_{\mu\nu}=\sum_m T^m_{\mu\nu} e^{\ii m\phi}$, and using  
\be
\delta(\phi-\phi(t)) = (2\pi)^{-1}\sum_m e^{\ii m(\phi-\phi(t))} \ , 
\ee 
the $T^m_{\mu\nu}$ components read (dropping the $m$-index for brevity)
\begin{subequations}
\label{eq:Tmunu21}
\begin{align} 
T_{00} & = \frac{\mu A\ \H^2}{2\pi\Sigma\sin\theta\, (\H-\omega p_\phi)}
 \ \delta(r-r(t)) \delta(\theta-\theta(t))
e^{-\ii m\phi(t)} \ , \\ 
T_{0i} & = \frac{\mu A\ (-\H) p_i}{2\pi\Sigma\sin\theta\, (\H-\omega p_\phi)}
 \ \delta(r-r(t)) \delta(\theta-\theta(t)) e^{-\ii m\phi(t)} \ , \\ 
T_{ij} & =
\frac{\mu A\ p_i p_j}{2\pi\Sigma\sin\theta\, (\H-\omega p_\phi)} \
\delta(r-r(t)) \delta(\theta-\theta(t)) e^{-\ii m\phi(t)} \,,
\end{align} 
\end{subequations}
where we have specified to the Kerr metric, $\sqrt{-g}=\Sigma\sin\theta$, and used 
$d\lambda/dt=A/(\H-\omega p_\phi)$ (see Eq.~\eqref{eq:ginv}).

Inspecting Eq.~\eqref{eq:T4} and Eq.'s~\eqref{eq:Tmunu21}, we define the 
following functions 
\begin{subequations}
\begin{align}
 L(t,r,\theta) & = \frac{\mu}{2\pi\Sigma\sin\theta}
 \frac{d\lambda}{dt} = \frac{\mu}{2\pi\Sigma\sin\theta}
 \frac{A}{\hat{H}-\omega p_\phi} \\
 M(t,r,\theta) & = m^{\mu *} p_\mu
 =  \dfrac{(r+\ii a \cos\theta) }{\sqrt{2} \Sigma}  
   \left( \ii a \sin\theta \ \hat{H} +  p_\theta - \ii \frac{p_\phi}{\sin\theta}
     \right) \\
 N(t,r,\theta) & = n^{\mu}   p_\mu
 = \dfrac{1}{2 \Sigma}  \left( -\H(r^2 + a^2) -\Delta p_r +  a p_\phi \right) \ ,
\end{align}
\end{subequations}
and use them to express the contractions of the stress-energy tensor with
the tetrad, 
e.g.~$T_{nm*} = L \ M  \ N \ \delta^2(...) \ e^{(...)}$.
The quantity $T_{4}$ in Eq.~\eqref{eq:T4} is then written in terms of
$L,M,N$ and their derivatives using the 
chain rule. In practice, in the code, the algebraic complexity of the
source is greatly reduced by computing and storing $L,M,N$ and
derivatives, and combining them.  

In order to compute the derivatives in Eq.~\eqref{eq:T4}
one needs to make a choice because 
the delta functions formally identify the background coordinates with
the particle's position coordinates. The result is theoretically
independent of the particular choice, but the explicit expressions 
can differ and, numerically, there may be differences due to
truncation errors (see discussion in~\cite{Nagar:2006xv}). Also, 
one has to perform the calculation consistently. We
leave all the background coordinates in the $L,M,N$ functions
unchanged and consider functions
of time only the reduced momenta $p_i(t)$, the Hamiltonian $\hat{H}(t)$
and the trajectory $q^i(t)$. 
The time dependence in the momenta and
the Hamiltonian is assumed to account for dissipative forces (radiation
reaction). The time derivatives of the $q^i(t)$ are
systematically substituted with the right-hand-side (r.h.s.) of the Hamiltonian equation
of motion, so they do not appear explicitly but only
$\dot{p}_i(t),\dot{\hat{H}}(t),\ddot{p}_i(t), \ddot{\hat{H}}(t)$ remain.

We want to mention here a difference between the calculation of the TE
source and the RWZE source. Differently from the TE, in the RWZE case it
is possible to remove 
explicit time derivatives from the source by using the (linearized)
Bianchi identities in the metric perturbation framework. Specifically
and referring to equations in Ref.~\cite{Martel:2005ir}, this can be
accomplished by substituting the Eq.~(4.21) into the even parity
source in Eq.~(4.27) and the Eq.~(5.12) into the odd parity
source in Eq.~(5.16). The substitution is used in the calculation
of~\cite{Nagar:2005ea} although not explicitly stated.

Let us comment on the use of HH coordinates. First the NP
operators in Eq.~\eqref{eq:T4} and the tetrad in
Eq.~\eqref{eq:Campanelli} have to be rewritten in the HH-coordinate
system. For iK coordinates explicit expressions from
\cite{Campanelli:2000nc} can be used as a starting point
to apply the transformation of Eq.~\eqref{eq:HH}. 
Because the particle's trajectory is usually computed in
BL coordinates we also have to apply the coordinate transformations to the
trajectory and to the momenta (i.e.~to $T^{\mu\nu}$). We emphasize that the
rewriting of the source term in HH-coordinates as sketched above is important
in our approach, even though the terms $T_{4,0}$ are invariant under coordinate
transformations (being tetrad scalars). The reason is that in the code 
the particle event at each time step must be located in hyperboloidal coordinates,
say $\rho(\tau)$, and any discrete representation of the delta
function (see Sec.~\ref{sec:num:delta}) involves a few grid points around
$\rho(\tau)$. The transformation $t=\tau+h(\rho)$
introduces a grid-point dependent BL-time $t$ at a given evolution slice $\tau$.
Using a trajectory in BL coordinates would mean to have a non-unique
particle's position at the evolution slices $\tau$ of the code. 
Re-writing the source term in the evolution coordinates removes the ambiguities and thus
greatly simplifies the source treatment.

\subsection{Gravitational strain}
\label{sec:TE:gw}

We describe the relation between the master variables of the Teukolsky equation and the
gravitational strain. At scri, the $s=-2$ master variable $\psi_m$ of the 2+1 TE written in HH (or RT)
coordinates is $r \psi_{4\ m}$, 
i.e. the $m$-mode of the Weyl scalar (in the Kinnersley tetrad) describing asymptotically
outgoing radiation multiplied by $r$. The $s=+2$ master variable
$\psi_m$ corresponds instead to the $m$-mode of the Weyl scalar
$r \psi_{0\ m}$ also (referring to the tetrad of eq.~\eqref{eq:Campanelli}). 

The GW strain $h=h_+ - \ii\, h_\times$ is found by integrating the asymptotic relation
\be
\ddot{h} = 2 \psi_4 \ 
\ee
for each $m$-mode. The integration gives $r\,h_m(u,\theta)$ along scri. 
We also compute the (spin weighted with $s=-2$) multipoles $r\,h_{\ell m}(u)$, defined through
\be
\label{eq:RWZ}
r\, h = \sum_{\l=2,m} \ r\, h_{\lm} {}_{-2}Y_\lm(\theta,\phi)
= \sum_{\l=2,m} \sqrt{\frac{(\l+2)!}{(\l-2)!}}\ \Psi_{\lm} \
{}_{-2}Y_\lm(\theta,\phi) \ ,
\ee
by mode-projecting $r\,h_m(u,\theta)$ in the $\theta$-direction. The complex quantities
$\Psi_\lm=\Psi^{\rm (e)}_\lm+\ii \Psi^{\rm (o)}_\lm$ in
Eq.~\eqref{eq:RWZ} are the RWZE variables~\cite{Nagar:2005ea}.
The energy flux at scri is given by
\be
\label{eq:EFlux-SCR}
\dot{E} = \frac{1}{16\pi}\int_{S_2} d\Omega\ |r\,\dot{h}|^2
= \frac{1}{16\pi}\sum_m \int_{-1}^1d\xi\ |r\,\dot{h}_m|^2 \ ,
\ee
where in the last expression we have used $\xi=\cos\theta$ 
and introduced the mode-decomposition
of $h$ to express the flux in terms of the 2+1 fields.
The angular momentum $\vec{J}=(J_x,J_y,J_z)$ flux is given by 
\be
\dot{J}_i = - \frac{1}{16\pi}\Re\left\{\int_{S_2} d\Omega\
  (r\,\dot{h})^* \mathcal{J}_i (r\,h) \right\} \ ,
\ee
where $\mathcal{J}_i$ are the spin 2 quantum mechanical operators, in
particular $\mathcal{J}_z=\p_\phi$. For equatorial
orbits $J_x=J_y= 0$, so the relevant quantity is
\be
\dot{J}_z = \frac{1}{16\pi}\Im\left\{\sum_m m
\int_{-1}^1d\xi\ (r\,\dot{h}_m)^* (r\,h_m)\right\} \ .
\ee
Similarly, the linear momentum $\vec{P}=(P_x,P_y,P_z)$ flux can be
computed from  
\be
\dot{P}_i = \frac{1}{16\pi}\int_{S_2} d\Omega\ n_i |r\,\dot{h}|^2 \ ,
\ee
where
$n_i=(\sin\theta\cos\phi,\sin\theta\sin\phi,\cos\theta)$. For
equatorial orbits $P_z=0$. 

The horizon-absorbed GW energy and angular momentum are defined using
the first law of black-hole mechanics~\cite{Hawking:1972hy,Bardeen:1973gs} 
$\frac{\kappa}{8\pi}\dot{A}_{\rm H}=\dot{M}_{\rm H}-\Omega_{\rm H}\dot{J}_{\rm
  H}$, where $\kappa=(r_+-M)/(r_+^2-a^2)$ is the surface gravity and 
$\Omega_{\rm H}=a/(2Mr_+)$ is the angular velocity of the
horizon. Considering the equations for the horizon generator dynamics,
the variation of the horizon mass and angular momentum can be expressed
as~\cite{Teukolsky:1974yv,Price:1986yy,Poisson:2004cw}, 
\be
\dot{M}_{\rm H} = \frac{1}{16\pi}\int dS\
\sigma^{AB}\mathcal{L}_t \gamma_{AB} \ \ , \ \
\dot{J}_{\rm H} = \frac{1}{16\pi}\int dS\
\sigma^{AB}\mathcal{L}_\phi \gamma_{AB} \ ,
\ee
where $dS$ is the horizon area element of the induced 2-metric, 
$\sigma^{AB}$ is the 'shear tensor' (see Eq.~(3.4) in \cite{Poisson:2004cw}),
and $\mathcal{L}_{t,\phi}$ are Lie derivatives with respect to the Killing
vectors of the background. A similar equation holds for the area
variation $\dot{A}_{\rm H}$, see Eq.~(4.24) in~\cite{Poisson:2004cw}.
The equations above are derived considering a particular coordinate
system on the horizon $(v,X^B)$ ($B=2,3$), where
$v=t+\int dr \Delta^{-1}(r^2+a^2)$ is the advanced time as in \eqref{eq:uHH_adv}
(connected to the ingoing Kerr time coordinate from \eqref{eq:iK} via $v=\tilde{t}+r$), 
and  $X^B=(\theta,\phi-\Omega_{\rm  H}v)$ in terms of BL coordinates.
Reference~\cite{Poisson:2004cw} specifies the final flux equations
for the 2+1 fields, which read
\be
\label{eq:EFlux-HRZ}
\dot{M}_{\rm H} = \frac{r_+^2+a^2}{4\kappa}\sum_m\left[
2 \kappa\int_{-1}^1 d\xi\ |f_{{\rm H}\, m}^+|^2
-\ii m\Omega_{\rm H}
\int_{-1}^1 d\xi\ \left( f_{{\rm H}\, m}^{+\ *}f_{{\rm H}\, m}^{-}
  -f_{{\rm H}\, m}^{+}f_{{\rm H}\, m}^{-\ *}\right)
\right] \ ,
\ee
\be
\label{eq:JFlux-HRZ}
\dot{J}_{\rm H} =-\frac{r_+^2+a^2}{4\kappa}\sum_m\ii m
\left[
\int_{-1}^1 d\xi\ \left( f_{{\rm H}\, m}^{+\ *}f_{{\rm H}\, m}^{-}
  -f_{{\rm H}\, m}^{+}f_{{\rm H}\, m}^{-\ *}\right)
\right] \ .
\ee
The complex quantities $f_{{\rm H}\, m}^\pm$ are defined as integrals of 
the $m$-mode components of the Weyl scalar $\psi_0$ at the horizon
\begin{align}
\label{eq:EFlux-HRZ-Phipl}
f_{{\rm H}\, m}^+(v,\theta) &=-e^{\kappa v}\int_{v}^\infty d v'
e^{-(\kappa-\ii m \Omega_{\rm H}) v'} \psi_{0\ m}(v',r_+,\theta)\\
\label{eq:EFlux-HRZ-Phimi}
f_{{\rm H}\, m}^-(v,\theta) &=-\int_{-\infty}^{v} d v'
e^{\ii m \Omega_{\rm H} v'} \psi_{0\ m}(v',r_+,\theta)
\ .
\end{align}
The Weyl scalar $\psi_{0\ m}$ is understood as the one
defined by the Hawking-Hartle tetrad. 
Our $s=+2$ TE master variable must be divided by $(4r_+(r_+^2+a^2)^2)$ 
because we use the tetrad in Eq.~\eqref{eq:Campanelli} and rescale by $r$.
Note also that $f_{{\rm H}\, m}^+(v)$ depends on the future behavior
of the field. As mentioned in~\cite{Poisson:2004cw}, the formalism is not
yet optimal for 2+1 simulations; a certain drawback for its use in our setup
will be pointed out in Sec.~\ref{sec:horabs}. Nonetheless,
we employ it for the first time (to our knowledge) in this work and leave
the development of a more practical method to the future.

\section{Numerical method}
\label{sec:num}

In Sec.~\ref{sec:disc}, we describe the numerical discretization of the TE 
(see also~\cite{Harms:2013ib}). In Sec.~\ref{sec:num:coords} RT
and HH coordinates are compared in their efficiency. Strategies for implementing the
Dirac $\delta$-functions are discussed in Sec.~\ref{sec:num:delta}. 

\subsection{Discretization} \label{sec:disc}
For numerical integration the 2+1 TE is written as a first-order in
time, and second-order in space system with reduction variables
$u=\{\psi,\ \p_\tau \psi\}$. The domain
$(x,\theta)\in[x_+,x_S]\times(0,\pi)$, where $x$ is the radial
coordinate and $x_+$ ($x_S$) is the location of the horizon (null
infinity), is uniformly discretized with $N_x\times N_\theta$ points. 
The spatial derivatives are
represented by finite differences up to eighth order of accuracy. The
stencils in the radial direction are 
centered in the bulk of the domain and lop-sided/sided at the
boundaries (we also tested ghost points at radial boundaries filled by extrapolation 
and found no advantage). 
The angular grid is staggered and ghost points are
employed to implement the boundary conditions on 
the axis. The ghost points are filled according to the parity
condition $\pi = (-1)^{m+s}$. Artificial dissipation operators are
implemented but not used for our results unless mentioned explicitly. 

A standard fourth order Runge-Kutta integrator is employed for time
advancing the solution. The time
step is chosen according to a CFL condition 
of type $\del t = C_{\rm CFL} \min(h_x,h_\theta)$, 
where $h_x$ is the grid
spacing in direction $x$ and the factor $C_{\rm CFL}$ 
accounts for the maximum coordinate speed of the PDE system. 
Even the most expensive simulations of this work (see Sec.~\ref{sec:inspl}) were
performed on a standard Linux desktop computer using a serial
implementation and the GNU C compiler. For example, EOB
insplunge simulations at the typical resolution $N_x\times
N_\theta=3600\times160$ took about 2 weeks to reach the final time
$5000\,M$. 

The angular integrals for $\ell$-mode projections are computed with the
Simpson rule, while integrals in time with the trapezoidal rule. The
integrals for the absorbed fluxes are calculated in post-processing;
accuracy is thus dependent on the sampling of the output in time and angular
direction. Note that the $f^+_{{\rm H}\, m}$
integrals at a given time $v_*$ are mostly determined by the master function at times 
$(v-v_*)\sim \kappa^{-1}$, due to the exponential function. Therefore 
we restrict the integrations to times $v$ such that
$e^{-\kappa(v-v_*)}>10^{-6}$. Since $\kappa \in[0.25,0.35]$ (for $|\ha| \in[0,0.9]$)
the interval is typically $v\in[v_*-50,v_*+50]$.  
We use Gaussian quadratures in the angular direction interpolating to
$\sim50$ Gauss-Lobatto output points, and manually vary the time
output sampling until results are satisfactory (see Sec.~\ref{sec:test:ciro}).

\subsection{Numerical efficiency of background coordinates}
\label{sec:num:coords}

\begin{figure}[t]
  \centering
   \includegraphics[width=0.49\textwidth]{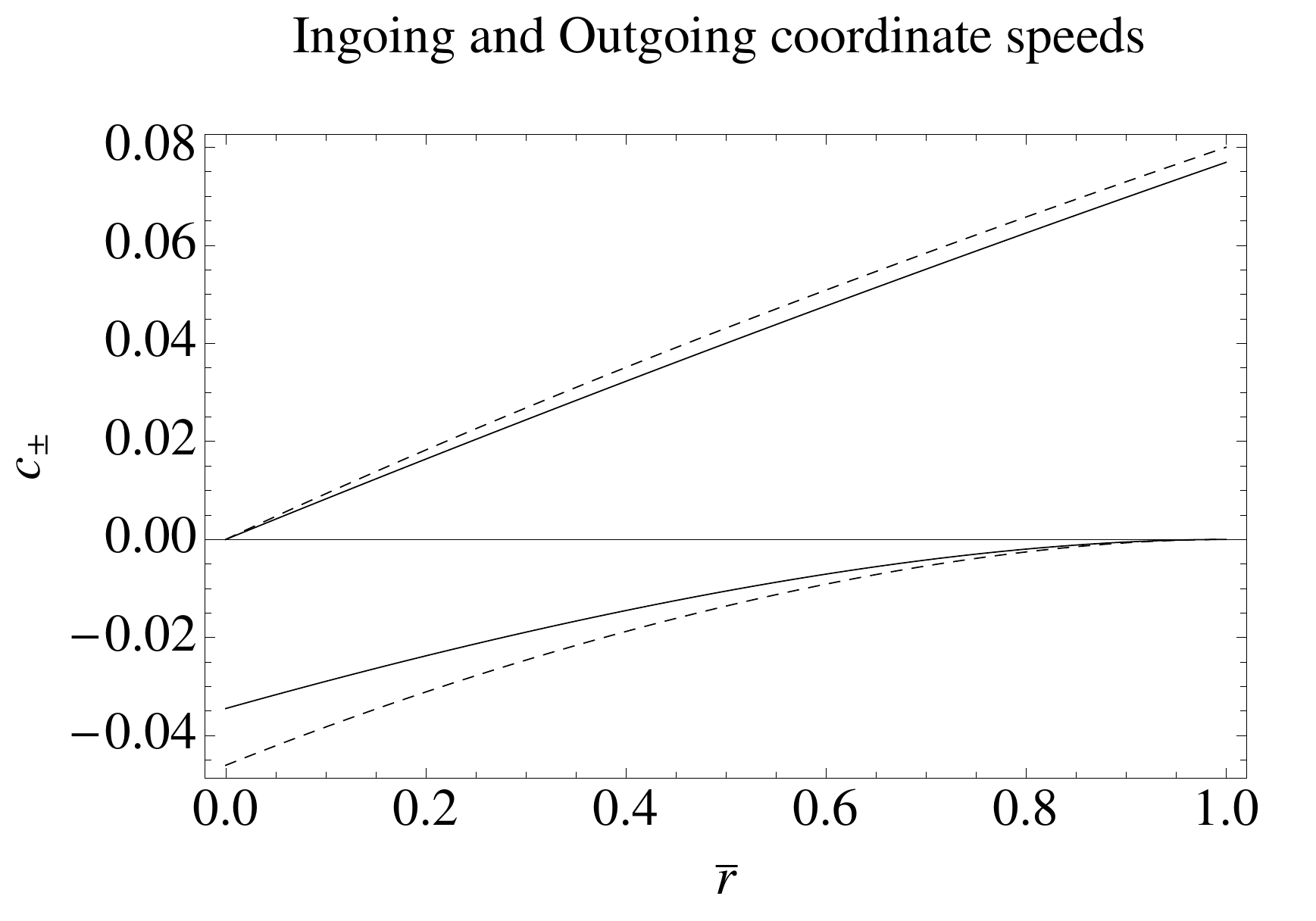}
   \includegraphics[width=0.49\textwidth]{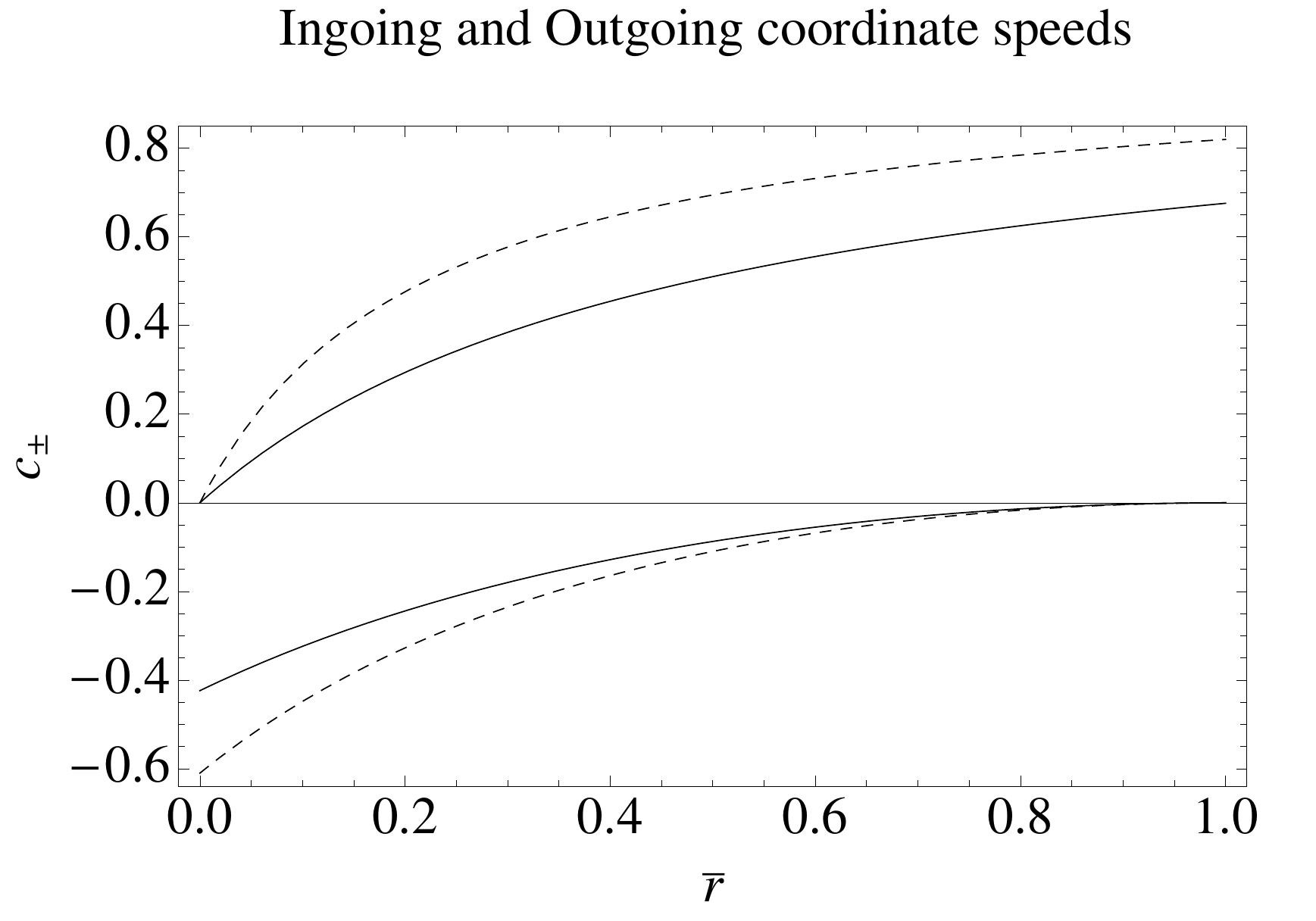}
   \caption{Visualization of ingoing ($c_-$, negative) and outgoing ($c_+$, positive) 
   radial coordinate speeds. Setting $a=0$, 
   on the left panel we use HH$_1$ (solid) and RT (dashed) coordinates and on the right panel 
   we use HH$_{10}$ (solid) and HH$_{20}$ (dashed) coordinates.
    The incoming (with respect to the domain) characteristic speeds 
    vanish at both boundaries (horizon and scri) so no boundary conditions are needed.	
    Note the radial domains are rescaled
    to $[0,1]$ for the comparison.}   
  \label{fig:coorspeeds}
\end{figure}

Both RT and HH coordinates have desirable properties for numerical
treatment: (i)~the horizon and scri are included in the computational
domain, (ii)~outgoing radial coordinate speeds $c_+$ vanish at the
horizon and ingoing radial coordinate speeds $c_-$ vanish at scri, so 
no particular treatment at the boundaries is needed (see Fig.~\ref{fig:coorspeeds}).

The HH$_S$ coordinates, however, are more flexible due to the presence of the
free parameter $S$, and can therefore be tuned to be more efficient than the RT
coordinates. To understand the effect of this parameter on the simulations, it is useful to
view hyperboloidal surfaces as mediating between characteristic and
Cauchy surfaces~\cite{Zenginoglu:2007jw}. 

Geometrically, $S$ is inversely correlated with the mean curvature of the foliation. 
Remember that the asymptotic mean curvature vanishes for Cauchy surfaces and is 
unbounded for characteristic ones. Therefore, a large $S$ gives a more
Cauchy-like behavior, whereas a small $S$ gives a more characteristic-like behavior. 

Numerically, the mean extrinsic curvature is related to the characteristic coordinate speeds and the spatial
wavelength of waves propagating across the grid \cite{Zenginoglu:2007jw}. A large $S$ implies a
low characteristic speed and a small spatial wavelength, whereas a small $S$ implies a
high characteristic speed and a large spatial wavelength. The trade-off is between the time step size
restricted by the CFL condition due to the characteristic coordinate speeds and spatial resolution 
restricted by the wavelengths to be represented. The free parameter allows us to find a good balance
between the time stepping and the spatial resolution requirements.

To find this balance, we compared the performances of RT and HH$_S$ coordinates with
$S=1,10,20$ for $a=0$ in wave scattering numerical experiments (no particle, 
as in~\cite{Harms:2013ib}), geodesic dynamics (Sec.~\ref{sec:test}), and EOB
insplunge simulations (Sec.~\ref{sec:inspl}) using the finite
difference algorithm described in Sec.~\ref{sec:disc}. Note that the RT slices have 
high curvature, which implies that there are hardly any waves 
to be resolved on the grid but the time stepping requirement becomes very restrictive. 
The HH$_1$ and the RT coordinates behave similarly with respect to numerical efficiency. 
Increasing $S$ allows us to use larger time steps, but the number of waves on the spatial grid
also increases. When $S$ is too large (above $S=20$), waves get blue shifted near the 
compactification boundary and artificial dissipation becomes necessary for stability. 
Artificial dissipation not only reduces numerical accuracy, but more importantly, it adds 
a significant computational cost. Therefore it is preferable to choose a value for $S$ that does
not require dissipation.

We found that $S=10$ provides a good balance between time stepping and spatial resolution 
requirements without artificial dissipation. The HH$_{10}$ coordinates give a speed up of $\sim2$ 
with respect to HH$_1$ and RT coordinates. We adopted HH$_{10}$ in all the 
simulations presented in the following.

\subsection{Representation of $\delta$ functions}
\label{sec:num:delta}

\begin{figure}[t]
  \includegraphics[width=0.49\textwidth]{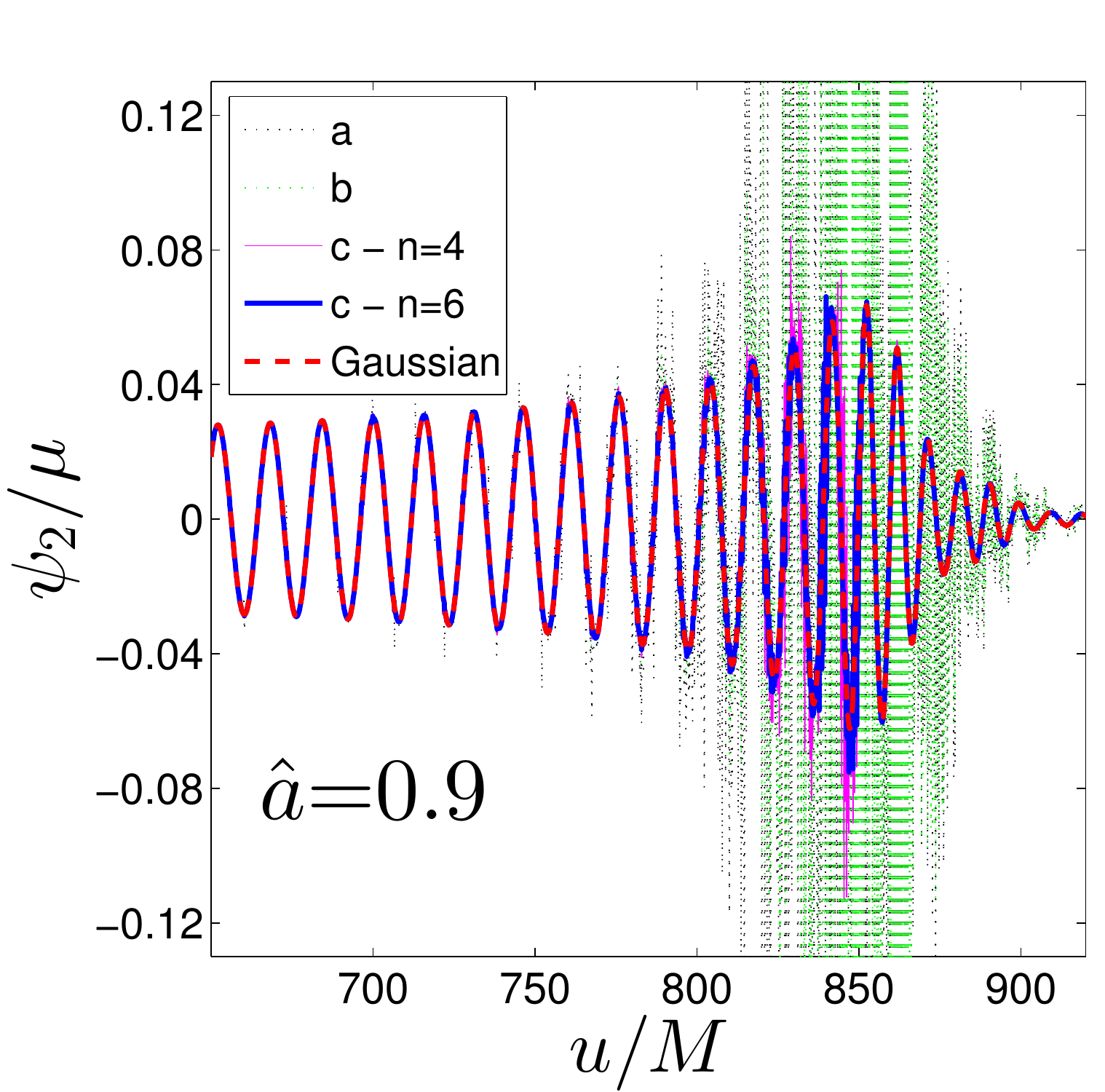}   
  \includegraphics[width=0.49\textwidth]{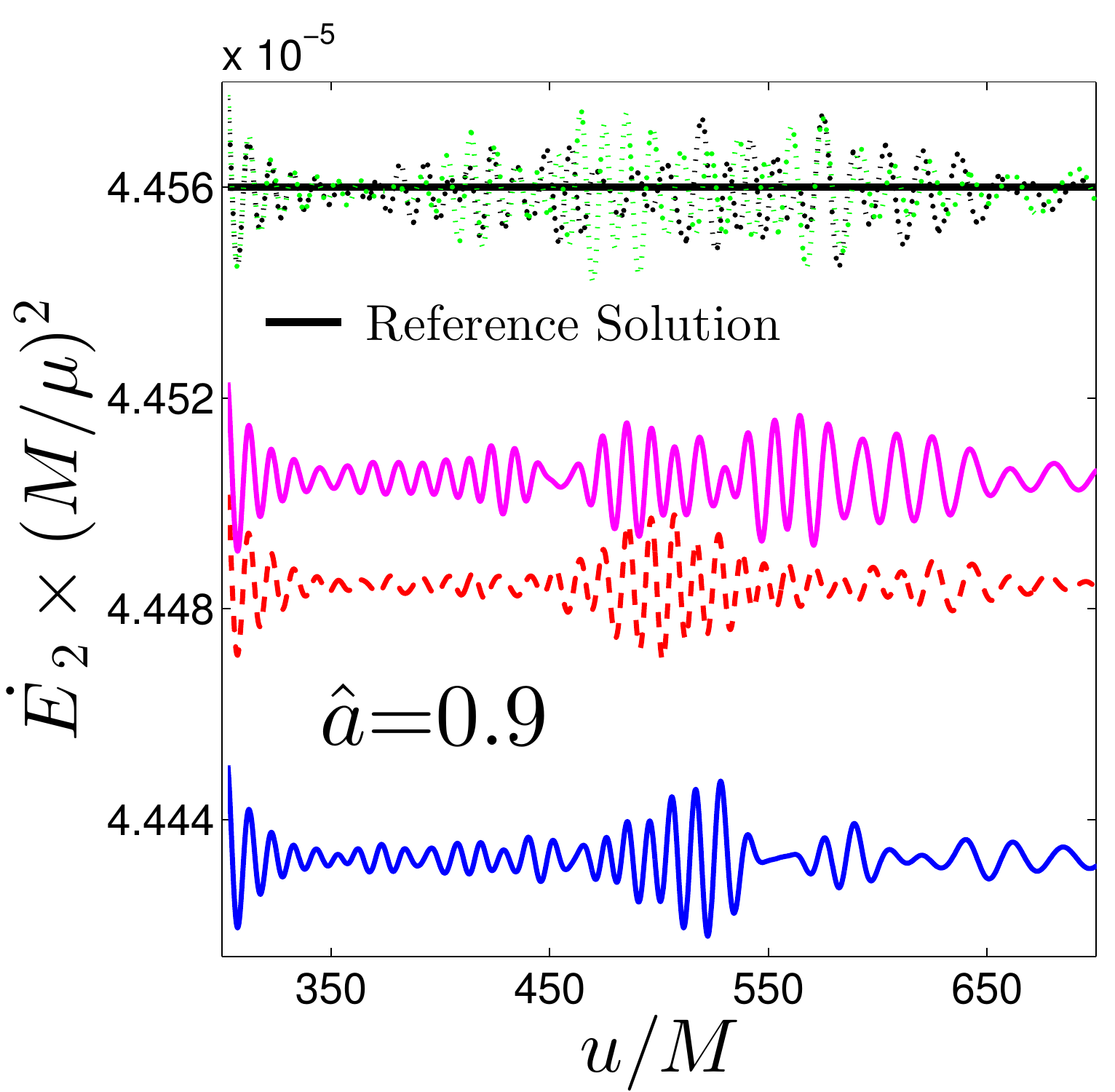}   
  \caption{Comparison of different $\delta$-function representations
    in the two different cases of an EOB insplunge (left panel) and a
    circular orbit (right panel). Note the two panels shows different
    quantities.  The left panel shows the evolution
    of the $m=2$ master variable $\Re(\psi_{2})$. The right panel shows
    the evolution of the GW fluxes.
    The left panel illustrates how the discrete representations lead to 
    instabilities in the simulation if few points
    are employed. Options (a), (b) and (c) with $n=4$ described in
    Sec.~\ref{sec:num:delta} are very noisy, while option (c) with $n
    \gtrsim 6$ gives results comparable to the smooth, analytical
    Gaussian representation.
    The right panel illustrates the accuracy of the discrete
    representations. The reference solution
    is taken from the frequency domain results of
    Ref.~\cite{Sundararajan:2007jg} and shown as a black solid line. 
    Option (c) with $n \gtrsim 6$ is
    less accurate than the Gaussian representation. Options (a) and
    (b) are instead very accurate in this case. 
    Thus we use a Gaussian for any trajectory moving in 
    the $(r,\theta)$ plane but the discrete representation of option
    (a) for circular orbit simulations.}
  \label{fig:Delta_representations}
\end{figure}

A key point in the numerical algorithm is the discrete representation
of the distributional $\delta$-functions appearing in the source term.
The main requirements are numerical stability and accuracy.
We implemented and tested two different methods.

The first method is the narrow Gaussian representation,
\be
\label{eq:delta_g}
\delta(x-q(t))\rightarrow
\delta_\sigma(x-q(t))=\frac{1}{\sigma\sqrt{2\pi}}\exp
\left[{-\frac{(x-q(t))^2}{2\sigma^2}}\right] \ , 
\ee
where $\sigma\sim n_\sigma h\ll M$, $n_\sigma\in\mathbb{N}$ as in
e.g.~\cite{Nagar:2006xv,Bernuzzi:2010ty}.  
This method is very simple, smooth, and analytical but, in principle,
computationally expensive since
(i)~the Gaussian must be well resolved on the grid ($n_\sigma\gtrsim4$); 
(ii)~exponential functions must be often evaluated during the time
evolution. 

The second method is a $2n$-points discrete $\delta$ as described in
\cite{Tornberg:2004JCoPh.200..462T,Engquist:2005JCoPh.207...28E,Sundararajan:2007jg}. The
prescriptions discussed therein comprise 
(a) an order $\mathcal{O}(h^2)$ with $n=1$, (b) an order $\mathcal{O}(h^4)$with $n=2$
and (c) an order $\mathcal{O}(h^2)$ with variable $2n$-points representation for the
$\delta$-function. Let us sketch 
the main ideas, for more details we refer to Sec.~III.A of
\cite{Sundararajan:2007jg}. Assume that the position of the particle $\alpha$ lies
between two grid points,  
$\alpha\in[x_k,x_{k+1}]$. Then, the discrete $\delta$ has support only
for $2n$-points $\delta_i$ around $\alpha$. These values are given
requiring that the integral properties of the $\delta$-function, e.g.
\be 
\label{eq:deltaintegral}
f(\alpha)=\int dx f(x)
\delta(x-\alpha)\approx\sum h f_i \delta_i \ ,
\ee
are preserved also on the discrete level.  
If by chance $\alpha=x_k$, setting 
$\delta_k=1/h$ and $\delta_i=0$ elsewhere solves the problem. In
general, 
$\alpha$ does not lie on a grid point so that interpolation has to be 
used. Linear interpolation leads to option (a) and (c), the more accurate
cubic interpolation yields option (b). The discrete representation 
uses (a) a 2-points support, (b) a 4-points support, and (c) a
variable $2n$-points support. 
For example, considering a 2-points support, linearly interpolating at
$\alpha$, and enforcing Eq.~\eqref{eq:deltaintegral} leads to option (a)
\be
\label{eq:delta_d}
\delta(x-q(t))\rightarrow \delta_{(a)\ j} = \begin{cases}
\gamma \, h^{-1} & ,j=k\\
(1-\gamma) \, h^{-1}  & ,j=k+1\\
0 & ,{\rm otherwise}\ ,
\end{cases}
\ee
where $\gamma= (x_{k+1} - \alpha)/h$.
Similar formulas can be derived for the first two derivatives \cite{Sundararajan:2007jg}. 
Overall, this method is expected to be computationally more efficient
than the Gaussian. However
(i)~high-order accuracy requires larger stencils, increasing the cost; 
(ii)~too narrow/lower order representation may lead to instabilities,
as we shall discuss next (see also \cite{Sundararajan:2007jg,Sundararajan:2008zm}).

We have tested the different $\delta$-representations for various
numerical setups in the cases of circular orbits and EOB insplunge
simulations. 
Summarizing our findings, the tests indicate that the
discrete $\delta$ method in its simplest representation, option (a),
is efficient and accurate for  
simulating a source that is effectively {\it not} moving on the computational domain,
e.g.~circular, geodesic orbits. In cases like inspiral-plunge,
instead, for the same computational cost the most accurate,
stable simulations are achieved with the 
Gaussian method. An example is given in
Fig.~\ref{fig:Delta_representations} where we plot for the different options on the left panel
the $m=2$ field $\Re(\psi_{2})$ for an EOB insplunge configuration
and on the right the GW energy fluxes from circular orbit experiments.
In case the source is effectively moving on the computational
domain (left panel) the discrete representations
(a), (b) and (c) with $n=4$ lead to instabilities towards merger.
A larger support makes the discrete representation smoother
and only option (c) with $n\gtrsim6$ gives results comparable to the smooth,
analytical Gaussian representation.
On the contrary, in case of circular orbits 
(right panel) options (a) and (b) are both more accurate and efficient.
The accuracy of the circular fluxes is here evaluated against a reference
solution taken from the frequency domain results of
Ref.~\cite{Sundararajan:2007jg} and shown as a black solid line.
Note that here the Gaussian is more accurate than option (c) with $n=6$.
Finally, because the insplunge motion considered later on in this work is fixed on
the $\theta=\pi/2$ plane, the best option is to use a combination of the
two methods: a Gaussian representation in radial direction and a
discrete representation in $\theta$ direction (we adopted option (b)).

\section{Dynamics of an inspiraling and plunging, nonspinning point-particle} 
\label{sec:dyn_RR}

Let us now discuss our prescription to model the dynamics of 
an inspiraling and plunging particle. The dynamics of the particle is
described using an Hamiltonian approach for the conservative part and
an analytical radiation reaction force that
accounts for the losses of angular momentum through GWs for the dissipative part. 
This analytic, EOB-resummed, but approximate, radiation reaction  
force will be discussed in detail and we will review two different choices that 
were proposed in the literature to define it. We will use the TE data to argue
that one of the two may be preferable for counterrotating orbits.

It is convenient to describe the Hamiltonian dynamics of a particle
(a short review is attached in~\ref{app:ham}) using dimensionless
quantities like $\hat{H}\equiv H/\mu$. We also denote  
``reduced'' quantities with respect to the background with a hat, 
e.g.  $\hat{t} \equiv t/M$, although they coincide
for $M=1$ as used in the simulations.
We focus on equatorial motion only, so that the spin of the black hole is
either aligned (corotating case) or antialigned (counterrotating case) with
the orbital angular momentum.
The particle Hamiltonian, Eq.~\eqref{eq:H_general}, specified to the equatorial motion reads, 
\be
\label{eq:Hkerr}
\hat{H}\equiv \hat{H}_{\rm SO}+\hat{H}_{\rm orb} = \omega p_\phi + \sqrt{A\left(1+\dfrac{p_\phi^2}{\varpi^2}\right)
                 + A\dfrac{\Delta}{\Sigma}p_\hr^2}\ ,
\ee
where $p_{\phi}\equiv P_{\phi}/(\mu M)$, $\hr = r/M$, $p_{\hr}\equiv P_{\hr}/\mu$
and the functions $(\omega,A,\Delta,\Sigma,\varpi)$ are given by Eqs.~\eqref{eq:Sigma}-\eqref{eq:varpi} 
specified to the equatorial plane ($\theta=\pi/2$). In Eq.~\eqref{eq:Hkerr} we
separate the Hamiltonian in a formally ``pure orbital'' part $\hat{H}_{\rm orb}$
and a ``pure spin-orbit'' part $\hat{H}_{\rm SO}$. This formal separation will be
used in Sec.~\ref{sec:inspl:09} below.
In the general case of nonconservative dynamics the radiation reaction force
enters in both the Hamiltonian equations for the momenta, yielding
both energy and angular momentum losses. The Hamilton's equations read
\begin{align}
\label{eq:pdot}
\dot{\hr}   & = \p_{p_\hr}\hat{H}, \\
\dot{\phi}   & = \p_{p_\phi}\hat{H} = \hOmg, \\
\dot{p}_{\hr}  & = -\p_\hr \hat{H} + \hat{{\cal F}}_{\hr}, \\
\dot{p}_\phi & = \hat{\cal{F}}_{\phi} \ ,
\end{align}
where the overdot stands for $d/d\hat{t}$. 
In the following, we neglect the radial flux contribution 
and set $\hat{{\cal F}}_{\hr}=0$. The reason is that a robust strategy to resum the 
$\hat{{\cal F}}_{\hr}$ in the strong-field regime is, at present, not available, 
even though the PN expansion of ${\hat{\cal F}}_{\hr}$ is known up to 2PN order~\cite{Bini:2012ji}. 
As a consequence, $\hat{{\cal F}}_{\hr}$ may be 
ill-behaved even for non-extremal values of $\ha$. Although $\hat{{\cal F}}_{\hr}$
has been considered at leading, Newtonian order in~\cite{Barausse:2011kb}, 
we believe it is still premature  to include this term in the analytical model of the dynamics.
We postpone to future work a detailed analysis of its properties needed to devise a robust
resummation procedure.

By contrast, the analytic expression of $\hat{\cal{F}}_{\phi}$ has been thoroughly
used in recent years. Nonetheless, a careful inspection of the literature
indicates that there are two ways of writing the flux during the plunge.
During the latter, the Kepler's constraint, $1=\hOmg^2 \hr^3$, is not
satisfied~\cite{Damour:2006tr,Damour:2006qz} (only valid for
the quasi-adiabatic, circular inspiral) and the two descriptions differ
in their way of relaxing the constraint.
We will contrast the original implementation of $\hat{\cal{F}}_\phi$ 
(see Ref.~\cite{Damour:2009kr} and references therein), here specified
to the test-mass case in Kerr spacetime, to the different one 
proposed in~\cite{Barausse:2011kb,Taracchini:2013rva,Taracchini:2014zpa}.   
The differences are essentially in the choices of the arguments 
of certain functions so to incorporate the non-Keplerian behavior during the plunge.

Let us discuss these differences in detail. In the circular approximation
the mechanical angular momentum loss is given by
\be
\hat{\cal F}_\phi = - \dfrac{1}{\nu \hOmg}\dot{E}\ ,
\ee
where the energy flux $\dot{E}$ is resummed according to the multipolar waveform 
resummation introduced in~\cite{Damour:2008gu} for nonspinning binaries and
extended in~\cite{Pan:2011gk} to the spinning case.
We consider multipoles up to $\ell=8$ and the energy flux is given by
\be
\label{eq:flux}
\dot{E}=\sum_{\ell=2}^8\sum_{m=1}^{\ell}\dot{E}_{\lm} = \dfrac{1}{8\pi}\sum_{\ell=2}^8\sum_{m=1}^{\ell}(m\hOmg)^2|
{\cal R} h_{\ell m}(x)|^2\ , 
\ee
where we used the usual PN-expansion variable $x=\hOmg^{2/3}$
and ${\cal R}$  is the distance~\footnote{Here, we use ${\cal R}$ instead of $r$,
  as in Eq.~\eqref{eq:RWZ}, to avoid confusion with the relative separation $r$.}  from the source. 
The multipoles $h_\lm(x)$ are written in factorized form as 
\be
\label{eq:h_lm}
h_{\lm}(x) =
h_\lm^{(N,\epsilon)}(x)\hat{S}^{(\epsilon)}T_{\lm}(\hOmg)\left[\rho_\lm(x)\right]^\ell
e^{\ii\delta_\lm}\ ,
\ee
where $\epsilon=(0,1)$ is the parity of $\ell+m$,
\be
h_\lm^{(N,\epsilon)}(x)=\dfrac{\nu}{\cal R}n_\lm^{(\epsilon)}c_{\ell+\epsilon}x^{\frac{\ell + \epsilon}{2}}Y^{\ell-\epsilon,-m}\left(\dfrac{\pi}{2},\phi\right)
\ee
is the Newtonian contribution, $Y^{\lm}(\theta,\phi)$ are
the scalar spherical harmonics, and the functions $n_\lm^{(\epsilon)}$ and $c_{\ell+\epsilon}$ are given 
in Eqs.~(4a), (4b) and (5) of Ref.~\cite{Damour:2008gu}. 
$T_\lm(\hOmg)$ is the tail factor, the source term
$\hat{S}^{(\epsilon)}$ is specified later, and $\rho_\lm$ and
$\delta_\lm$ are the residual amplitude and phase corrections. We will
specify their structure below.
Note here that the tail factor $T_{\lm}(\hOmg)$ explicitly depends on the orbital frequency $\hOmg$, 
while all other factors depend on $x=\hOmg^{2/3}$ because of Kepler's constraint
for circular orbits. On a Kerr background, the Kepler constraint may be written as
\be
\hr^3\left(1+\ha \hr^{-3/2}\right)^2 \hOmg^2 = \hr_\Omega^3\hOmg^2=1 \ ,
\ee
where we defined 
\be
\hr_\Omega = \hr\left(1 + \ha \hr^{-3/2}\right)^{2/3} \ .
\ee
Since the circular Kepler constraint is not satisfied during the plunge, the usual 
practice is to modify the standard ${\hat{\cal F}}_\phi$ derived along circular orbits 
so that it is not imposed explicitly in the analytical expressions during the plunge
(though it is automatically recovered during the quasi-circular inspiral). In the literature, 
one finds two prescriptions to construct ${\hat{\cal F}}_\phi$'s such that the 
circular Kepler's constraint is relaxed during the plunge.

\begin{figure}[t]
  \centering
   \includegraphics[width=0.32\textwidth]{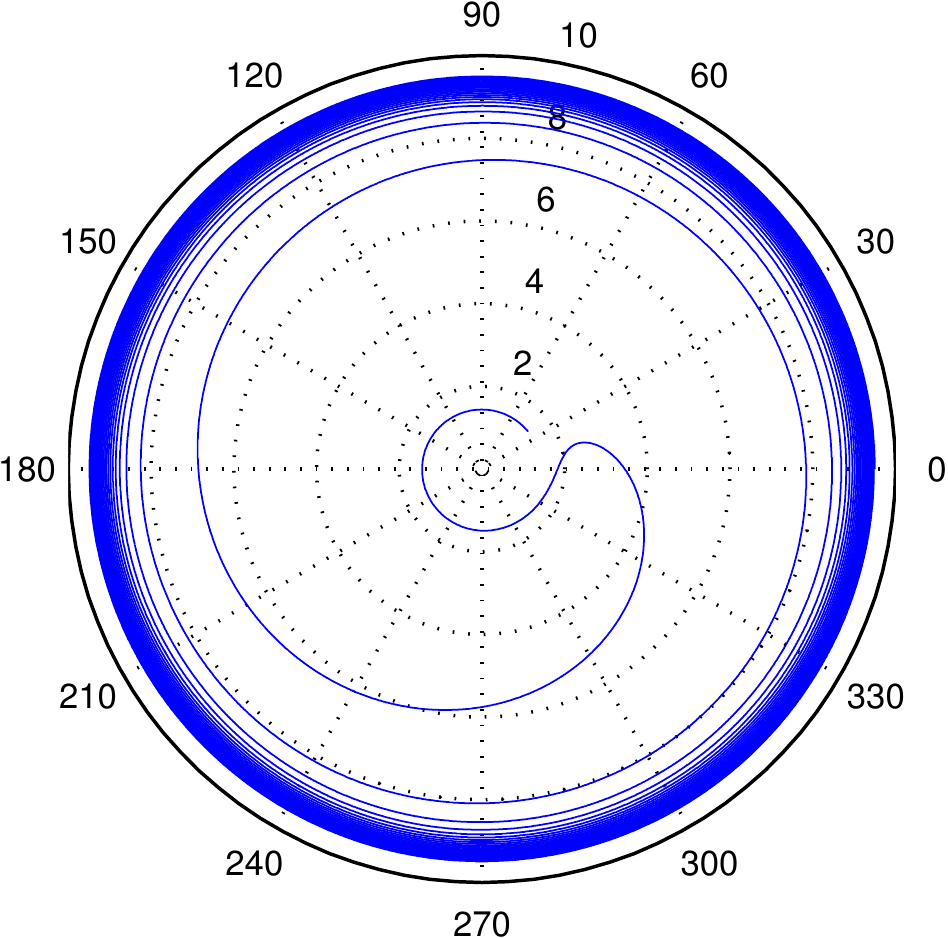}
   \includegraphics[width=0.32\textwidth]{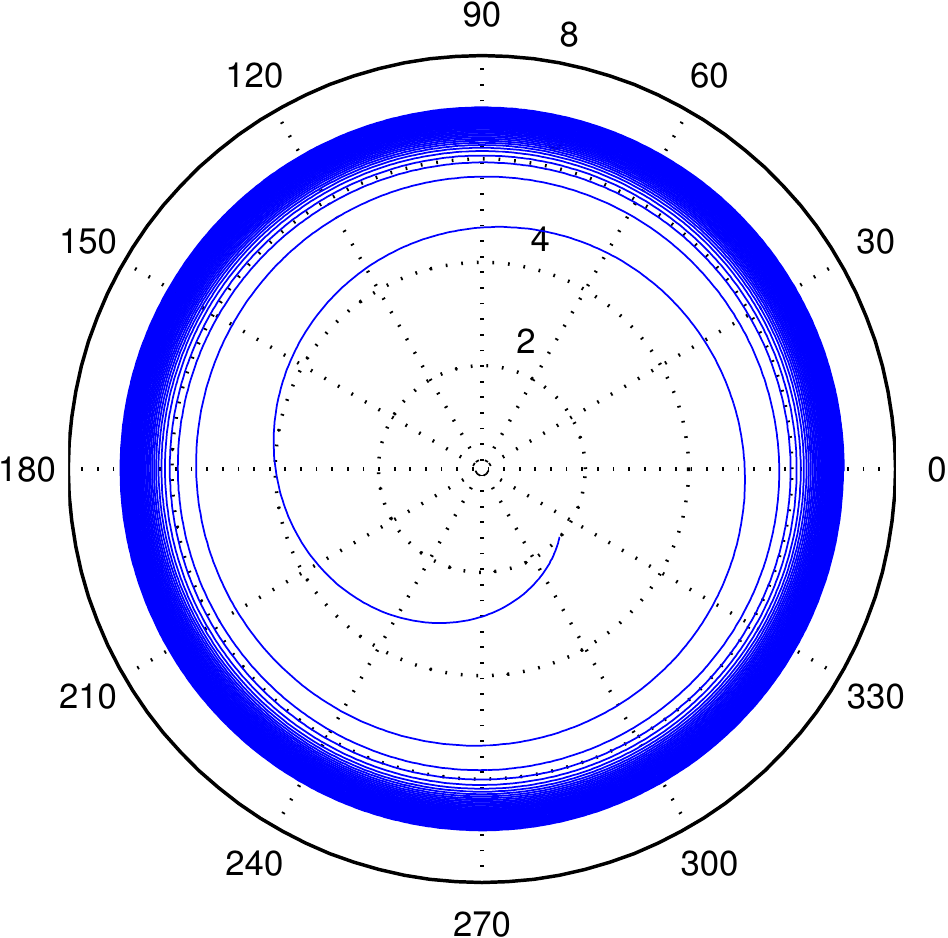}
   \includegraphics[width=0.32\textwidth]{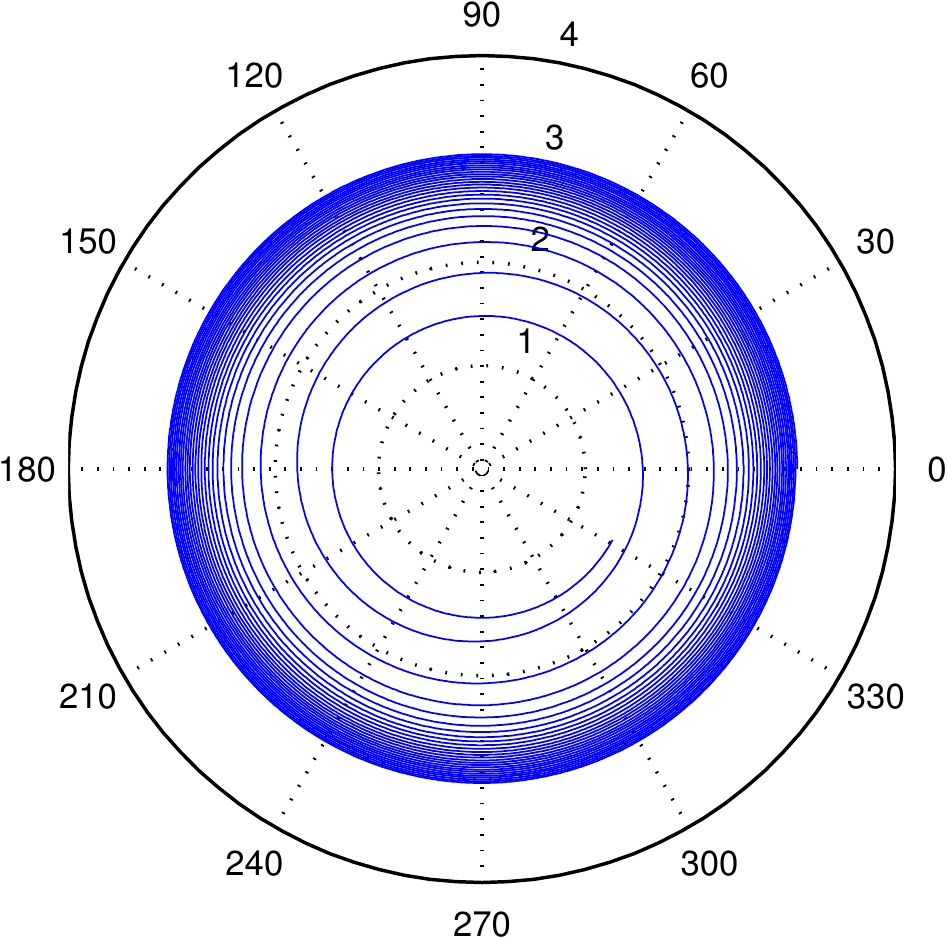}
  \caption{Examples of insplunge trajectories for $\ha=-0.9$ (left),
    $a=0$ (center), and $\ha=+0.9$ (right). The last stable circular
    orbits (LSO) are located, respectively, at
    $\hr_{\rm LSO}\simeq8.697,6.0,2.424$. The light rings (LR) are located at
    $\hr_{\rm LR}\simeq3.910,3.0,1.558$.}  
  \label{fig:traj} 
\end{figure}

(A) -- {\it the ${v_\phi}$-prescription:}
The prescription introduced in~\cite{Damour:2007xr,Damour:2009kr} is to impose that the argument 
$x$ entering in the functions $h_\lm^{(N,\epsilon)}(x)$ and $\rho_\lm(x)$
is the non-Keplerian (squared) tangential velocity, $x=v_\phi^2$,  where 
\be
\label{eq:vphi}
v_\phi = r_\Omega \hOmg \ .
\ee
When the black hole is spinning, semi-integer powers of $x$ appear in the $\rho_\lm$'s.
This introduces a subtlety when the particle is counterrotating with respect to the
black hole, so that, to keep the correct sign, one should see the $\rho_\lm$'s
as functions of $\sqrt{x}$ (with the correct sign) and not of $x$.
The sources $\hat{S}^{(\epsilon)}$ are computed along the dynamics and we use 
$\hat{S}^{(0)}=\hat{H}$ and $\hat{S}^{(1)}=p_\phi v_\phi$ (the Newton-normalized 
orbital angular momentum).
The $\rho_\lm$-functions are given by Eqs.~(29a)-(29i) of Ref.~\cite{Pan:2010hz},
augmented by the full 5PN-accurate terms computed in Ref.~\cite{Fujita:2010xj}. 
Note that the PN-based calculations have been pushed to 22~PN for the nonspinning 
case~\cite{Fujita:2012cm} and to 20~PN for the spinning case~\cite{Shah:2014tka}, 
so that more analytical information is in principle available and will be incorporated in the model in the future.

(B) --{\it the $v_{\Omega}$-prescription:}
Following Ref.~\cite{Barausse:2011kb}, the first difference with (A) is 
that the argument of the $h_\lm^{(N,\epsilon)}(x)$ 
prefactors for the $(2,1)$ and $(4,4)$ multipoles differs from the
general prescription introduced above. The straightforward rewriting 
of  Eqs.~(13a) and (13b) of Ref.~\cite{Barausse:2011kb} in our 
notation~\footnote{Reference \cite{Barausse:2011kb} uses the notation 
$V_\phi^\ell$, where the $\ell$ is a {\it label}, and
{\it not} an exponent, to denote $x^{(\ell +\epsilon)/2}$.} gives
\begin{align}
x & = v_\phi^2 \qquad\qquad\qquad\, (\ell,m)\neq(2,1),\;(4,4)\\
\label{eq:x21x44}
x &= r_{\Omega}^{-\frac{2}{\ell+\epsilon}}
v_{\phi}^{\frac{2(\ell+\epsilon-2)}{\ell+\epsilon}} \qquad (\ell,m) = (2,1),\;(4,4) \ , 
\end{align}
with $v_\phi$ defined by Eq.~\eqref{eq:vphi}. Explicitly for $(\ell,m)=(2,1)$
($\epsilon=1$) and $(\ell,m)=(4,4)$ ($\epsilon=0$) the expression Eq.~\eqref{eq:x21x44}
yields
\begin{align}
\label{x21}
x & = \hOmg^{2/3} = v_\Omega^2                     \qquad\qquad (\ell,m)=(2,1), \\
\label{x44}
x & = r_\Omega^{1/2} \hOmg = r_\Omega^{1/2} v_\Omega^3 \qquad\; (\ell,m)=(4,4),
\end{align}
where we have introduced the {\it circular} velocity  $v_{\Omega}=\hOmg^{1/3}$.
A second difference to (A) is that the odd-parity source is normalized to the 
inverse of $v_{\Omega}$ instead of $v_{\phi}$; namely one uses 
$\hat{S}^{(1)}\equiv p_{\phi} v_{\Omega}$. In other words, the ``Newtonian'' 
angular momentum used to normalize the odd-parity source is computed
imposing the Kepler's constraint.
Finally, the argument of the $\rho_{\lm}$ functions given by Eqs.~(29a)-(29i) of 
Ref.~\cite{Pan:2010hz} is taken to be $v_\Omega$ and not $v_\phi$. 
Note that here the $\rho_{\lm}$'s contain all the 5PN-accurate
nonspinning terms. This is done for consistency with case (A) and
differs from Ref.~\cite{Barausse:2011kb} where 
only part of the 5PN-accurate information was retained.

In summary, prescription (A) is ``less Keplerian'' than prescription (B) because 
of the explicit relaxation of the Kepler's constraint when choosing the argument
of the various functions. A priori, one expects (A) to give a more accurate representation
of the ``actual'' flux when the deviations from circularity are larger;
notably when the black hole spin is high and antialigned with
$p_\phi$. However, (A) and
(B) should be essentially equivalent when the plunge is more circularized, that is,
when the black hole spin is high and aligned with $p_\phi$.
In this work, we consider (A) as our {\it standard} way of implementing the 
non-Keplerian behavior in an analytical expression of the radiation reaction. 
On top of its simplicity and uniformity of implementation
(the circular velocity $v_\Omega=\hOmg^{1/3}$ is ubiquitously replaced by $v_\phi$
and Kepler's constraint is always relaxed), it offers a more consistent
analytical representation of the loss of mechanical angular momentum during 
the plunge. 

Before contrasting the two flux prescriptions with TE fluxes, 
we show in Fig.~\ref{fig:traj} examples of 
particle trajectories, as obtained with (A), for $\ha=-0.9$ (left) $\ha=0$ (center) 
and $\ha=+0.9$ (right). The qualitative differences of the dynamics as a function 
of the rotation parameter $\ha$ are mainly due to the spin-orbit coupling in the 
Kerr potential for geodesics. The latter determines the positions of the last 
stable orbit (LSO), $\hr_{\rm LSO}(\ha)$, and the light ring (LR), $\hr_{\rm LR}(\ha)$. 
Compared to the nonrotating case ($\ha=0$), positive spins (aligned to the
orbital angular momentum) move the LSO and LR closer to each other and closer to the horizon, 
e.g.~$\hr_{\rm LSO}(\ha>0)<\hr_{\rm LSO}(0)$. As a consequence, the 
plunge becomes progressively ``more circular'' as $\ha\to1$, and the radial 
momentum of the particle attains a minimum for $\ha=1$. Negative spins (antialigned 
with $p_\phi$) move the LSO and the LR farther from each other, and farther from the horizon, 
i.e.~$\hr_{\rm LSO}(\ha<0)>\hr_{\rm LSO}(0)$. Retrograde plunges are
characterized by a turning point, $\hOmg=0$, after which 
$\hOmg$ changes sign until the particle locks to the horizon. 
This effect is due to frame dragging and happens at $\hr<\hr_{\rm LR}$. 
The radial momentum of the particle during plunge, as a function of
the rotation parameter, attains a maximum for $\ha=-1$.

\begin{figure}[t]
  \centering
   \includegraphics[width=0.48\textwidth]{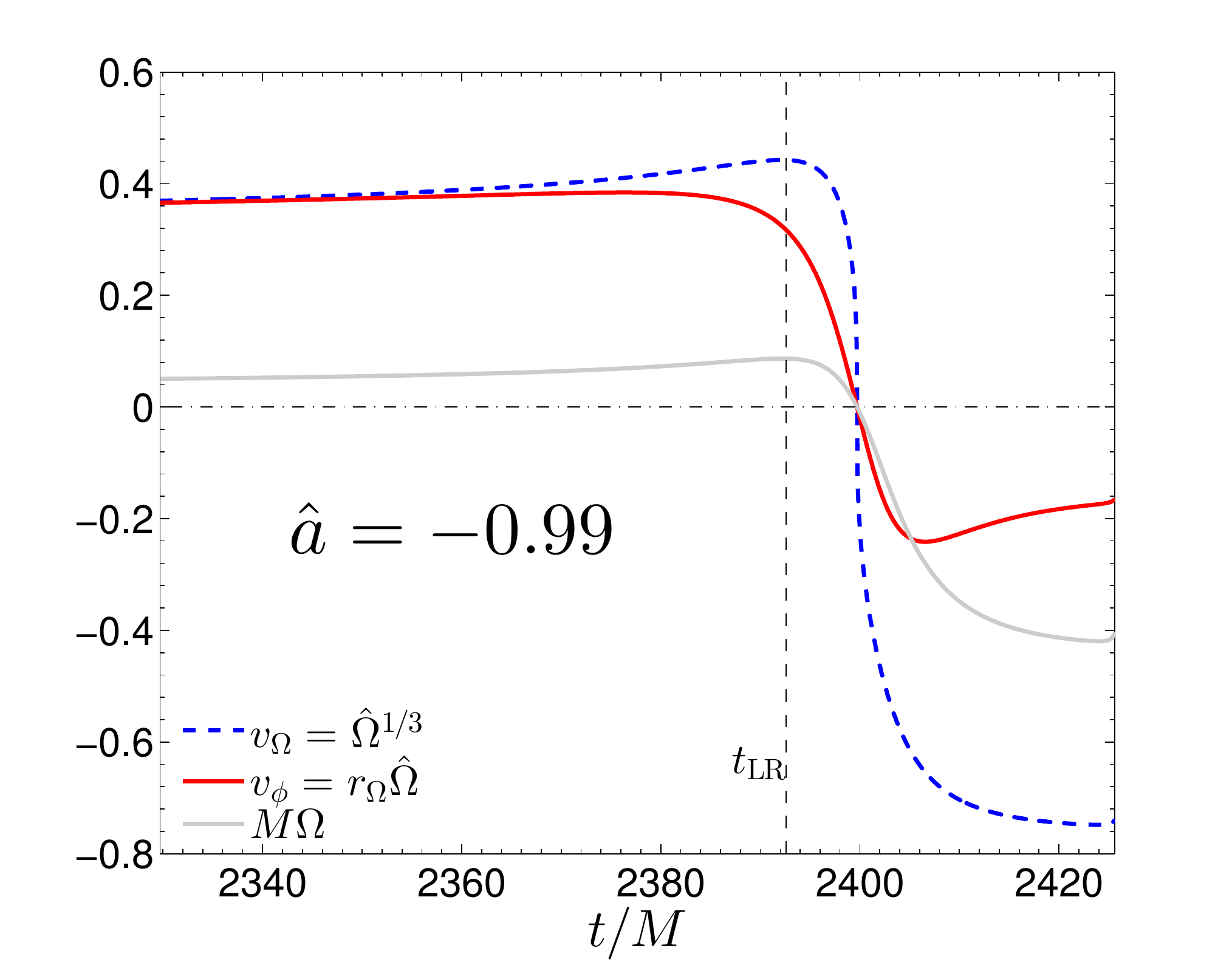}
   \includegraphics[width=0.49\textwidth]{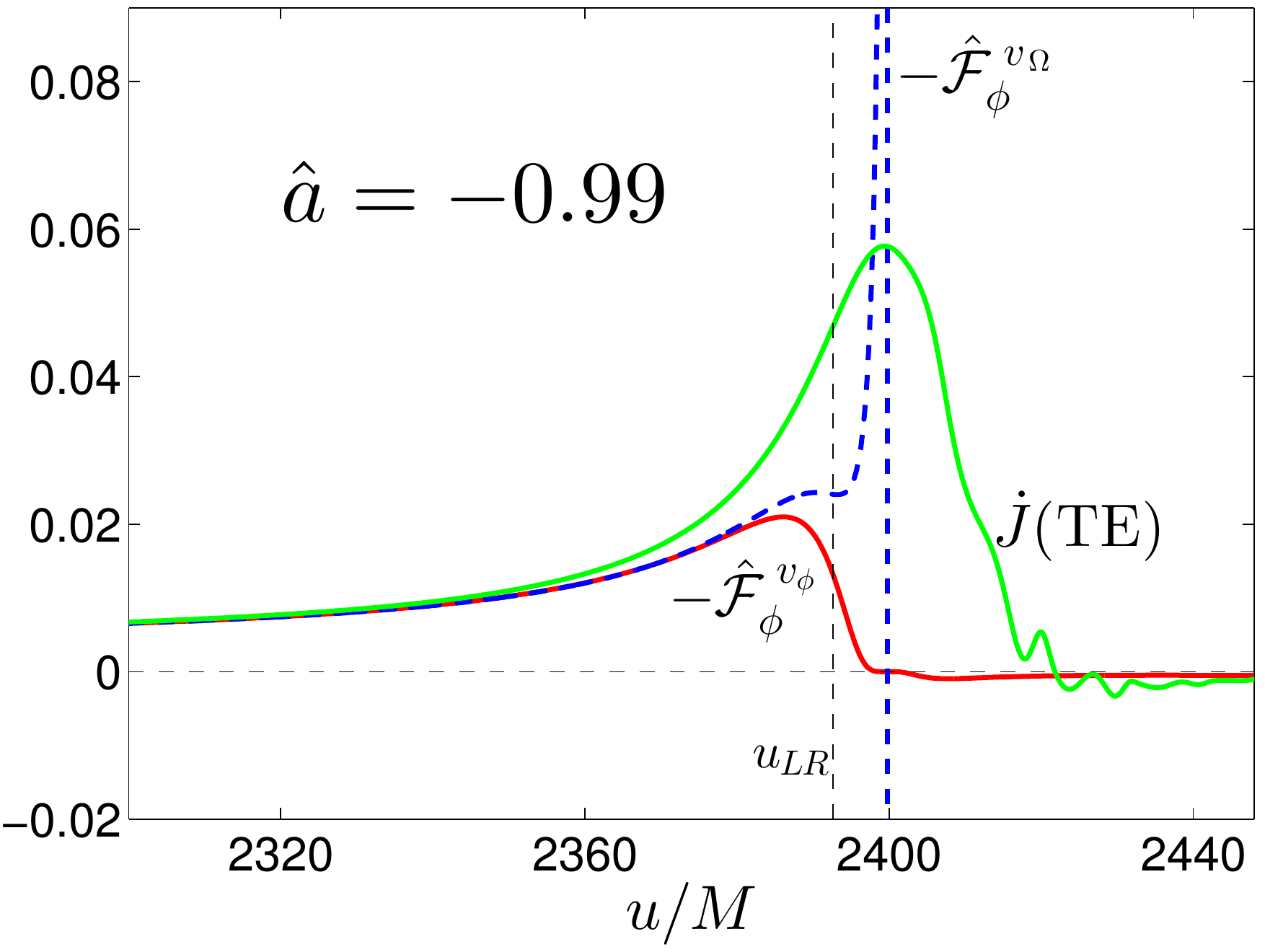}
  \caption{Comparing two different prescriptions for the mechanical angular momentum
   loss, $-\hat{\cal{F}}_\phi^{v_\phi}$ and $-\hat{\cal{F}}_\phi^{v_\Omega}$,
   with the TE angular momentum flux $\dot{J}^{\rm TE}$ (all values normalized with $(M/\mu^2)$). 
   We set $\ha=-0.99$ and focus on late-plunge and merger phase.
   The left panel compares $v_\Omega$ and $v_\phi$; the right panel compares the fluxes.  
   The vertical dashed line marks the light-ring crossing. 
   The prescription $\hat{\cal{F}}_\phi^{v_\phi}$ captures the correct qualitative 
   behavior of the TE flux. Remaining differences are due to next-to-quasi-circular
   effects not included in the modeling of $\hat{\cal{F}}_\phi$ and to QNMs ringdown
   oscillations.}
  \label{fig:flux_EOB} 
\end{figure}

Now that we have discussed qualitatively the structure of insplunge trajectories,
let us motivate why we use prescription (A) for the
radiation reaction force. We focus on the case $\ha=-0.99$
and perform a consistency comparison between EOB and TE fluxes.
The particle inspirals from $\hr_{0}=9.5$. The LSO is at $\hr_{\rm LSO}=8.97$, the 
light-ring at $\hr_{\rm LR}=3.99$ and the outer horizon at $\hr_+=1.14$.
The left panel of Fig.~\ref{fig:flux_EOB} compares the time evolution of 
$\hat{\Omega}$, $v_\Omega$, and $v_\phi$ close to merger; the light-ring crossing
time is identified by the dashed vertical line. The right panel of the figure
presents, versus time, the triple comparison between: (i) the mechanical angular 
momentum loss $-\hat{\cal{F}}^{v_\phi}_{\phi}$ computed via prescription 
(A); (ii) $-\hat{\cal F}_{\phi}^{v_\Omega}$ obtained from prescription (B); 
(iii) the TE angular momentum flux computed from the waves extracted 
at future null infinity along dynamics (A).
Figure~\ref{fig:flux_EOB} highlights that the differences in the
analytical fluxes only occur in the very late plunge phase, very
close to the LR crossing (when $\hr=\hr_{\rm LR}$) at $t_{\rm LR}/M=2392.6$. 
As a consequence, the plunge dynamics is unaffected by the differences 
between prescriptions (A) and (B).
This is why in Fig.~\ref{fig:flux_EOB} we report only one TE curve: 
the dynamics is practically the same and so are the TE fluxes. 
However, the comparison of Fig.~\ref{fig:flux_EOB} indicates that $-\hat{\cal F}^{v_\phi}_{\phi}$ 
is qualitatively closer to the numerical flux than $-\hat{\cal F}^{v_\Omega}_{\phi}$
around and after the light-ring crossing. The main reason for this difference resides 
in the behavior of $\hOmg$ and its effect on the $(2,1)$ and $(4,4)$ Newtonian contributions 
to the analytical flux, that are proportional to certain powers of Eqs.~\eqref{x21} and~\eqref{x44}.
The difference between $-\hat{\cal{F}}^{v_\phi}_\phi$ and  $-\hat{\cal{F}}^{v_\Omega}_\phi$
is then easily explained from the different behavior of $v_\Omega$ and
$v_\phi$. Referring to the left panel of Fig.~\ref{fig:flux_EOB}, one sees that, after the $\hOmg=0$ point,
$v_\Omega$ keeps decreasing monotonically, while $v_\phi$ is limited and eventually 
its derivative changes sign. Since $\dot{E}_{21}\propto x^6$ and $\dot{E}_{44}\propto x^7$
one understands the origin of the unphysical growth of the $-\hat{\cal F}_\phi^{v_\Omega}$ when
$x=v_\Omega^2$ for $(2,1)$ and by $x=r_\Omega^{1/2}v_\hOmg^3$ for $(4,4)$,
since one gets  $\dot{E}_{21}\propto \hOmg^4$ and  $\dot{E}_{21}\propto r_\Omega^{7/2}\hOmg^7$.
By contrast, the milder behavior of $v_\phi$ when $\hOmg\to 0$
produces limited fluxes  $(\dot{E}_{21},\dot{E}_{44})$ and a globally
more consistent behavior of $-\hat{\cal F}_\phi^{v_\phi}$. 
It is also worth noting the consistency between $-\hat{\cal F}_\phi^{v_\phi}$ and $\dot{J}$
after the turning point $\hOmg=0$, with both quantities having approximately the same
magnitude. On the contrary, $-\hat{\cal F}_\phi^{v_\Omega}$ is very different there 
because of the vertical tangent of $v_\Omega$ at $\hOmg=0$.
The residual differences between $-\hat{\cal F}^{v_\phi}_{\phi}$ and $\dot{J}$ seen in the right
panel of Fig.~\eqref{fig:flux_EOB} are due to: (i) next-to-quasi-circular terms depending 
explicitly on $p_{r_*}$ that are not included in the analytical
modeling and (ii) effects of the ringdown. Though we will quantify the amount
of noncircularity of the dynamics versus $\ha$ in Sec.~\ref{sec:inspl} and connect
it with the multipolar structure of the waveform around the light-ring crossing, 
the proper modeling of next-to-quasi-circular 
effects in the EOB waveform and flux and ringdown
is out of the scope of this paper. 

When the black hole spin is aligned with the orbital angular momentum, we shall
show below, in Sec.~\ref{sec:inspl:fluxes}, that the differences between 
$-\hat{\cal F}_\phi^{v_\Omega}$ and $-\hat{\cal F}_\phi^{v_\phi}$ become much 
smaller and practically negligible as $\ha\to 1$. Actually, when $\ha\to 1$ 
one finds that they are equally inaccurate with respect to TE fluxes, because 
of missing higher PN (spin-dependent) terms in the expansion of the $\rho_\lm$'s.

\section{Code tests: accuracy of gravitational waveforms and fluxes}
\label{sec:test}

Before discussing in detail the structure of the multipolar waveforms for $\ha\neq0$ and 
their properties when $|\ha|\to 1$, we present several tests of our new
computational infrastructure.
In Sec.~\ref{sec:test:ciro} we first consider circular, equatorial
orbits for different values of $\ha$, calculate the GW energy fluxes
emitted to future null infinity and to the horizon, and compare 
with the results of Ref.~\cite{Sundararajan:2007jg,Taracchini:2013wfa}. 
In Sec.~\ref{sec:test:radial} we 
discuss the radial geodetic plunge (no radiation reaction) for $\ha=0$
and compare with the RWZE results of~\cite{Bernuzzi:2010ty}. 
This test also gives us the opportunity to discuss 
the numerical treatment of the source during the plunge. 
Self-convergence is studied in Sec.~\ref{sec:test:conv} for 
insplunge waveforms. We show that the expected
convergence rate is attained already at low resolutions and the
absolute phase errors are small. Note that this is a challenging 
test of the main physical application of the method.
In Sec.~\ref{sec:test:inspl} we compare the multipolar insplunge waveform for
a nonrotating background computed with the 2+1 {\it Teukode} and with the 1+1 RWZE 
code from~\cite{Bernuzzi:2010xj,Bernuzzi:2011aj,Bernuzzi:2012ku}.    
All the fluxes shown are normalized by the appropriate powers of $M,\mu$.

\subsection{GW fluxes from circular orbits}
\label{sec:test:ciro}

\begin{table}[t]
\centering
\caption{GW energy fluxes at scri, $\dot{E}^\infty_m$,  and at the horizon, $\dot{E}^H_m$, 
  for circular, equatorial orbits at various $\hr_0$ for $m=2,3$ 
  and background rotations $\ha=0.0,0.9$. The values are normalized by $(M/\mu)^2$. 
  Radii below the LSO are marked with $*$.
  The resolution used for the shown results is $N_x\times N_\theta=2400\times200$.  
  The horizon fluxes are computed with 
  two different methods: the usual frequency domain formula
  \cite{Teukolsky:1974yv} applicable in our time domain setup
  because of circular orbits, and the time domain formula in 
  Eq.~\eqref{eq:EFlux-HRZ} (in brackets).
  $\Delta \dot{E}_m^{\infty,H}/\dot{E}_m^{\infty,H}$ are the percentual 
  relative differences to the frequency domain 
  values of~\cite{Sundararajan:2007jg,Taracchini:2013wfa}. Note that 
  our results include {\it all} the $\ell$-mode contributions, while the reference  
  solution truncates the sums at $\ell=8$.}
\label{tab:Efluxciro} 
\begin{tabular}{lll|cc|cc}
  \hline
  \hline
  $\ha$ & $m$ &$\hr_0$ & $\dot{E}^{\infty}_m$ \color{white}{${A^A}^A$}  
  & $\Delta \dot{E}^\infty_m/\dot{E}^\infty_m[\%]$ 
  & $\dot{E}^{H}_m$  & $\Delta\dot{E}^H_m/\dot{E}_m^H[\%]$ \\
  \hline
  0    &  2 & 4* &  8.580479 e-03  &  8.33e-03  &  5.64953e-04 \, (5.64849e-04)   &  2.03e-02\\
  0    &  2 & 6  &  7.368338 e-04  &  3.58e-04  &  2.62484e-06 \, (2.62443e-06)   &  3.91e-03\\
  0    &  2 & 8  &  1.650495 e-04  &  1.52e-03  &  1.09970e-07 \, (1.09953e-07)   &  4.00e-05\\
  0    &  2 & 10 &  5.373492 e-05  &  2.75e-03  &  1.13139e-08 \, (1.13122e-08)   &  2.14e-03\\
  \hline
  0.9  &  2 & 4  &  2.661563 e-03  &  2.57e-03  & -5.28423e-05 \, (-5.28346e-05)  &  4.72e-03 \\
  0.9  &  2 & 6  &  4.621241 e-04  &  2.81e-03  & -3.98467e-06 \, (-3.98441e-06)  &  1.30e-03 \\
  0.9  &  2 & 8  &  1.254217 e-04  &  3.44e-03  & -5.68006e-07 \, (-5.67988e-07)  &  9.13e-04 \\
  0.9  &  2 & 10 &  4.455909 e-05  &  3.49e-03  & -1.19689e-07 \, (-1.19702e-07)  &  1.36e-03 \\
  \hline
  \hline
  0    &  3 & 4* &  2.710318 e-03  &  7.95e-03  &  6.92585e-05 \, (6.92581e-05)   &  5.34e-03\\
  0    &  3 & 6  &  1.459721 e-04  &  1.22e-02  &  5.41814e-08 \, (5.41814e-08)   &  8.86e-03\\
  0    &  3 & 8  &  2.449258 e-05  &  1.31e-02  &  8.61375e-10 \, (8.61376e-10)   &  1.17e-02\\
  0    &  3 & 10 &  6.434177 e-06  &  1.34e-02  &  4.69154e-11 \, (4.69155e-11)   &  1.28e-02\\
  \hline
  0.9  &  3 & 4  &  6.466345 e-04  &  1.37e-02  & -3.00663e-06 \, (-3.00675e-06)  &  8.98e-03 \\
  0.9  &  3 & 6  &  8.042190 e-05  &  1.34e-02  & -1.17094e-07 \, (-1.17111e-07)  &  1.11e-02 \\
  0.9  &  3 & 8  &  1.717198 e-05  &  1.35e-02  & -1.00392e-08 \, (-1.00421e-08)  &  1.20e-02 \\ 
  0.9  &  3 & 10 &  5.043443 e-06  &  1.34e-02  & -1.40038e-09 \, (-1.40118e-09)  &  1.21e-02\\
\hline
\hline
\end{tabular}
\end{table}

\begin{table}[t]
  \centering
    \caption{GW energy fluxes for a circular, equatorial orbit at $\hr_0=6$ for $\ha=0.9$ 
    in the $l=m=2$ mode at different finite extraction radii, for
    waves extrapolated using Eq.~\eqref{eq:extrap_formulas} and $K=2$, and for
    waves at null infinity. The values are normalized by $(M/\mu)^2$. }
    \label{tab:Eflux_finite} 
    \begin{tabular}[t]{c||c|c|c|c|c|c|c|c}
      \hline
      \hline
      $\hr$ & $100$ & $200$ & $300$ & $500$ & $740$ & $1000$ & Extrp. ($K=2$) & $\scri$ \\
      \hline
      $\dot{E}_{22}\times 10^{4}$ & 4.546 & 4.595 & 4.604 & 4.608 & 4.610 & 4.610   & 4.611 & 4.611 \\
      \hline
      \hline
    \end{tabular} 
\end{table}

This test considers circular equatorial trajectories at various radii
$\hr_0=4,6,8,10$, for $\ha=0,0.9$ and $m=2,3$. GW energy fluxes at scri
are compared with the frequency domain results kindly provided to us by 
Scott Hughes using an improved version of his frequency domain 
code from~\cite{Sundararajan:2007jg,Taracchini:2013wfa}.

GW fluxes at the horizon are calculated in two 
different ways: (i)~using the formulas presented in
Sec.~\ref{sec:TE:gw} and performing in postprocessing the integral of
Eq.~\eqref{eq:EFlux-HRZ}, and (ii)~using the frequency domain formulas of
\cite{Teukolsky:1974yv}. The latter calculation is possible because we
are considering circular orbits. It is performed as a check of
method (i) to verify its accuracy and robustness. The data of the horizon
fluxes are compared with the corresponding ones also provided to us by 
Scott Hughes~\cite{Taracchini:2013wfa}. The relevant results are collected in 
Table~\ref{tab:Efluxciro}. Using $N_x\times N_\theta=2400\times200$ we reproduce  
three digits of the frequency domain data, that is, the agreement is 
around $0.01\%$. We mention that already a resolution of $1200\times100$ ( runtime
$\sim0.5$ hours on a standard desktop machine) suffices to obtain agreement
up to two digits. These results rely on the waveform extraction at
scri. For completeness, Table~\ref{tab:Eflux_finite} lists, for 
the $\ell=m=2$ mode, the energy flux computed from waveforms extracted at finite radii
as well as the value obtained by extrapolation (see Eq.~\eqref{eq:extrap_formulas} below),
that coincides with the one computed from scri waveforms. 

Similar results are obtained for the horizon fluxes. Table~\ref{tab:Efluxciro} 
shows that the time domain calculation in post-processing is indeed accurate. 
Note that the agreement with the frequency domain data is obtained 
also below the last stable orbit, e.g. $\hr_0 = 4$ for $\ha = 0$.

\subsection{Geodetic radial infall for $\ha=0$: TE vs. RWZE waveforms}
\label{sec:test:radial}

\begin{figure}[t]
  \includegraphics[width=0.49\textwidth]{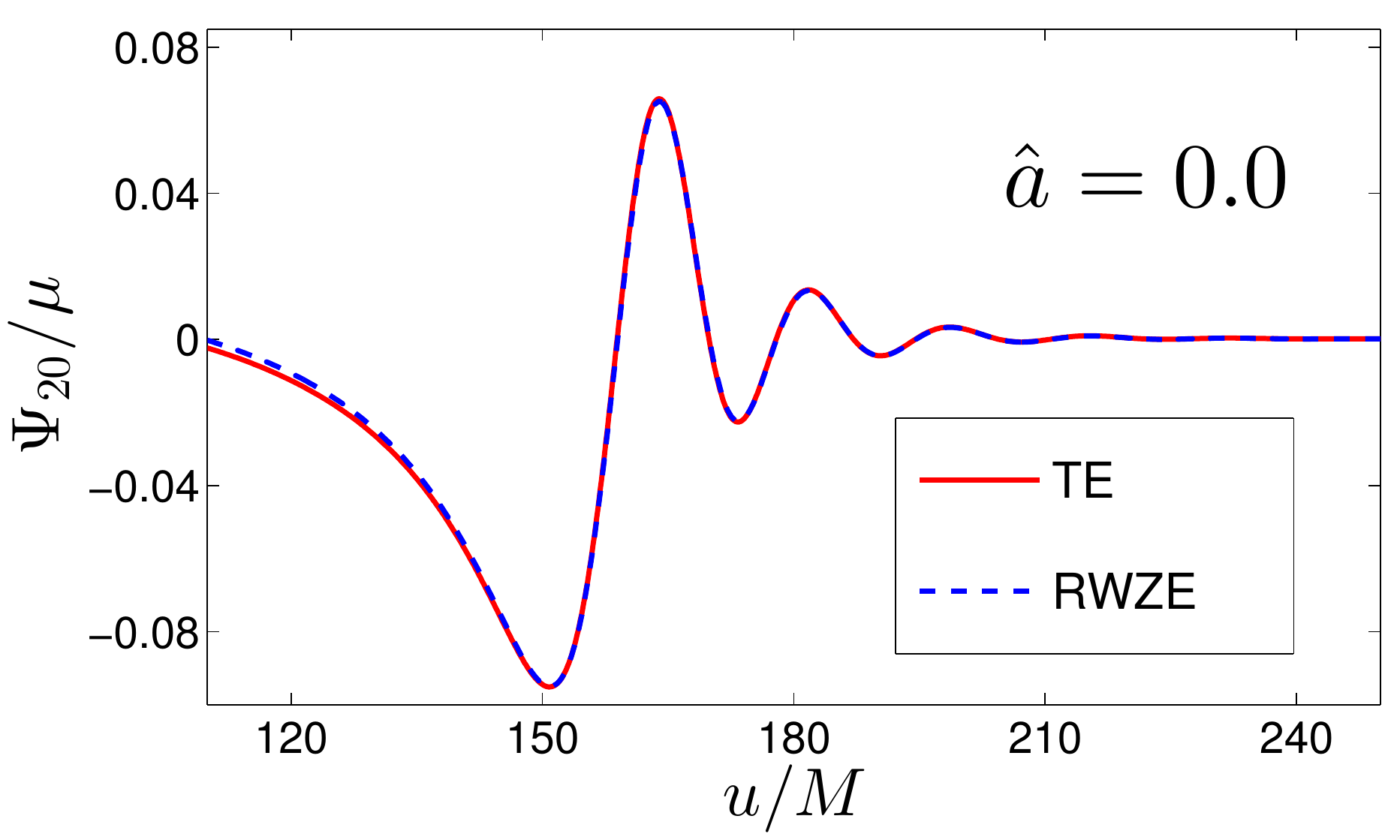}    
  \includegraphics[width=0.49\textwidth]{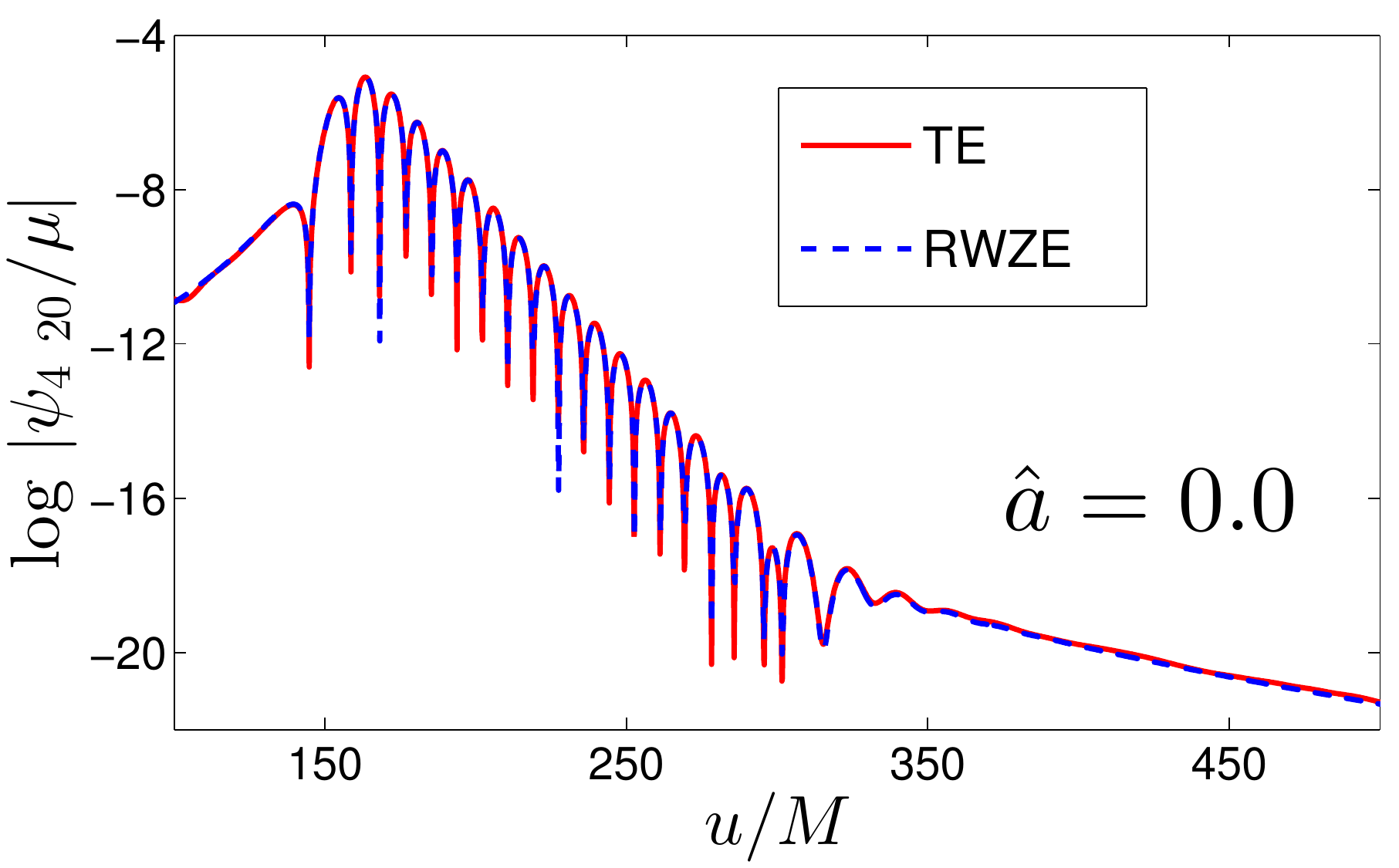}   
  \caption{Comparison of 2+1 TE/1+1 RWZE code waveforms for geodetic radial
    infall dynamics. The particle is falling from $\hr_0=25$ along the
    $x$-axis onto a Schwarzschild black hole. 
    Left: The $\ell=2$, $m=0$ waveform $\Psi_{20}(u)$ (RWZE variable) at $\scri$
    ({\it Teukode}) and large finite radius (RWZE code).
    Right: The $\ell=2$, $m=0$ component of the Weyl scalar $\psi_{4\ 20}$ at $\scri$ ({\it Teukode})
    and large finite radius (RWZE code) in logarithmic scale.
    } 
  \label{fig:RAD_a00}
\end{figure}

This test considers a radial trajectory from $\hr_0=25$ along the
$x$-axis in Schwarzschild background ($\ha=0$). For simplicity, we
focus only on the $\ell=2$, $m=0$ multipole. 
But contrarily to previous calculations of the waveform from the radial geodetic
plunge along the z-axis~\cite{Martel:2001yf}, also the multipoles with $m\neq 0$ are
switched on since the particle is moving in the 
equatorial plane.

The TE waveforms are compared with the RWZE ones computed as described
in~\cite{Bernuzzi:2010ty}. Waveforms from the RWZE code are extracted at large finite radius 
$\hr\sim2200$, TE waveforms are extracted at $\scri$. 
The initial data for the RWZE code solve the 
linearized Hamiltonian constraint~\cite{Martel:2001yf}.  
On the contrary, initial data for the TE are not solving the
constraints and we trivially set $\psi=\dot{\psi}=0$. This produces an
initial burst of junk radiation which is radiated away after
$\sim200M$. In Fig.~\ref{fig:RAD_a00} we compare the outcome
of the two codes in the $\ell=2$, $m=0$ RWZE variable $\Psi_{20}(u)$ (left) and 
the Weyl mode $\psi_{4\ 20}(u)$ (right). 
Despite the differences in the setup, we find visual
agreement; quantitative differences are below a few percent. In
particular the $\psi_{4\ 20}(u)$ variables (right panel) agree also
during the tail phase and both codes capture the correct tail decay.
The simulation of the correct tail phase requires artificial
dissipation. The tail cannot be captured in the 2+1 RWZE variable
because the integration to $\Psi_{20}$ of the TE data produces
inaccuracies. 

Let us discuss the behavior of the particle at the horizon. In BL
coordinates the source smoothly ``switches-off'' when approaching the
horizon due to a red-shift effect driven by the term $d\lambda/dt\to0$
in Eq.~\eqref{eq:Tmunu}. If, in addition, the computational domain
does not include the horizon (e.g. the tortoise coordinate is employed) no particular
treatment is needed for the source approaching the horizon, see
e.g.~\cite{Sundararajan:2010sr}. The situation is different in
horizon penetrating coordinates. We observe a
red-shift ``shrinking'' effect but the source does not approach zero
towards the horizon, instead it reaches a finite limit. 
When the particle has reached the last point before the horizon,
we remove the remaining half of the Gaussian by simply advecting
it out of the computational domain.
This procedure is somehow unphysical but does not affect the waveform
since it involves very few points close to the horizon. Furthermore (i)~the
inspiral-plunge waveform is mainly determined by the particle crossing
the light-ring and entering the potential well, (ii)~the ringdown part
of the waveforms is essentially particle-independent.

\subsection{Self-convergence of insplunge waveforms}
\label{sec:test:conv}

\begin{figure}[t]
  \includegraphics[width=0.49\textwidth]{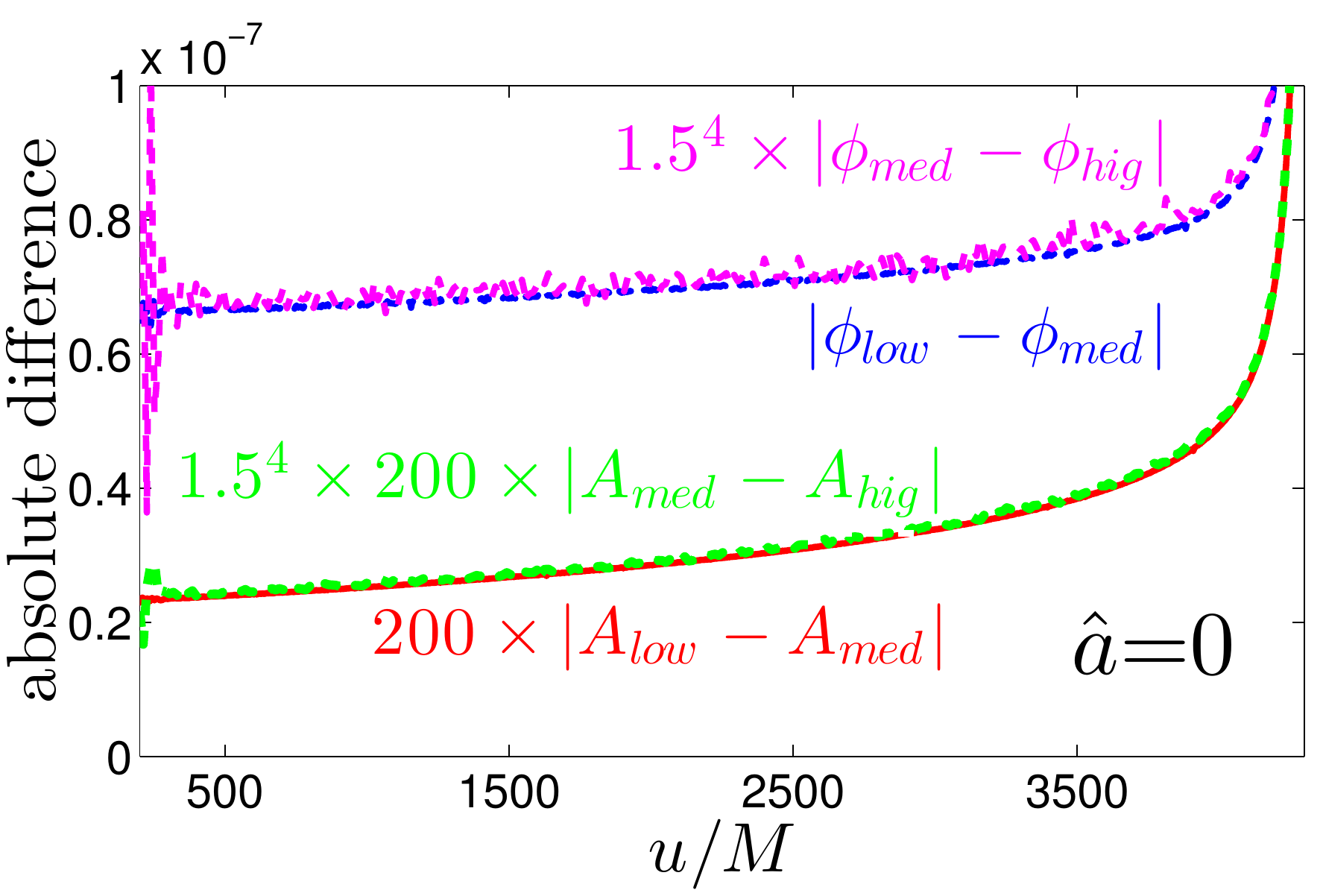}   
  \includegraphics[width=0.49\textwidth]{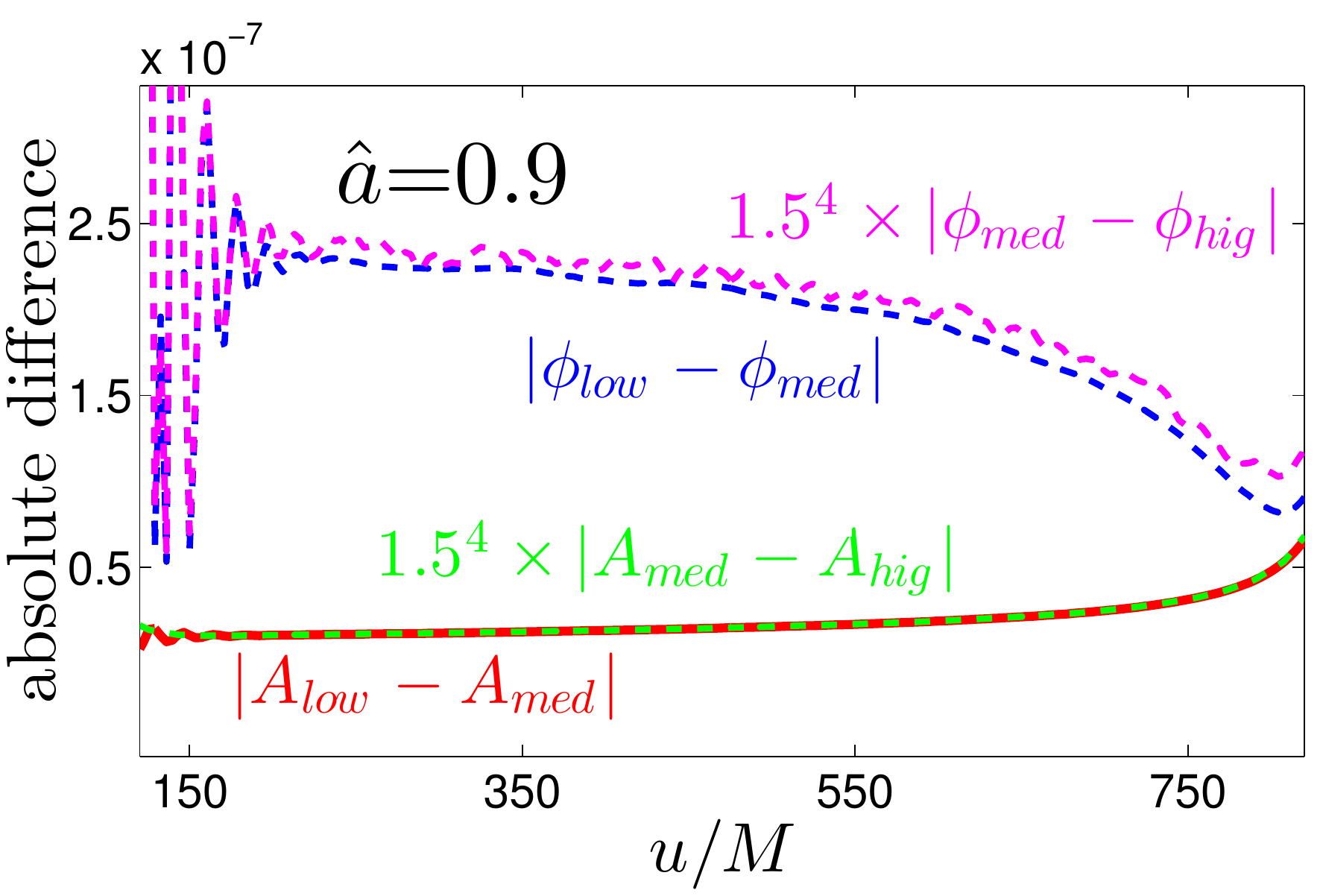}   
  \includegraphics[width=0.49\textwidth]{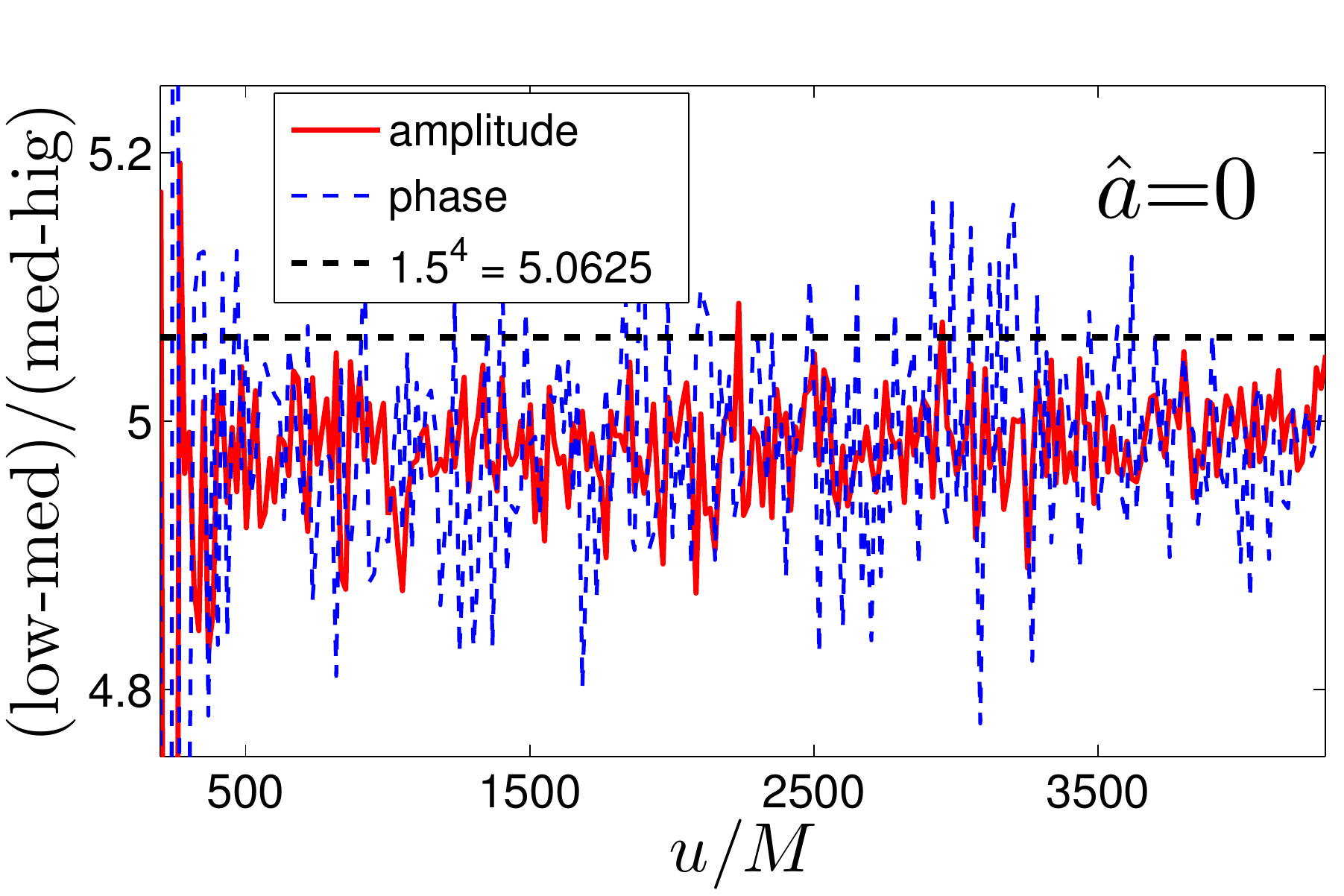}   
  \includegraphics[width=0.49\textwidth]{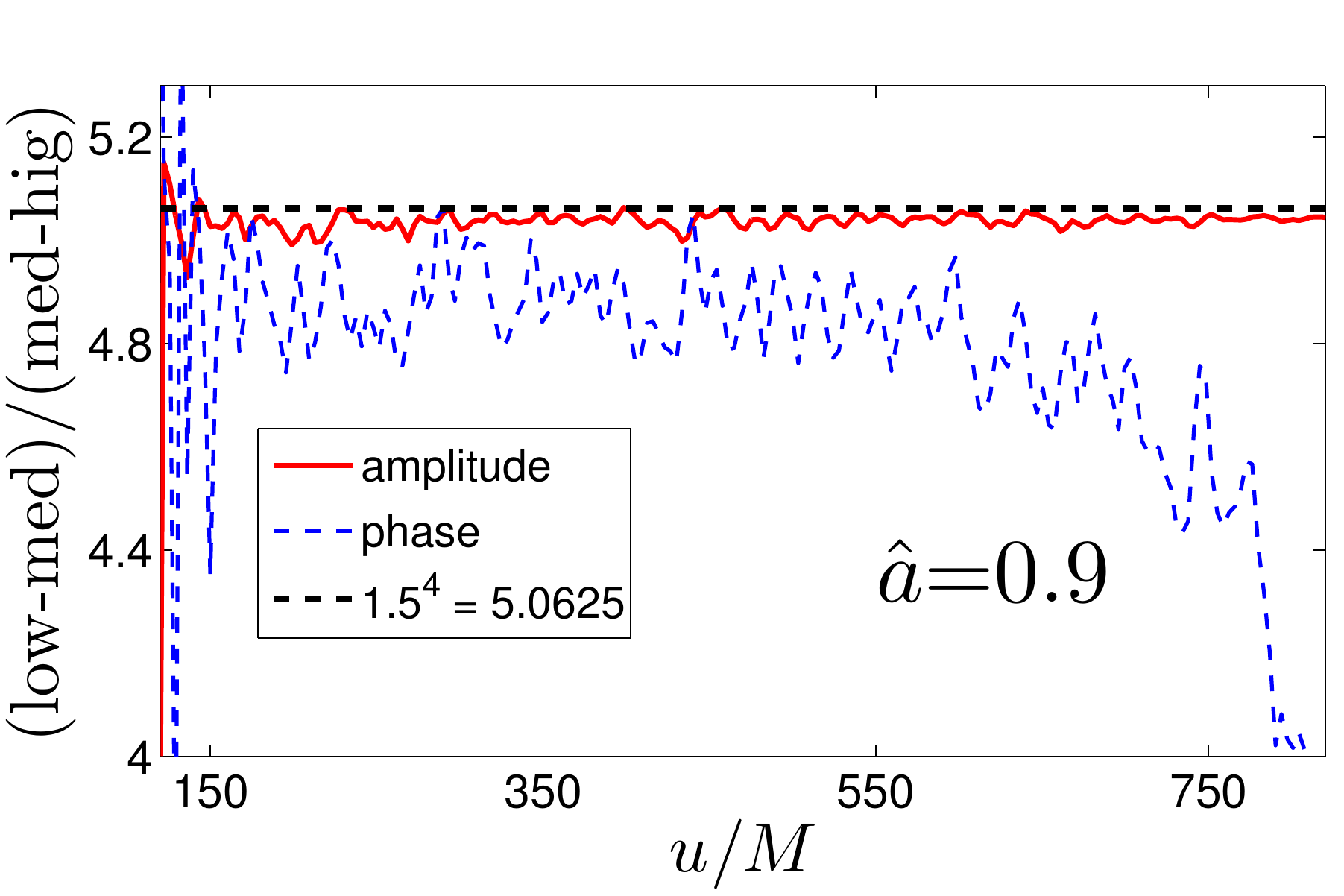}  
  \caption{Self-convergence for the amplitude and phase at the extraction
    point $(\rho,\theta)=(10,\pi/2)$ for insplunge waveforms. Left:
    $\ha=0$, right: $\ha=0.9$.
    The triplet uses radial resolutions $N_x=(400,600,900)$,
    with $n_\sigma=(4,6,9)$ and $N_\theta=30$. 
    The expected scaling for the error is $1.5^4\approx5.0625$ for
    $4$th-order finite differences. 
    Top panels: absolute differences in phase $\Delta \phi$ and
    amplitude $\Delta A$ between various resolutions. The
    differences between medium and high resolution are rescaled by the
    expected factor assuming convergence and lay on top of the
    differences between low and medium resolution. For visualization 
    the differences in amplitude are rescaled by an arbitrary factor
    $200$ in the left panel.
    Bottom panels: ratios of absolute differences.
    Convergence is thus observed already at very low resolutions.
    }
  \label{fig:conv}
\end{figure}

To test the accuracy of the code, Fig.~\ref{fig:conv} shows  
self-convergence tests for $\ha=0$ (left) and $\ha=0.9$ (right) obtained
at resolutions $N_x=(400,600,900)$ (all with $N_\theta=30$), using $4$th 
order finite differencing and a minimal number of points for the Gaussian,
i.e.~$\sigma=n_\sigma h_x$ with $n_\sigma=(4,6,9)$.
The absolute differences in phase and amplitude between the low and medium
resolutions are at the level of $\Delta\phi\sim10^{-7}$ and $\Delta
A\sim10^{-8}$ (top panel). In both cases we obtain the expected 4th order convergence 
up to merger (bottom panels). Convergence is slightly worse in the 
$\ha=0.9$ case. The plots are noisy in the ringdown phase (not shown),
where the fields exponentially decrease by several orders of magnitude. 
Hence, we expect larger relative errors during the ringdown. For
the science runs of the next section a much higher resolution is employed.

\subsection{TE vs. RWZE insplunge waveforms for $\ha=0$}
\label{sec:test:inspl}

\begin{figure}[t]
  \includegraphics[width=\textwidth]{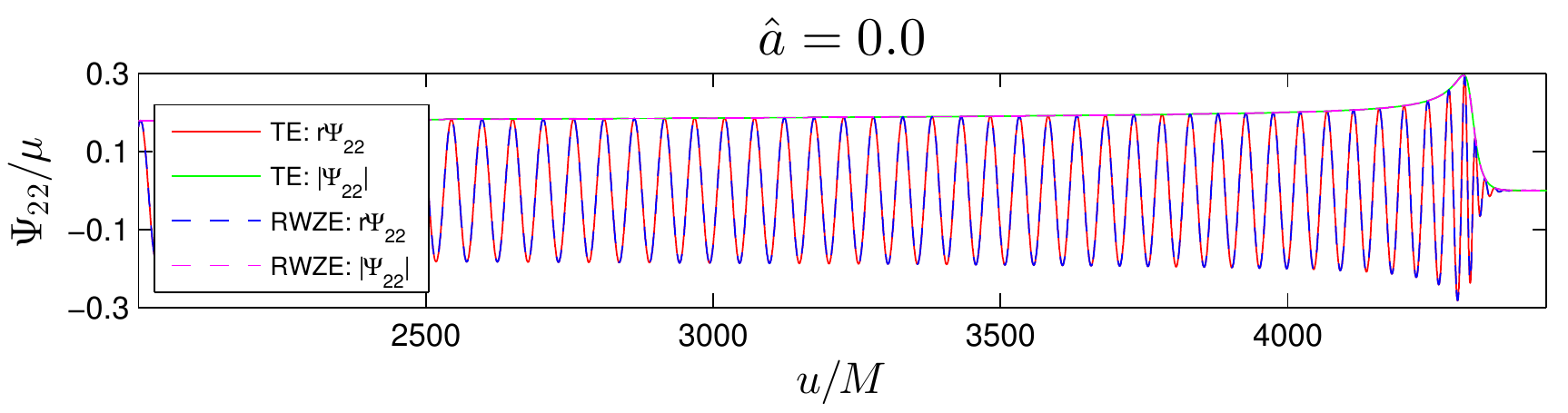}    \\
  \includegraphics[width=0.49\textwidth]{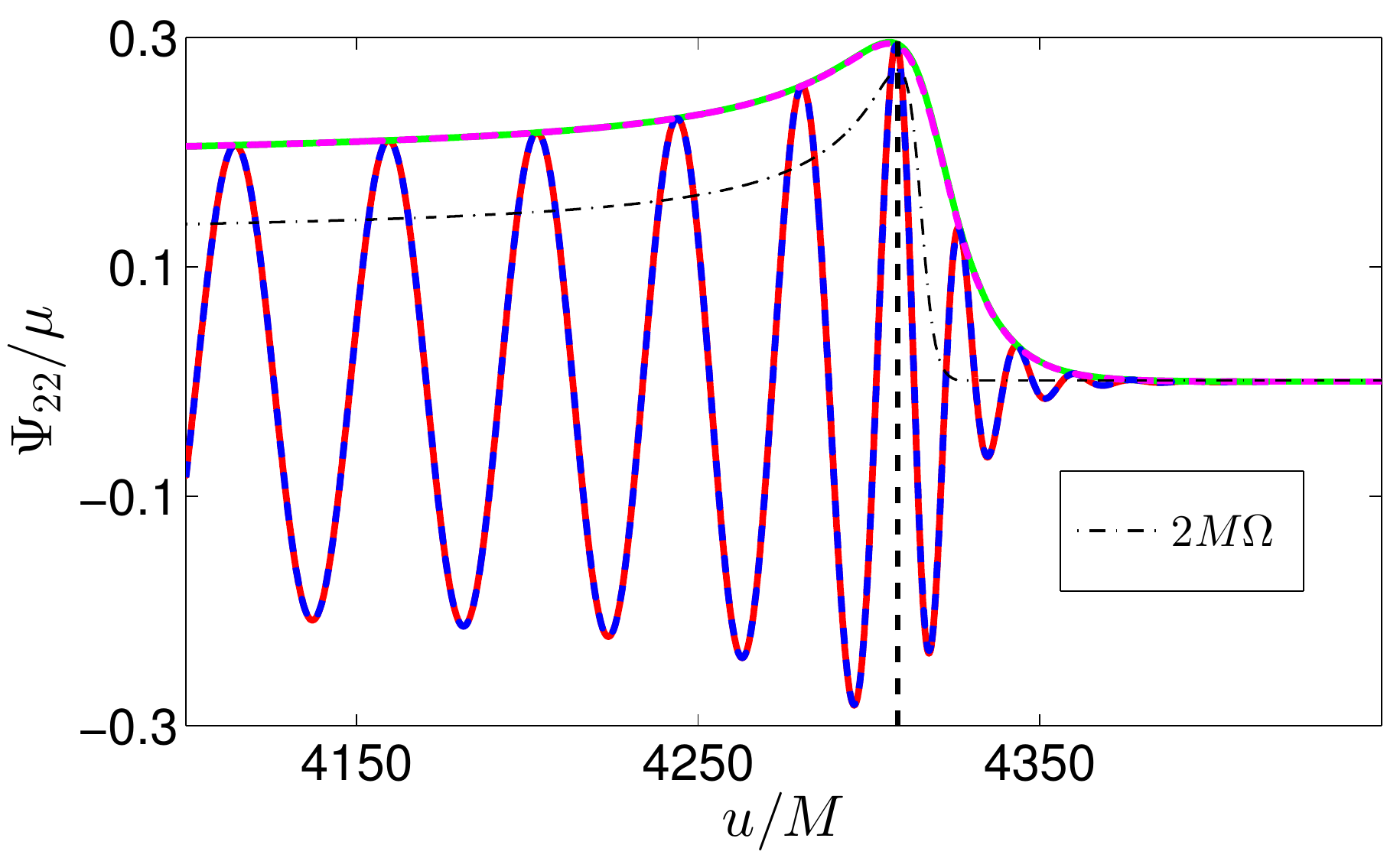}   
  \includegraphics[width=0.49\textwidth]{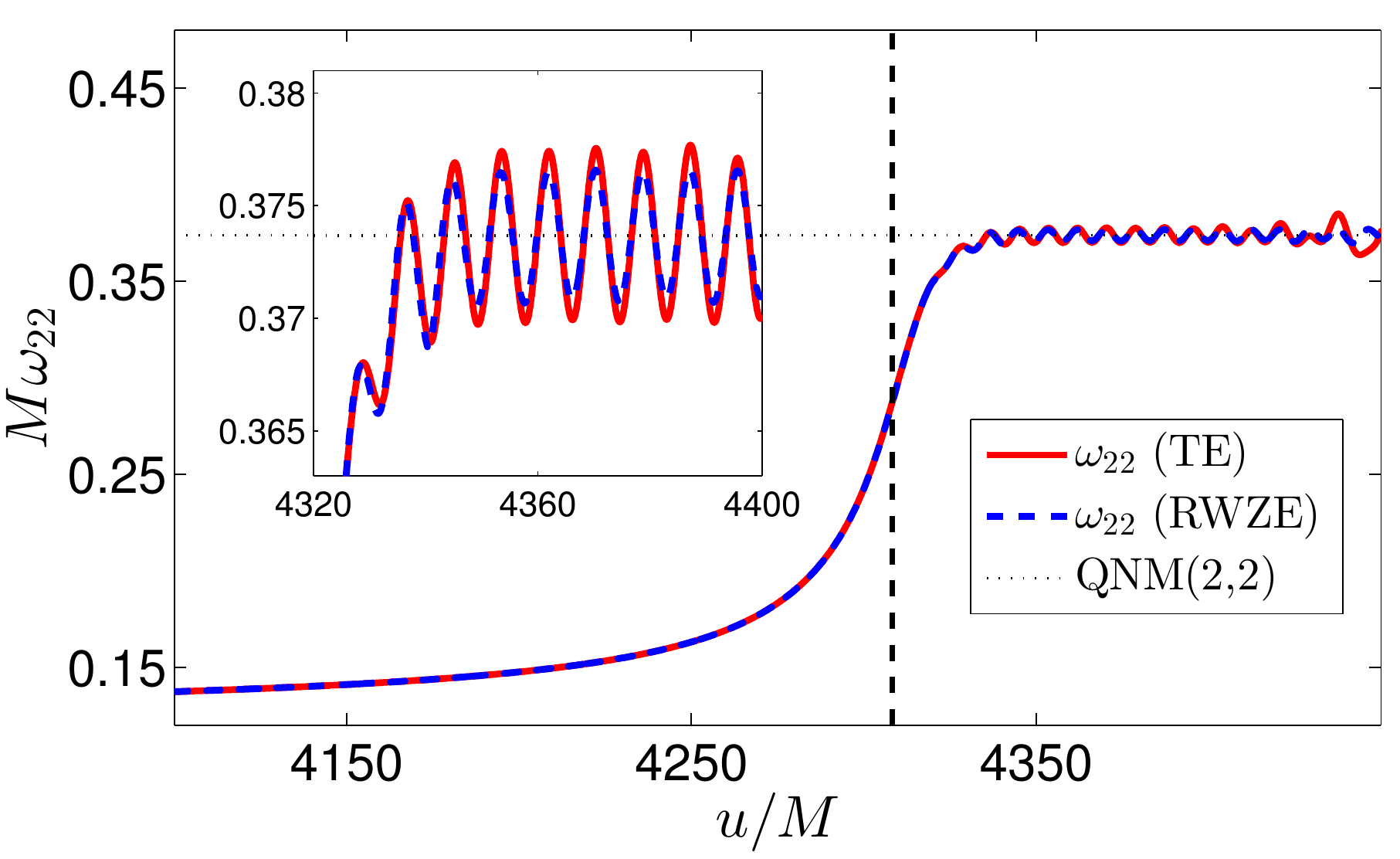}   
  \caption{Case $\ha=0$: comparison between the 2+1 TE and 
    the 1+1 RWZE code for insplunge-ringdown waveforms with $\mu/M=10^{-3}$. Shown is the 
    real part of the $\Psi_{22}$ metric  waveform extracted at $\scri$ 
    together with its amplitude and frequency. The vertical lines mark the time of the 
    light-ring crossing ($t_{\rm LR}=u_{\rm LR}=u_{\Omega^{\rm max}}=4308.39M$). Twice the orbital
    frequency $2\hOmg$ is represented with a dash-dotted black line (bottom left panel). 
    The dotted horizontal line in the right panel marks the fundamental 
    QNM frequency~\cite{Berti:2009kk}.} 
  \label{fig:comparison_qualitative_EOB_a00}
\end{figure}

\begin{figure}[t]
  \includegraphics[width=0.49\textwidth]{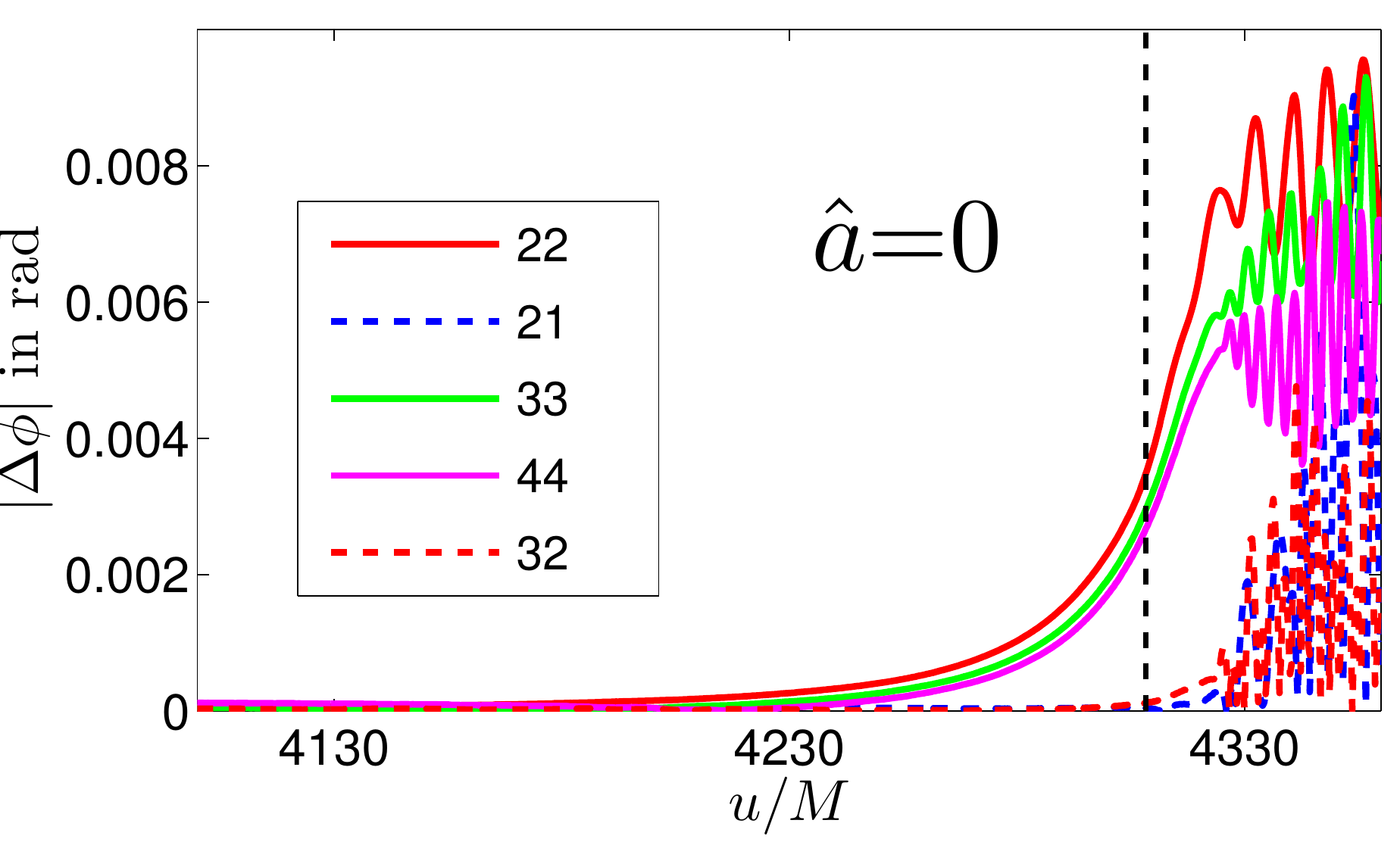} 
  \includegraphics[width=0.49\textwidth]{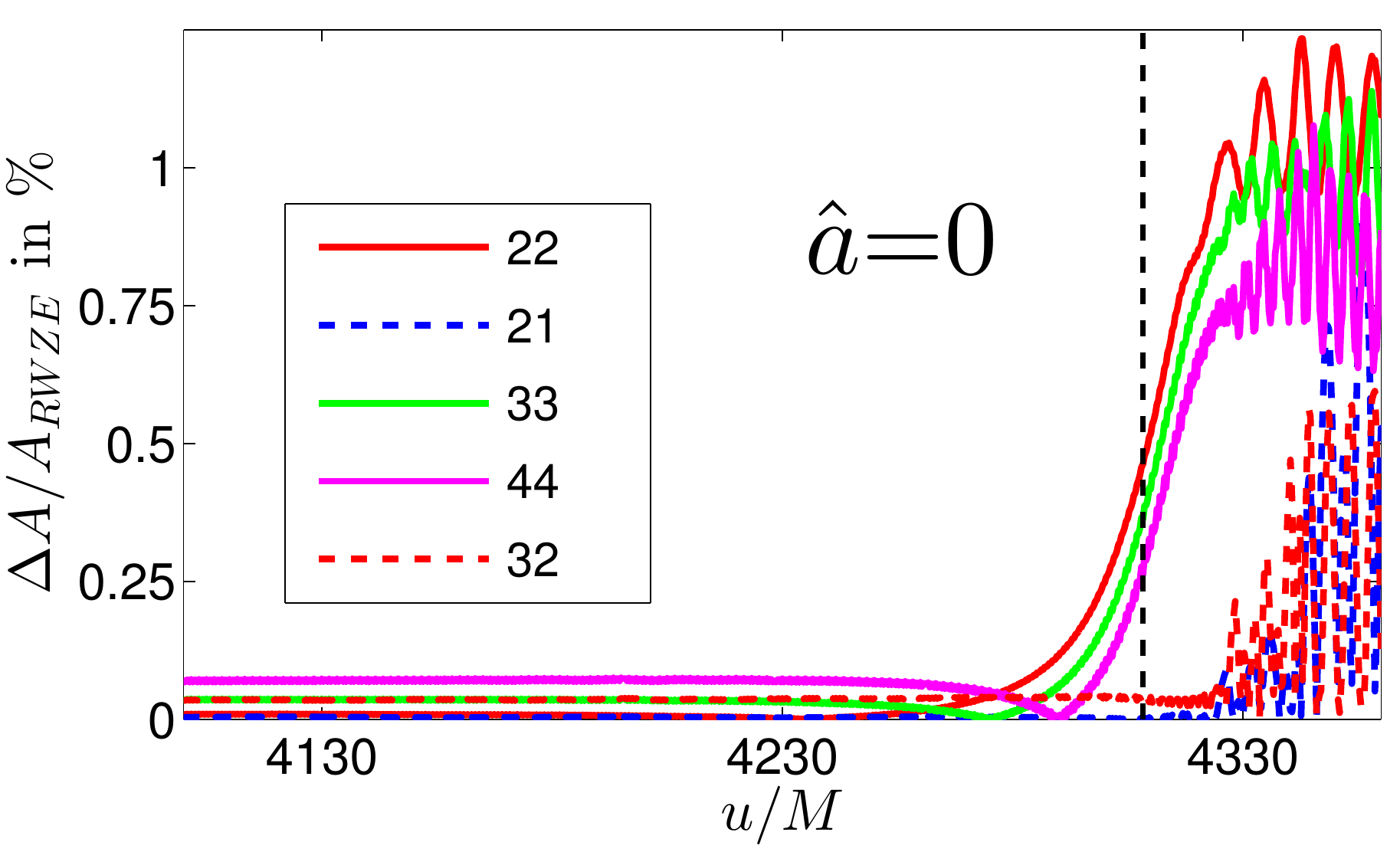}   
  \caption{Case $\ha=0$: comparison between the 2+1 TE and the 1+1 RWZE code for 
    insplunge-ringdown waveforms with $\mu/M=10^{-3}$. Absolute phase (left) and 
    relative amplitude (right) differences for the dominant multipoles at $\scri$.
    The absolute phase differences are $\lesssim2\times10^{-3}$~rad until the
    time of the light-ring crossing ($u_{\rm LR}=u_{\Omega^{\rm max}}=4308.39M$) 
    and remain $\lesssim0.01$ during the ringdown ($u>4308.39M$).
    The relative amplitude differences are at the order of $\sim0.25\%$
    until $u_{\rm LR}$ and remain $\lesssim1.25\%$ afterwards.}
  \label{fig:comparison_quantitative_EOB_a00}
\end{figure}

\begin{figure}[t]
  \includegraphics[width=0.49\textwidth]{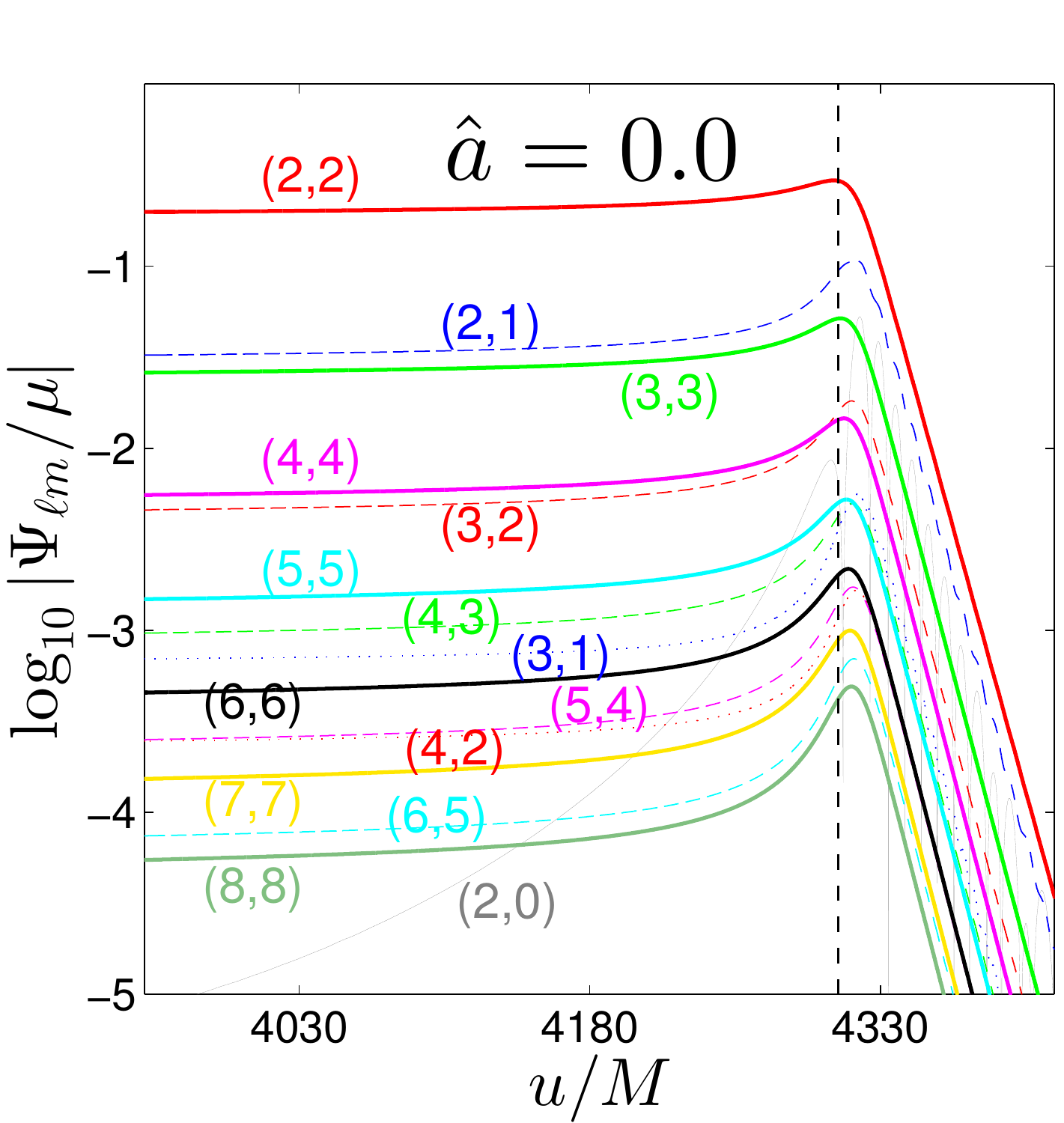}   
  \includegraphics[width=0.49\textwidth]{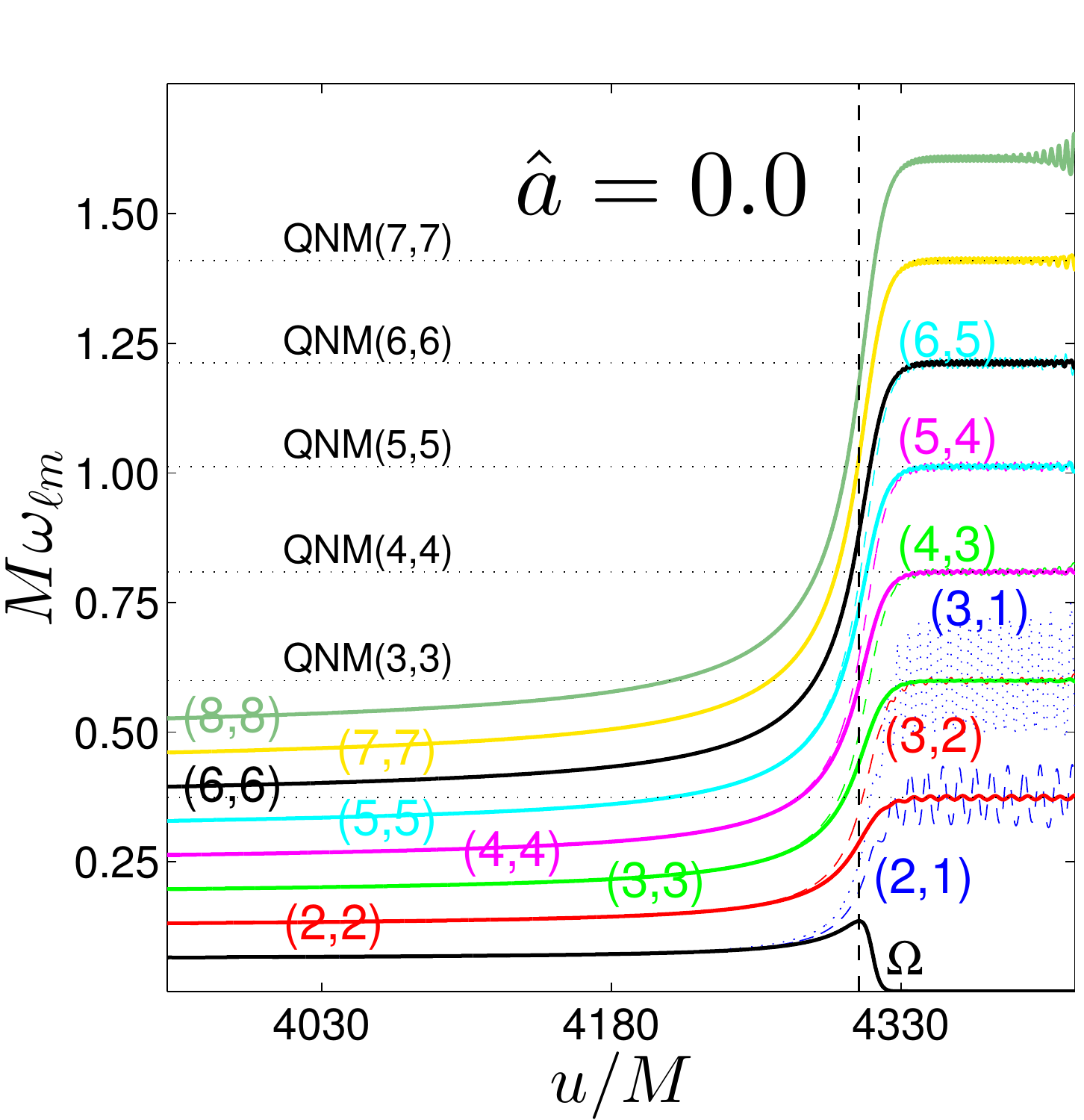}   
  \caption{Multipolar structure around merger of the $\Psi_\lm$ insplunge waveforms
    with $\mu/M=10^{-3}$ and $\ha=0$ obtained from the 2+1 TE. 
    Amplitudes (left) and GW frequencies (right). The vertical lines 
    correspond to the maximum of the orbital frequency $u_{\Omega^{\rm max}}=u_{\rm LR}=4308.39M$
    (the orbital frequency $\hat{\Omega}$ is also shown as a black line in the right panel).
    The horizontal lines in the right panel mark the QNM frequencies 
    of the black hole~\cite{Berti:2009kk}.} 
  \label{fig:EOB_a00_multipolar structure}
\end{figure}

\begin{table}[t]
  \centering
  \caption{Properties of multipolar waveforms at merger for $\ha=0$.
    The retarded time at the crossing of the light ring $u_{\rm LR}$, 
    coincides with the time of the maximum of the orbital frequency,
    $u_{\rm LR}=u_{\Omega^{\rm max}}=4308.39M$
   (for $\ha\neq0$ they can differ, see Table~\ref{tab:test_flux} below)
    , with  $M\Omega^{\rm max}= 0.136$.
    The peak of each multipolar amplitude divided by $\mu$,  
    $\hat{A}_\lm^{\rm max}\equiv A_\lm^{\rm max}/\mu$ occurs at another time 
    $u_{A_\lm^{\max}}\neq u_{\Omega^{\max}}$. The table lists the differences
    $\Delta t_\lm=u_{A_\lm^{\max}}-u_{\Omega^{\rm max}}$. For completeness we also 
    state the peak values $\hat{A}_\lm^{\max}$ and the frequencies at that time $M\omega_\lm^{A_\lm^{\max}}$.
    Values in brackets refer to 1+1 RWZ simulations of~\cite{Bernuzzi:2010xj}.
    The table confirms that the peak of the $\ell=m=2$ multipole occurs
    before the light ring crossing as pointed out in~\cite{Bernuzzi:2010xj}.} 
  \label{tab:peaks_0} 
  \begin{tabular}[t]{ccccc}
\hline
\hline
\vspace{-3.5mm}
\\

    $\ell$ & $m$ & $\Delta t_\lm$ & $\hat{A}_\lm^{\max}$ &
    $M\omega_\lm^{A_\lm^{\max}} $ \\
    \hline
    2 & 2 &  -2.38 (-2.56) &  0.29589  (0.29472)       &  0.27335 (0.27213)  \\ 
    2 & 1 &  9.41  (9.37)  &  0.10694  (0.10692)       &  0.29067 (0.29064) \\ 
    3 & 3 &  1.11  (1.00)  &  0.051673  (0.051456)     &  0.45462 (0.45321) \\ 
    3 & 2 &  6.85  (6.84)  &  0.018170  (0.018174)     &  0.45181 (0.45174) \\ 
    3 & 1 &  10.55 (10.54) &  0.0056954  (0.0056872)   &  0.41176 (0.41129) \\ 
    4 & 4 &  2.90  (2.82)  &  0.014581  (0.014523)     &  0.63541 (0.63400) \\ 
    4 & 3 &  7.22  (7.21)  &  0.0049634  (0.0049653)   &  0.63686 (0.63668) \\ 
    4 & 2 &  9.54  (9.51)  &  0.0016570  (0.0016543)   &  0.62603 (0.62533) \\ 
    5 & 5 &  4.18  (4.12)  &  0.0052278  (0.0052093)   &  0.81811 (0.81672) \\ 
    5 & 4 &  7.63  (7.63)  &  0.0017267  (0.0017277)   &  0.82170 (0.82148) \\ 
    6 & 6 &  5.20  (5.14)  &  0.0021703  (0.0021636)   &  1.00027 (1.00013) \\ 
    6 & 5 &  8.09  (8.09)  &  0.00069673  (0.00069726) &  1.00079 (1.00077) \\   
    \hline
    \hline
  \end{tabular} 
\end{table}

We compare the 2+1 TE data for $\ha=0$ with the 1+1 RWZE data 
of~\cite{Bernuzzi:2010xj,Bernuzzi:2011aj}. We use exactly the same
dynamics for both sets of simulations, with $\hr_0=7$. 
Figure~\ref{fig:comparison_qualitative_EOB_a00} 
shows the RWZE $\Psi_{22}$ complete
(inspiral-plunge-ringdown) waveform extracted at $\scri$ as computed 
from two simulations with $\mu/M=10^{-3}$. The data are in excellent visual
agreement also during the ringdown. A quantitative comparison for various
multipoles is shown in Fig.~\ref{fig:comparison_quantitative_EOB_a00},
that reports phase and amplitude differences. 
Note that no time/phase alignment is required to perform such a
comparison since both sets of waveforms are extracted at scri and
generated from the same dynamics.
Phase differences are
$\Delta\phi_\lm=|\phi^{\rm TE}_\lm-\phi^{\rm RWZ}_\lm|\lesssim10^{-3}$~rad until
the time of the light ring crossing ($u_{\rm LR}=u_{\Omega^{\rm max}}=4308.39M$) 
and remain below $0.01$ during the ringdown ($u > u_{\rm LR}$). 
The relative amplitude differences $\Delta A_\lm/A_\lm$ are at the order of
$0.25\%$ until $u_{\rm LR}$ and remain $\lesssim1.25\%$ during
the ringdown. 

These differences are larger than those estimated from
self-convergence tests, so they have systematic origin. 
The two codes are independent, use different 
coordinate systems, solve different equations, and differ in many implementation
details. Furthermore, the {\it Teukode} is a 2+1 code and the 2+1 RWZ
waveforms are reconstructed from $\psi_4$ by integration. 
Remarkably, these systematic differences in phase and amplitude are
small enough to be  negligible for many practical purposes. 
We note in particular that the $\Delta\phi_\lm$ are
significantly smaller than the differences between RWZE and EOB
waveforms found in~\cite{Bernuzzi:2010xj,Damour:2012ky}. Thus, the result of
that analysis is robust and confirmed here with an independent
waveform data set. In particular, in~\cite{Bernuzzi:2010xj} it was
reported for the first time that the peak of the $\ell=m=2$ multipole
is located earlier in time than the peak of the orbital frequency
(for $\ha=0$ the peak coincides with $u_{\rm LR}$). This observation 
is confirmed also by the 2+1 TE data, see Table~\ref{tab:peaks_0}.

The multipolar amplitudes and frequencies near merger are shown 
in Fig.~\ref{fig:EOB_a00_multipolar structure}. The data
agree with the analysis of the multipolar structure performed
in~\cite{Bernuzzi:2010ty} within the 1+1 RWZE approach. Note in
particular the oscillations in the quasi normal mode (QNM) frequencies, e.g.
in modes $(2,1)$,$(2,2)$ and $(3,1)$. As explained in~\cite{Damour:2007xr,Bernuzzi:2010ty}, 
these oscillations arise from the interference between positive ($m>0$) 
and negative ($m<0$) QNM frequencies. 

We analyzed also the effect of finite-radius extraction
on the waveforms; these results are reported in
Sec.~\ref{sec:inspl:finiter} together with $\ha\neq0$ data.

\section{From quasi-circular inspiral to merger and ringdown: dynamics
  and waveforms for $\ha\neq 0$}
\label{sec:inspl}

In this Section we discuss the structure of the waveforms emitted from
inspiraling and coalescing configurations with spin $\ha\neq 0$. 
The underlying dynamics of the particle is computed according 
to Sec.~\ref{sec:dyn_RR}. 
The purpose of this Section is threefold: 
(i) check the consistency of the (two) analytical expressions of the
mechanical angular momentum loss with the angular momentum flux
computed from the waves, Sec.~\ref{sec:inspl:fluxes};  
(ii) characterize quantitatively the multipolar waveform around merger,
as obtained with $\hat{\cal F}_\phi^{v_\phi}$, for $|\ha|\leq0.9$ in 
Sec.~\ref{sec:inspl:09} and for nearly-extremal configurations $0.9<|\ha|\leq0.9999$ 
in Sec.~\ref{sec:inspl:09999}, (iii) quantify waveform 
extrapolation errors in Sec.~\ref{sec:inspl:finiter}.

All simulations in this work refer to $\mu/M=10^{-3}$ and post-circular initial
data. Simulations are done for different values of 
black hole spins $\ha\in[-0.9999,+0.9999]$ and $m =0,1,...,8$.
For $|\ha|<0.99$ the initial separations $\hr_0$ were chosen 
such that the systems perform about $\sim 25$ orbits before merger.
This implies that the inspiral is
``very-strong-field'' when the spins are highly positive (with $\hr_0\sim 3$),
and relatively ``less-strong-field'' when spins are highly negative (with $\hr_0\sim 10$).
For nearly-extremal simulations the pronounced potential-well close to the black hole
``traps`` the junk radiation for long times (e.g. $1000M$ for $\ha=0.9999$), so that we 
needed bigger separations/longer inspirals for $|\ha|\geq0.99$.
We collect in Table~\ref{tab:test_flux} in~\ref{app:DataTables}
detailed information about the configurations we consider. 

The simulations discussed below use $6$th-order finite
differences. The resolution employed is $N_x\times N_\theta=(3600\times160)$. 
This setup is chosen to guarantee that the truncation errors, as
estimated from self-convergence tests (Sec.~\ref{sec:test:conv}), are
around $10^{-10}$, at least for the
dominant multipoles and for most of the configurations. 


\subsection{Checking the consistency of the analytical radiation reaction}
\label{sec:inspl:fluxes}

\begin{figure}[t]
  \includegraphics[width=0.49\textwidth]{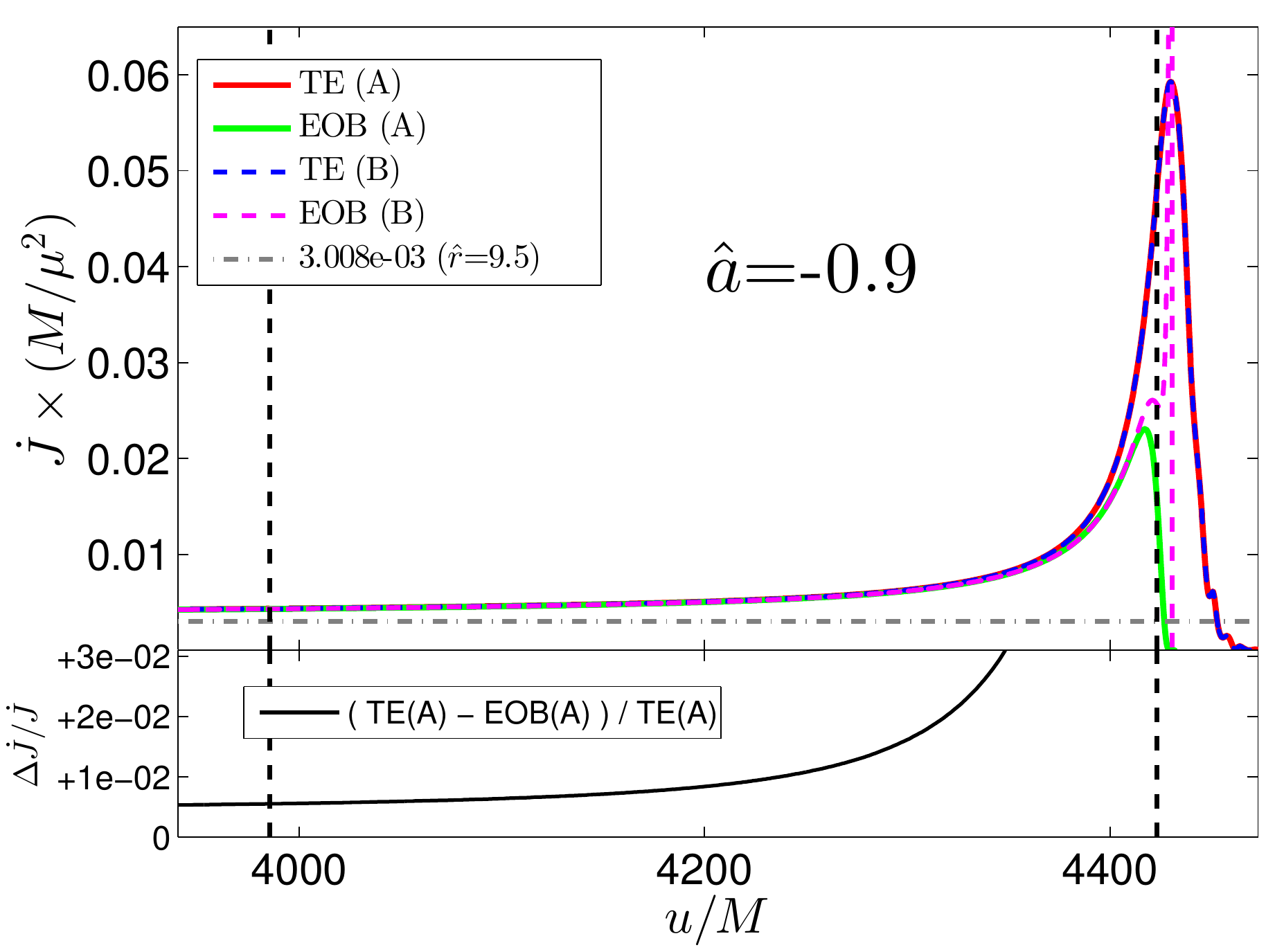}
  \includegraphics[width=0.49\textwidth]{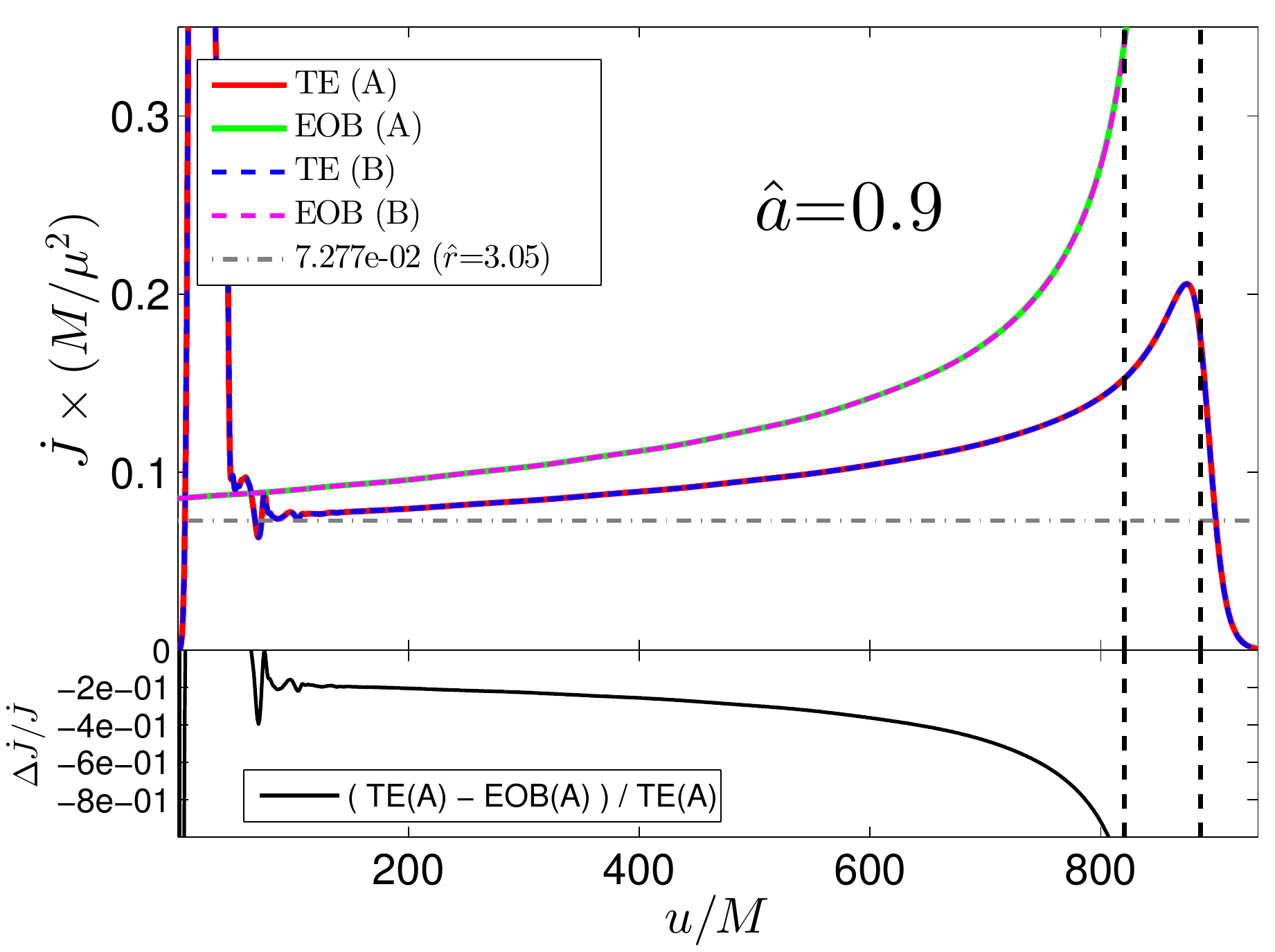}
   \caption{Consistency of numerical and EOB fluxes for $\hat{a}=-0.9$ (left) and
     $\hat{a}=+0.9$ (right). The top panels show the two analytical EOB-flux-prescriptions
     (A) and (B) (discussed in Sec.~\ref{sec:dyn_RR}) and the respective numerical fluxes
     produced using those prescriptions. The vertical lines correspond to the LSO and the LR crossing. 
     The horizontal line is the circular flux corresponding to the initial separation.
     The bottom panels show the difference between the numerical and 
     the analytical flux when using prescription (A) 
     (the respective line for (B) would lie on top in the plotted sector). 
     Looking at the bottom left panel the analytical prescriptions for $\hat{a}=-0.9$ match
     the numerical fluxes within $\lesssim1\%$ until well beyond the
     LSO. Instead, for $\hat{a}=0.9$ the analytical information is less accurate ($\sim100\%$ off at LSO).
     The numerical fluxes (A) and (B) are visually the same in both plots though
     the flux prescriptions differ significantly at LR. }  
  \label{fig:EOB_fluxes_TKvsDNBB0}
\end{figure}

\begin{figure}[t]
  \includegraphics[width=0.49\textwidth]{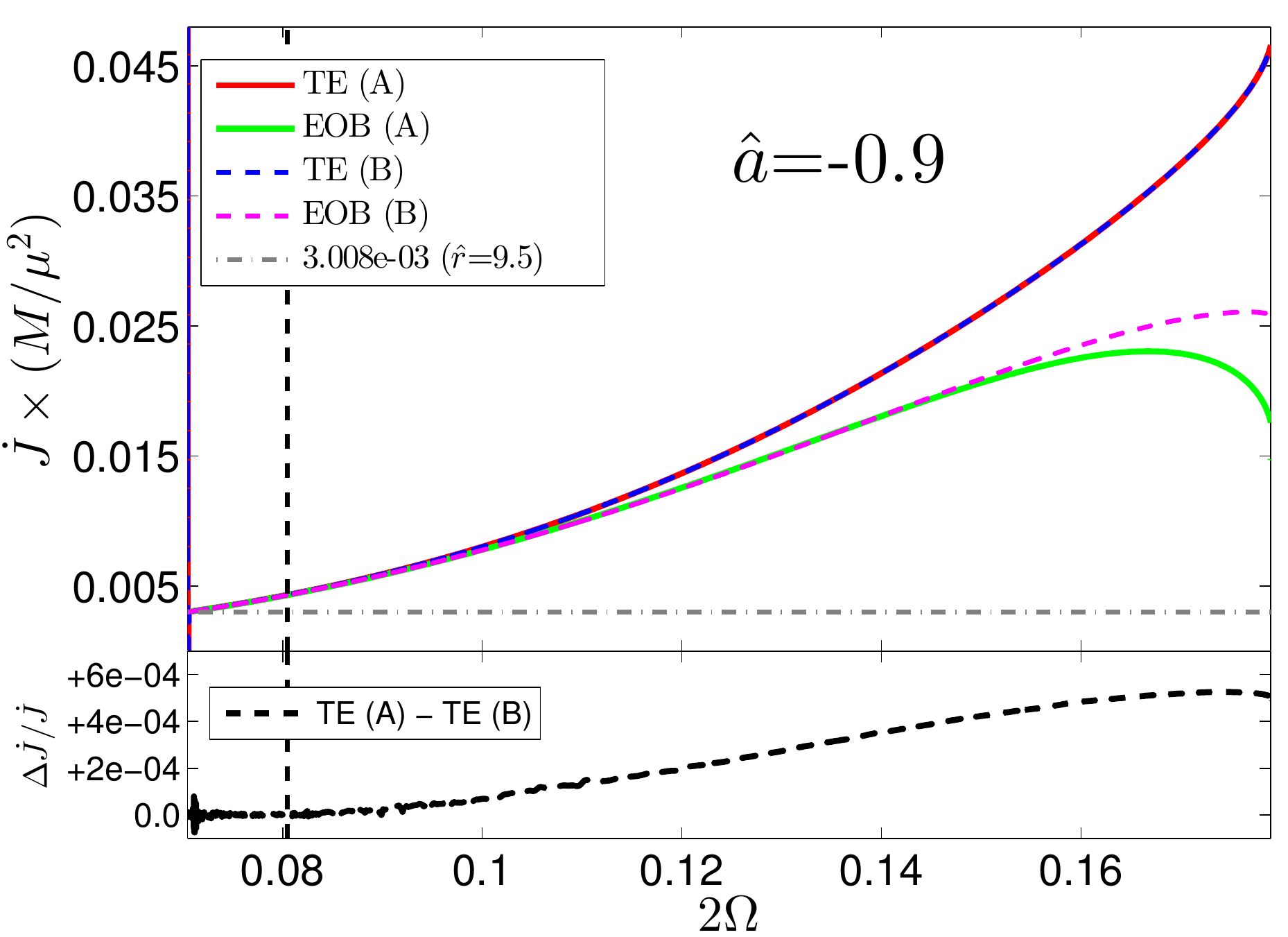}
  \includegraphics[width=0.49\textwidth]{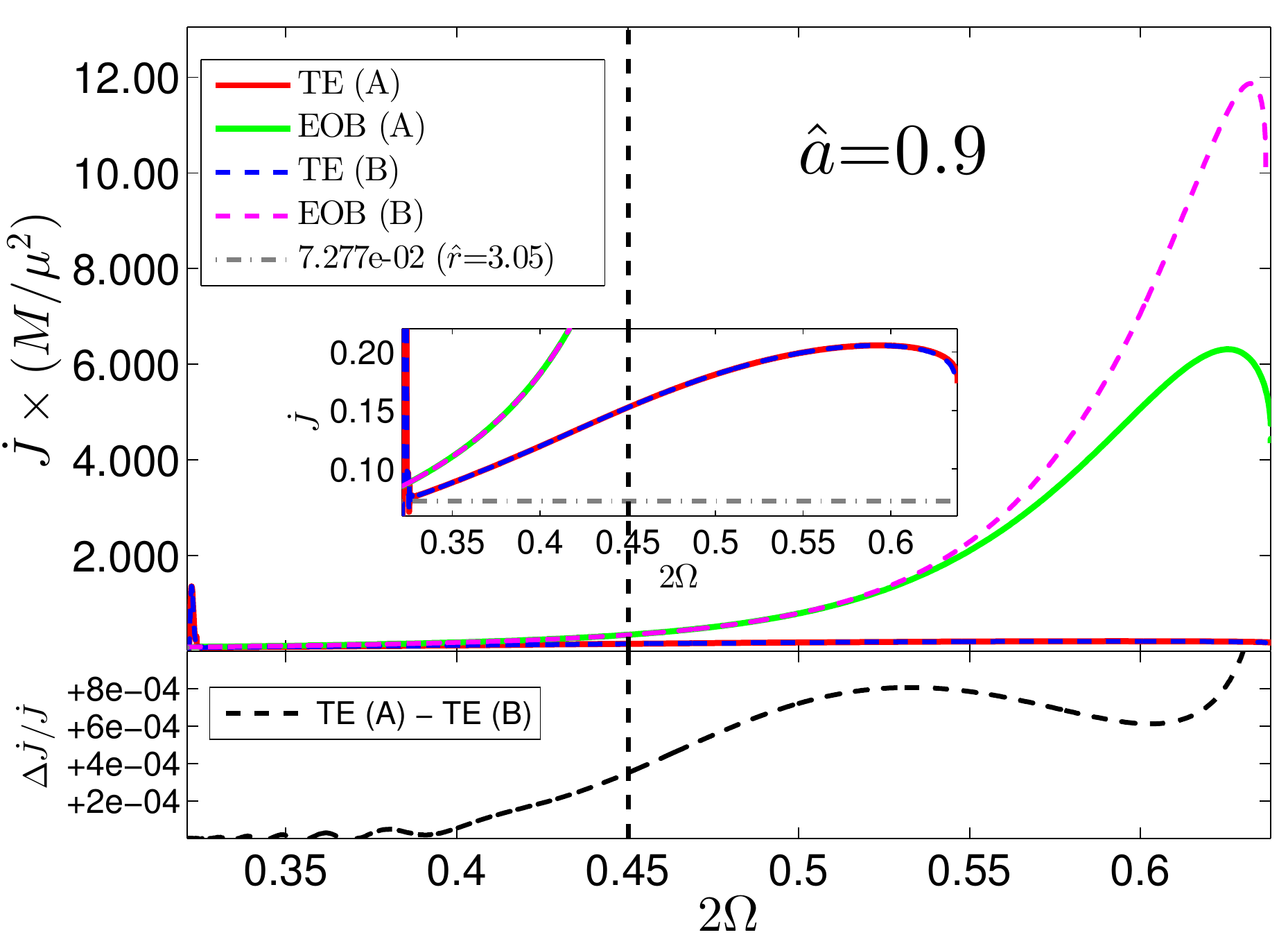}
     \caption{Consistency of numerical and EOB analytical fluxes for $\ha=-0.9$ (left)
     and $\ha=+0.9$ (right). The two major panels show the EOB-fluxes as obtained 
     with prescriptions (A), i.e. $-\hat{\cal{F}}_\phi^{v_\phi}$,  
     and (B), i.e. $-\hat{\cal{F}}_\phi^{v_\Omega}$, (discussed in
     Sec.~\ref{sec:dyn_RR}) and the two respective numerical fluxes 
     as functions of twice the orbital frequency $\hat{\Omega}=M\Omega$.
     The bottom panels show the
     difference between the numerical flux as produced with prescription (A) and (B).
     The vertical lines correspond to the LSO, the plots
     terminate at the LR. The horizontal line is the circular flux
     corresponding to the initial separation, which was chosen to 
     provide $\sim25$ orbits before merger. Note that all 
     orbits are spent before the LSO,
     $\hat{\Omega}<\hat{\Omega}_{\rm LSO}$.}     
  \label{fig:EOB_fluxes}
\end{figure}

We compare the two prescriptions for the radiation reaction
$\hat{\mathcal{F}}_\phi$ described in Sec.~\ref{sec:dyn_RR} with the
numerical fluxes of the TE, to complement the analysis in Sec.~\ref{sec:dyn_RR}.
For both $\hat{\cal{F}}_\phi^{v_\phi}$  and $\hat{\cal{F}}_\phi^{v_\Omega}$
we compute an insplunge trajectory and perform a TE simulation. The
consistency between the analytical flux and the numerical one (or
``exact'', for the given dynamics) is a crucial test of the
consistency of the analytical model. Note that the comparison presented 
here is meaningful because in our setup the TE waveforms are extracted 
at scri. In the nonspinning case the accuracy of the 5PN-accurate, resummed, 
analytical radiation reaction has been tested in~\cite{Damour:2007xr,Bernuzzi:2010ty,Bernuzzi:2011aj}. 
In these references it was shown that (for $\mu/M \leq 10^{-3}$) 
the 5PN-accurate information yields a radiation reaction consistent up
to a few percent even below the LSO crossing.

When $\hat{a}\neq 0$ things are more complicated since the accuracy
of the analytical flux, that is based on a limited PN knowledge of the
$\rho_\lm$'s, actually depends on the value of $\hat{a}$. 
As already pointed out in previous works, the analytical information we are employing here 
is not sufficient to guarantee agreement between
analytical and numerical fluxes for $\hat{a}>0.7$. 
Practically speaking, the analytical radiation reaction
is too large and yields a smaller number of GW cycles up to merger 
than what it should be. 
This discrepancy was pointed out in Ref.~\cite{Pan:2010hz},
and thoroughly analyzed in follow-up
works~\cite{Yunes:2010zj,Barausse:2011kb,Taracchini:2013wfa,Taracchini:2014zpa}. 
In particular, Ref.~\cite{Taracchini:2013wfa} proposed to fit several
high-order coefficients entering the $\rho_\lm$'s to the circular flux
in order to improve the behavior of the purely analytical $\rho_\lm$'s as
$\ha\to 1$ (, see their Fig.~13).
Here, for simplicity, we rely exclusively on
analytical information, though we do not expect
good consistency between the analytical and numerical fluxes 
for large, positive values of the black hole spin.
In Sec.~\ref{sec:sf_flux} we present a method to calculate the {\it
  consistent} GW flux at linear order in $\nu$.

Let us compare the GW fluxes for spins $\ha=\pm0.9$ (with $\hr_0$ as in Table~\ref{tab:test_flux}).
Figure~\ref{fig:EOB_fluxes_TKvsDNBB0} displays the comparison in the time domain. In the top panels we contrast 
$\dot{J}$ with $-\hat{{\cal F}}_\phi^{v_\phi}$ and $-\hat{\cal{F}}^{v_\Omega}_\phi$ 
while the bottom panels show the relative difference between $\dot{J}$ and $-\hat{{\cal F}}_\phi^{v_\phi}$. 
The two vertical lines on the plot indicate, from left to right, the LSO and the LR crossing.
As expected, we find excellent agreement between the analytical and numerical
flux when $\ha=-0.9$ (also beyond the LSO and almost up to merger), whereas the agreement
is poor for $\ha=+0.9$ (the analytical flux is off by a factor 2 at LSO crossing) due
to the lack of higher PN information in the $\rho_\lm$'s. This discrepancy holds for both 
implementations of the non-Keplerian behavior we discussed in Sec.~\ref{sec:dyn_RR}. 
Instead, the differences in the prescriptions manifest in the behavior of the fluxes 
for $\ha=-0.9$ around the light ring crossing time. $-\hat{\cal{F}}^{v_\Omega}_\phi$
(dashed magenta line in Fig.~\ref{fig:EOB_fluxes_TKvsDNBB0}) deviates from 
$-\hat{\cal{F}}^{v_\phi}_\phi$, consistently with the $\ha=-0.99$ case analyzed above,
with $-\hat{{\cal F}}_\phi^{v_\phi}$ remaining close, in shape, to $\dot{J}$.

\begin{figure}[t]
  \centering  
  \includegraphics[width=0.32\textwidth]{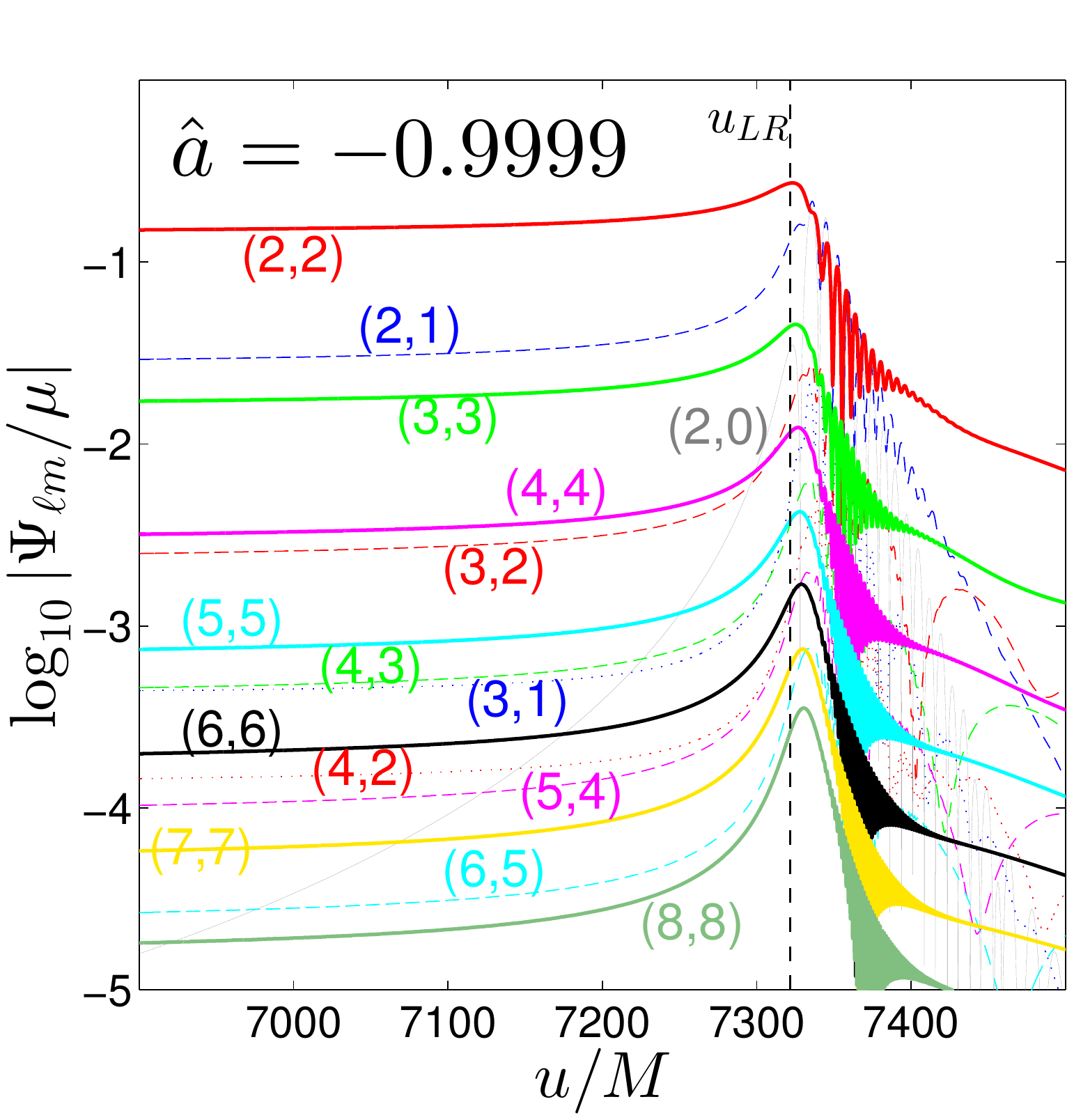}
  \includegraphics[width=0.32\textwidth]{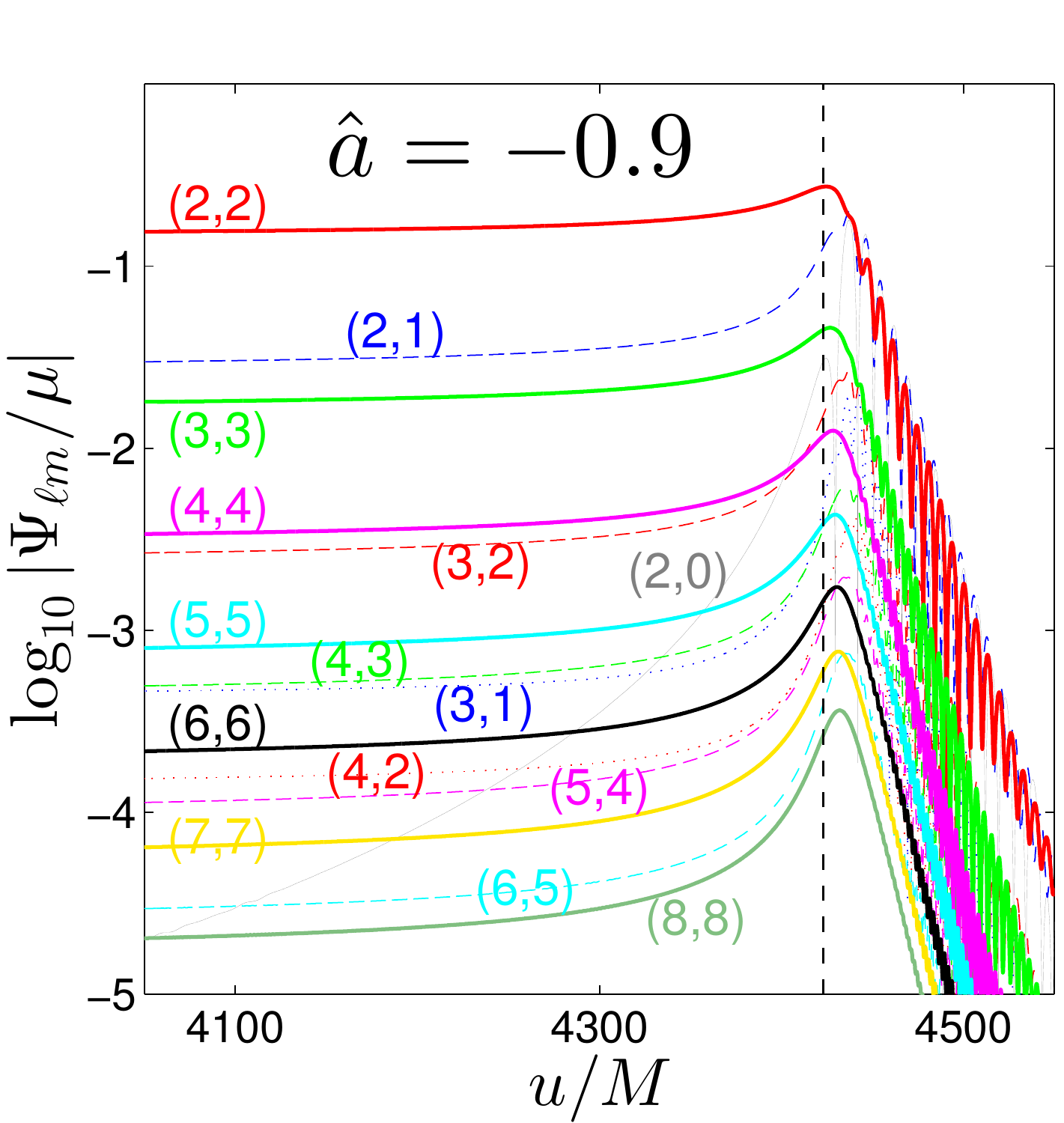}   
  \includegraphics[width=0.32\textwidth]{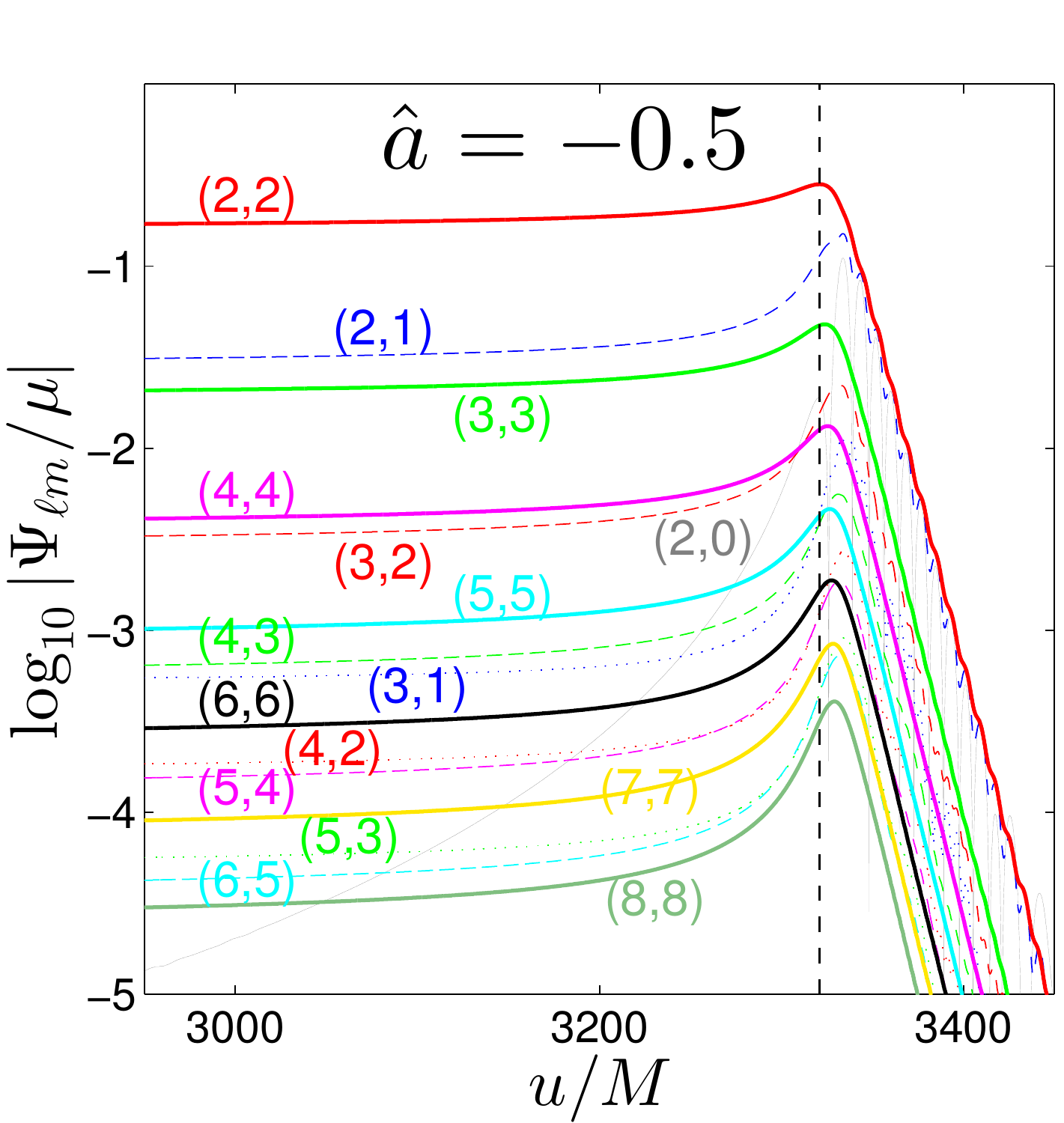}   \\
  \includegraphics[width=0.32\textwidth]{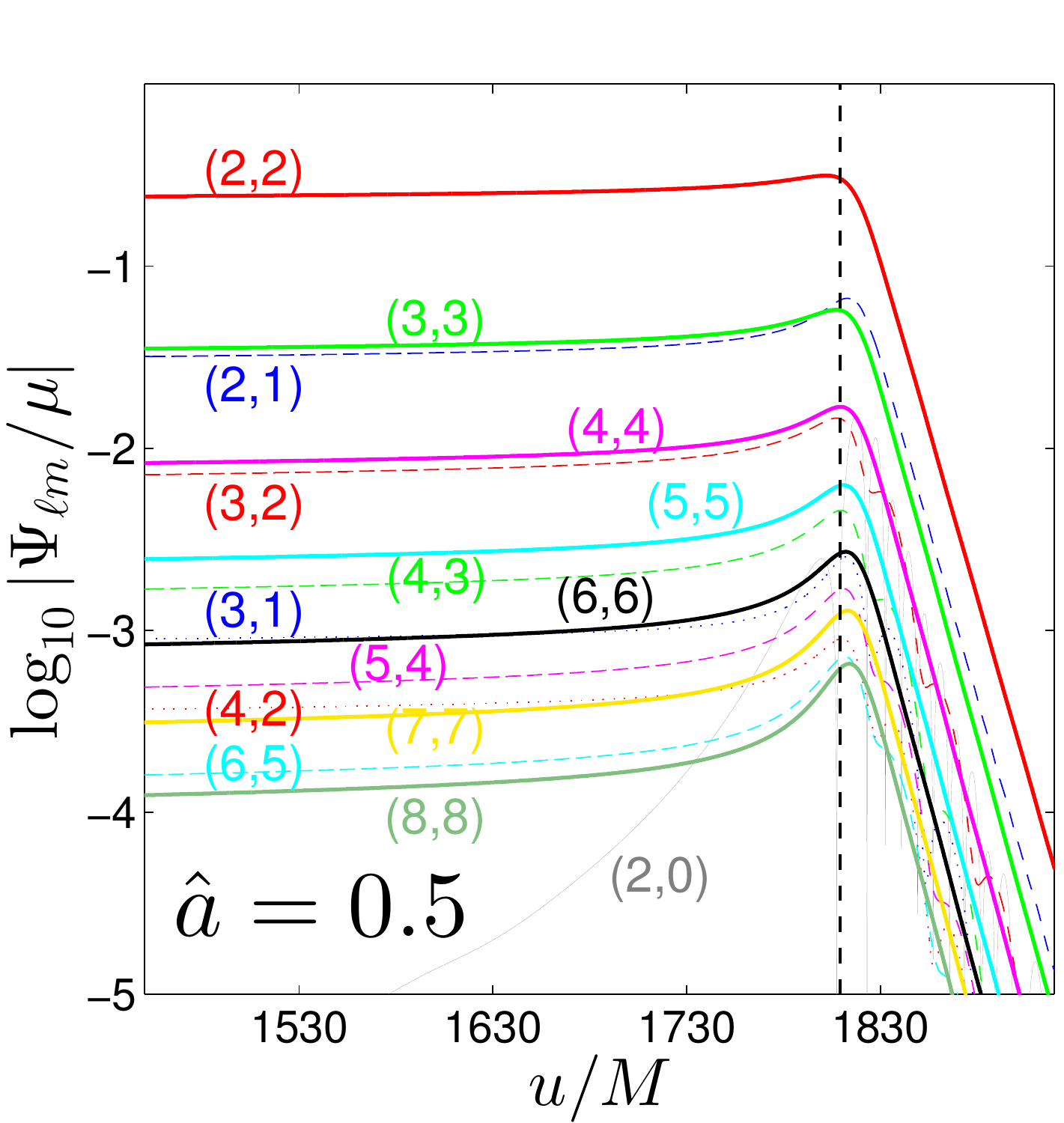}   
  \includegraphics[width=0.32\textwidth]{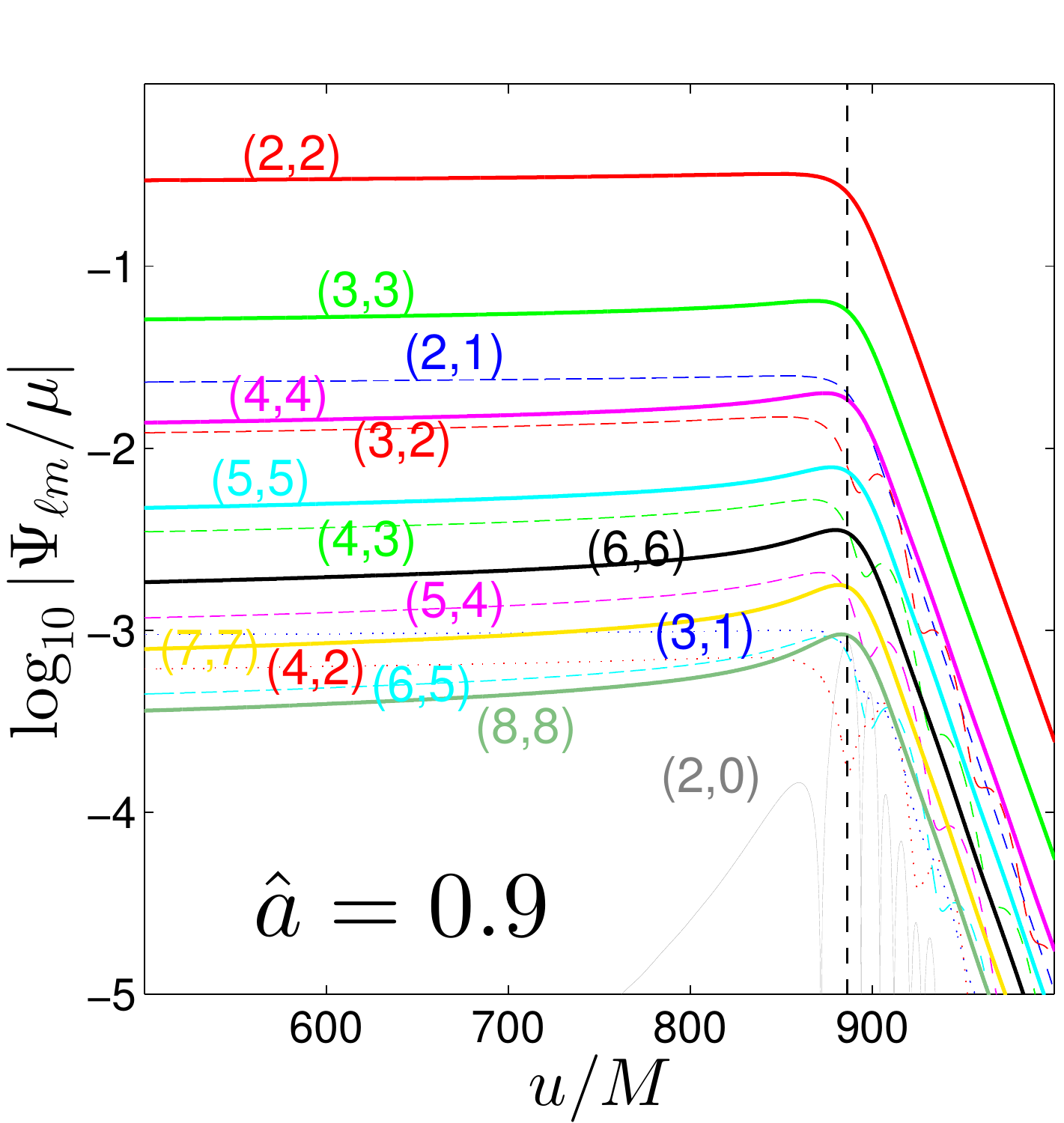}   
  \includegraphics[width=0.32\textwidth]{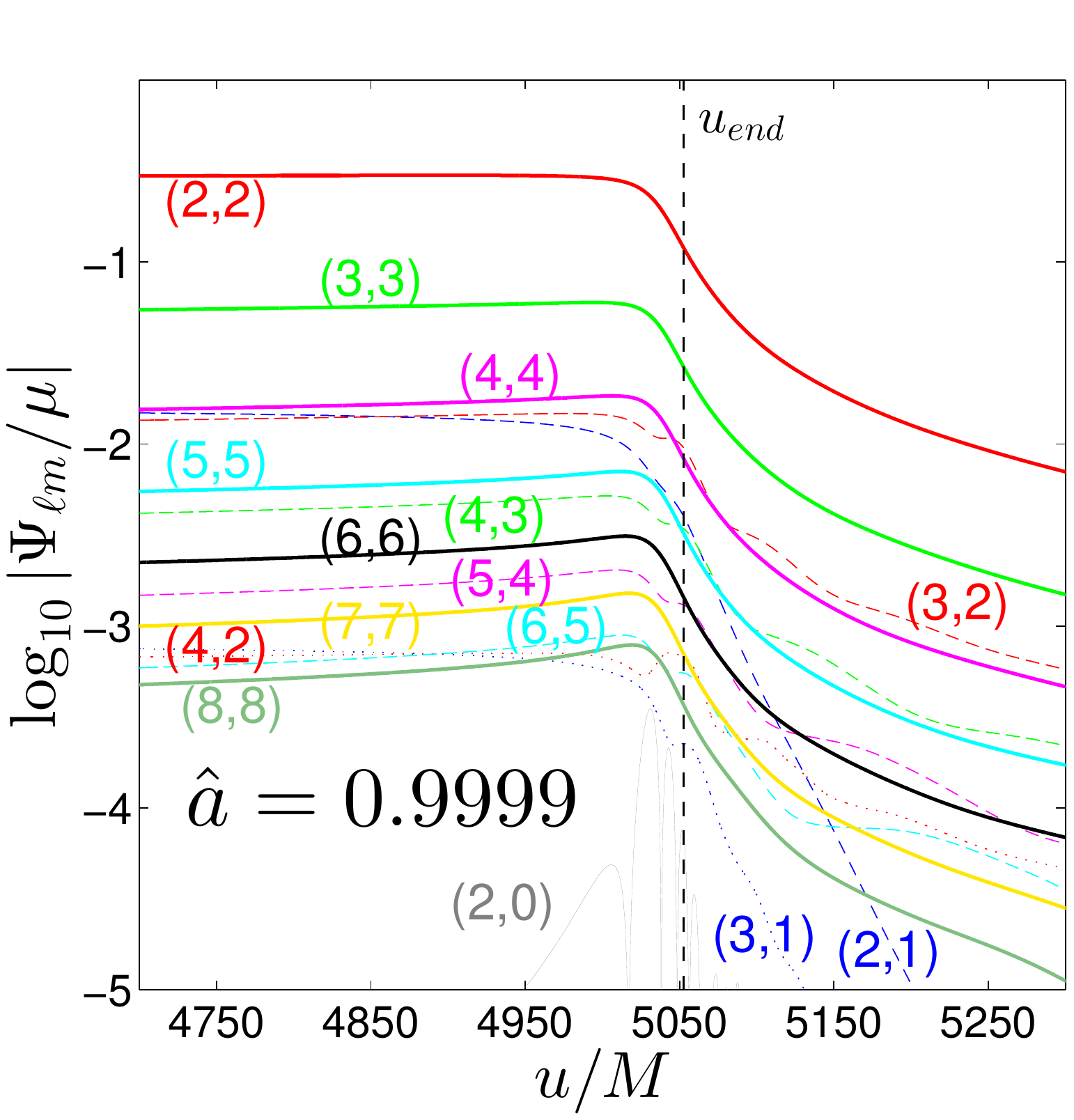}   
  \caption{Multipolar amplitudes of the $\mu/M=10^{-3}$ insplunge waveforms
    for certain values of $\ha$.
    The vertical line on each plot marks the 
    crossing of the LR, except for $+0.9999$, where it marks the end 
    of the trajectory (in this case our trajectory stops slightly before the
    horizon, see discussion in text).
     }
  \label{fig:EOB_multipolar_structure_amp}
\end{figure}

\begin{figure}[t]
  \centering
  \includegraphics[width=0.49\textwidth]{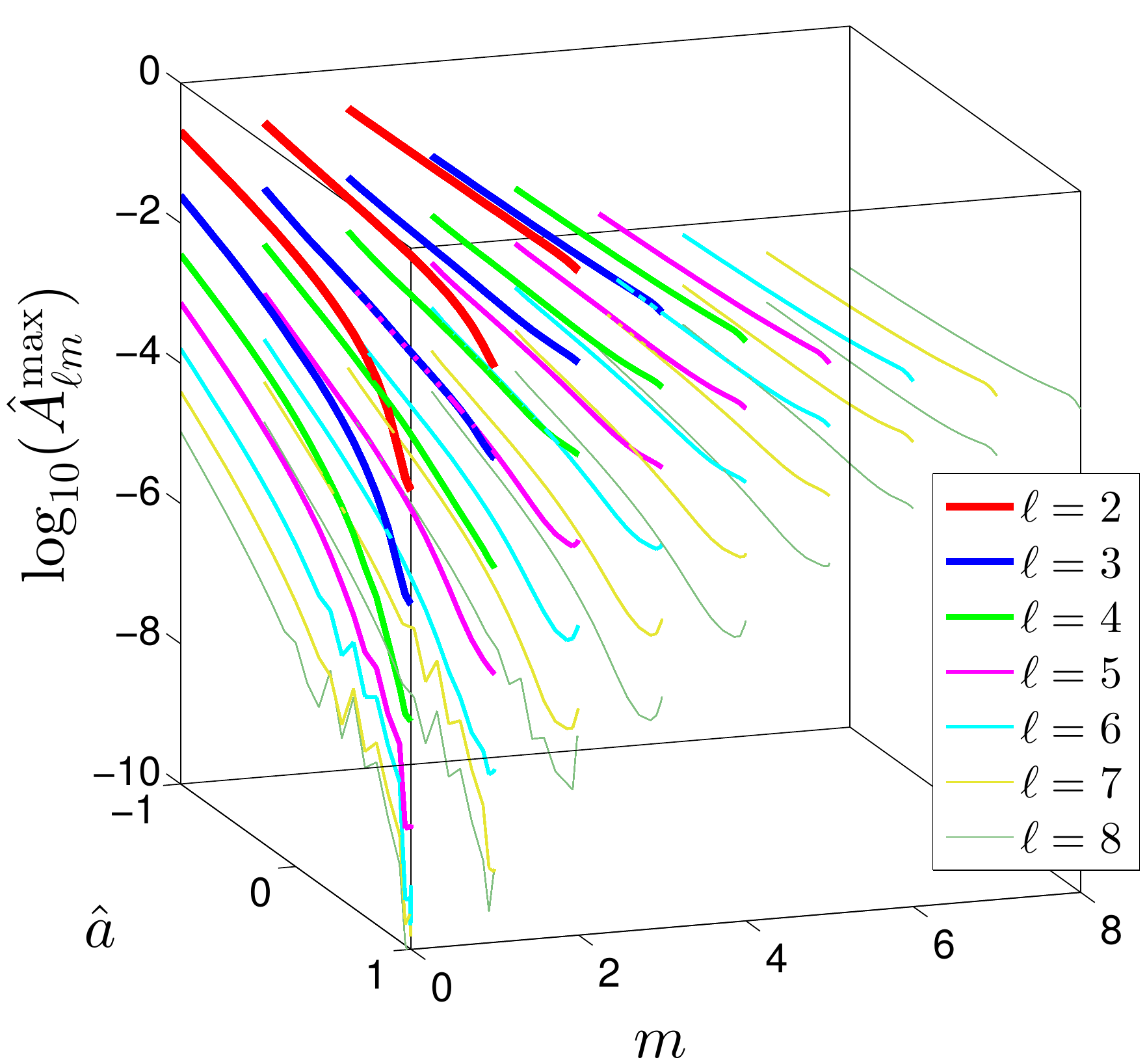}
  \includegraphics[width=0.49\textwidth]{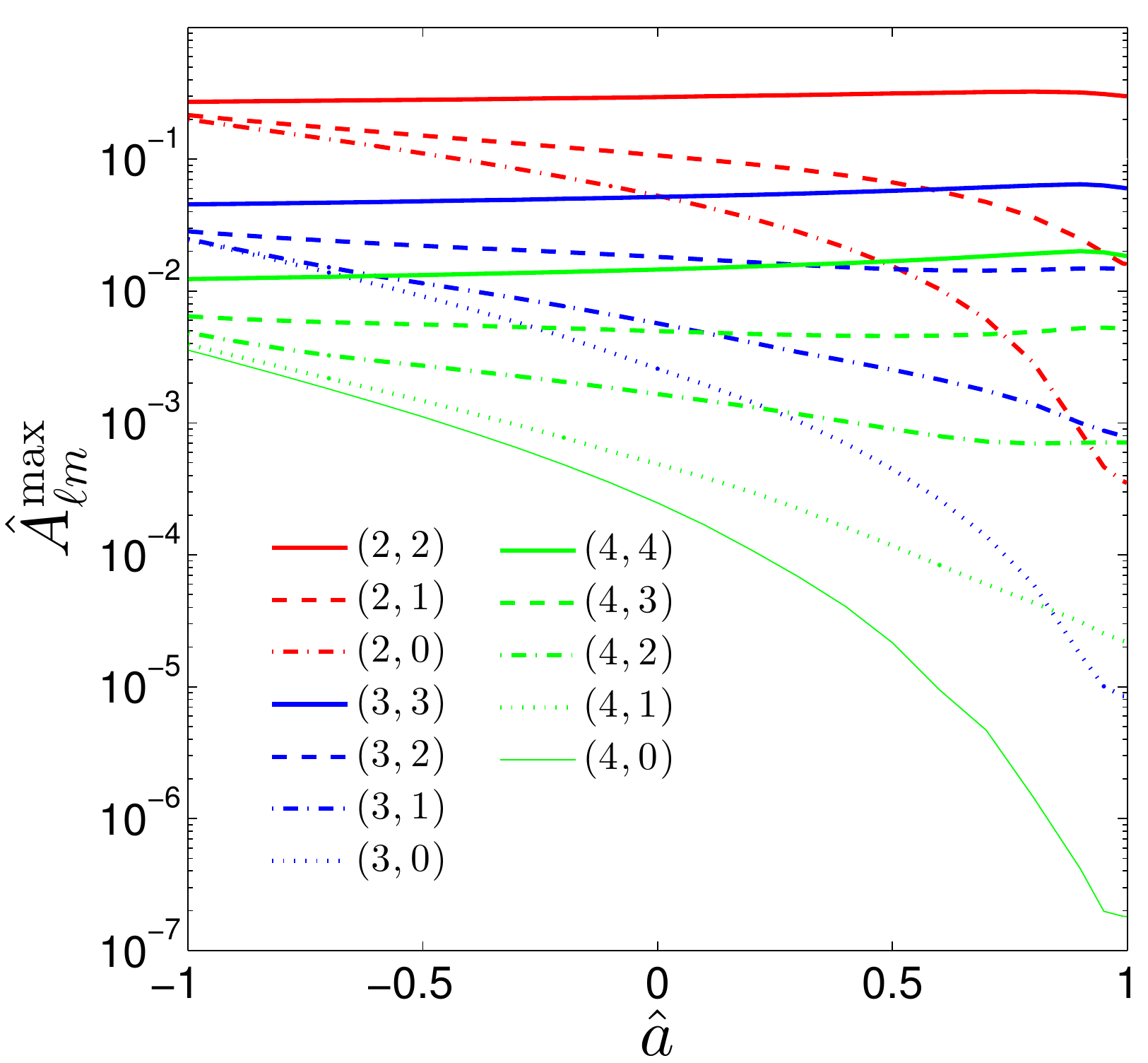}
  \caption{Peaks of multipolar amplitudes as functions of the black
    hole spin. 
    Left: 3D visualization of all the data up to
    $\ell=m=8$. $\ell$-modes are identified with colors, the spin
    dependence is marked by solid lines.
    Right: peaks of the dominant multipoles as functions of the
    spin. This plot is a 2D cut of the left panel (restricting to $\ell\leq4$).
    The $m=0$ and $m=1$ modes have the largest amplitudes for $\ha\to-1$. 
    The diagonal modes have weak dependence on the spin and attain a
    maximum for spin values $0.8<\ha<1$.} 
  \label{fig:multipolar_peaks}
\end{figure}

Figure~\ref{fig:EOB_fluxes} illustrates the same comparison
under a different and complementary point of
view: the fluxes are plotted versus the orbital frequency $\hat{\Omega}=M\Omega$.
Note that the $\sim25$ orbits (most of the simulated time) have  
frequencies $\hat{\Omega}<\hat{\Omega}_{\rm LSO}$ (the LSO is marked by vertical lines). 
For $\ha<0$ (left panel), the analytical
prescription (A) ((B)) deviates from the numerical flux by about $1$\% ($1$\%)
at the LSO and $60$\% ($30$\%) at the LR. This is roughly
comparable to the nonspinning case~\cite{Bernuzzi:2011aj}.
The unphysical feature of prescription (B) for $\ha<0$ as
discussed in Sec.~\ref{sec:dyn_RR} happens at $r<r_{\rm LR}$ and does therefore not 
affect the waveforms: both analytical fluxes generate almost
identical numerical fluxes. Differences between the two TE fluxes are
below $0.2\%$ during the whole insplunge.
The reason for this is simply that during the plunge 
(i.e. after the LSO crossing) the radiation reaction does not 
play a significant role (quasi-geodetic plunge.) Although the 
standard flux can still be improved, we consider the 
results for $\ha\leq0$ quite satisfactory for studying
large-mass-ratio mergers.  
For $\ha=0.9$ the figure illustrates clearly how both analytical 
prescriptions systematically deviate from the numerical fluxes already 
at the beginning of our (rather strong-field) inspiral. Comparing 
the deviations for different $\ha>0$ runs we find that higher spins suffer 
from larger deviations. For instance, for $\ha=0.5$ both 
flux prescriptions deviate from the numerical outcomes by about $2$\% at 
the LSO and more than $60$\% at the LR; for $\ha=0.9$ the deviation 
is about a factor two already at the LSO. The reason is that, differently 
from $\ha<0$, the plunge dynamics for $\ha\to1$ is very circular and the radiation 
reaction plays a significant role also in the strong-field regime. 
We conclude that an urgent step for future analytical developments 
(beyond using the effective fits of~\cite{Taracchini:2013wfa}) 
is the inclusion of higher PN corrections to the $\rho_\lm$'s 
residual amplitude corrections~\cite{Shah:2014tka}.

\subsection{Waveforms for $\ha\neq0$: multipolar hierarchy at merger}
\label{sec:inspl:09}

\begin{figure}[t]
  \centering
  \includegraphics[width=0.49\textwidth]{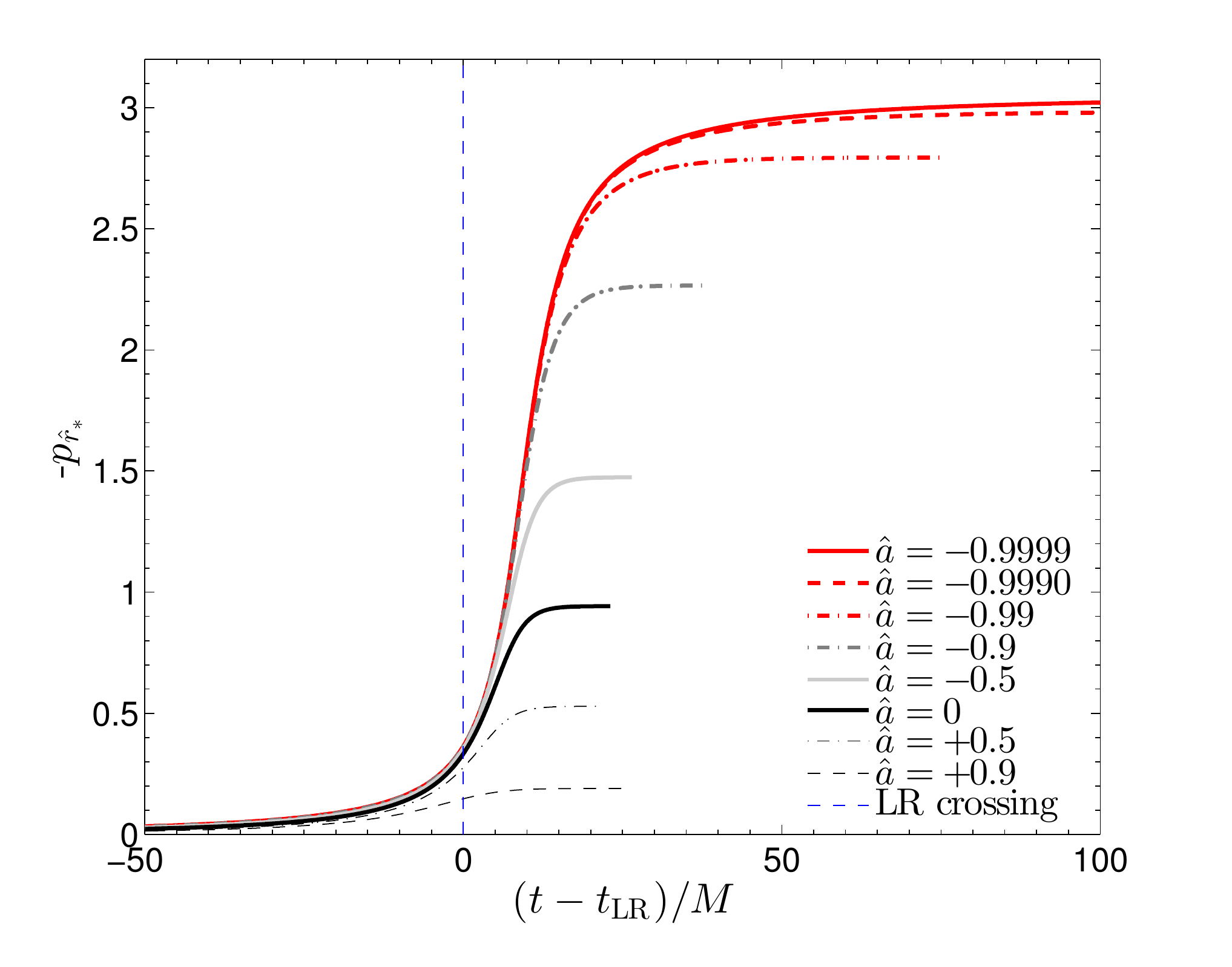}
  \caption{Time evolution of the radial momentum $p_{r_{*}}$ around the light-ring crossing time
  as a function of $\ha$. The magnification of $p_{r_{*}}$ around merger when $\ha\to -1$ 
  is responsible for the sharpness of the peak of the multipolar waveform and 
  the increased importance of subdominant multipoles.} 
  \label{fig:prstar}
\end{figure}

Gravitational waveforms for a particle inspiraling and plunging into a Kerr
black hole have been computed for the first time by Sundararajan et al.~\cite{Sundararajan:2010sr}
and then used and updated in~\cite{Barausse:2011kb,Taracchini:2014zpa}. In particular, 
the recent Taracchini et al.~\cite{Taracchini:2014zpa} paper (that
appeared while the current study was being finalized) presented a
detailed analysis of the ringdown waveforms for 
modes $(\ell,m)=(2,2), (2,1),(3,3),(3,2),(4,4),(5,5)$. Our study
confirms previous findings but also extends/complements them because it 
(i)~computes TE data for all multipoles up to $\ell=m=8$; 
(ii)~considers higher values of the black hole spin; 
(iii)~presents a detailed analysis of $m=0$ modes; 
and (iv) explicitly connects the structure of the multipolar 
waveform around merger with the noncircularity of the plunge.

\begin{figure}[t]
  \centering
  \includegraphics[width=0.32\textwidth]{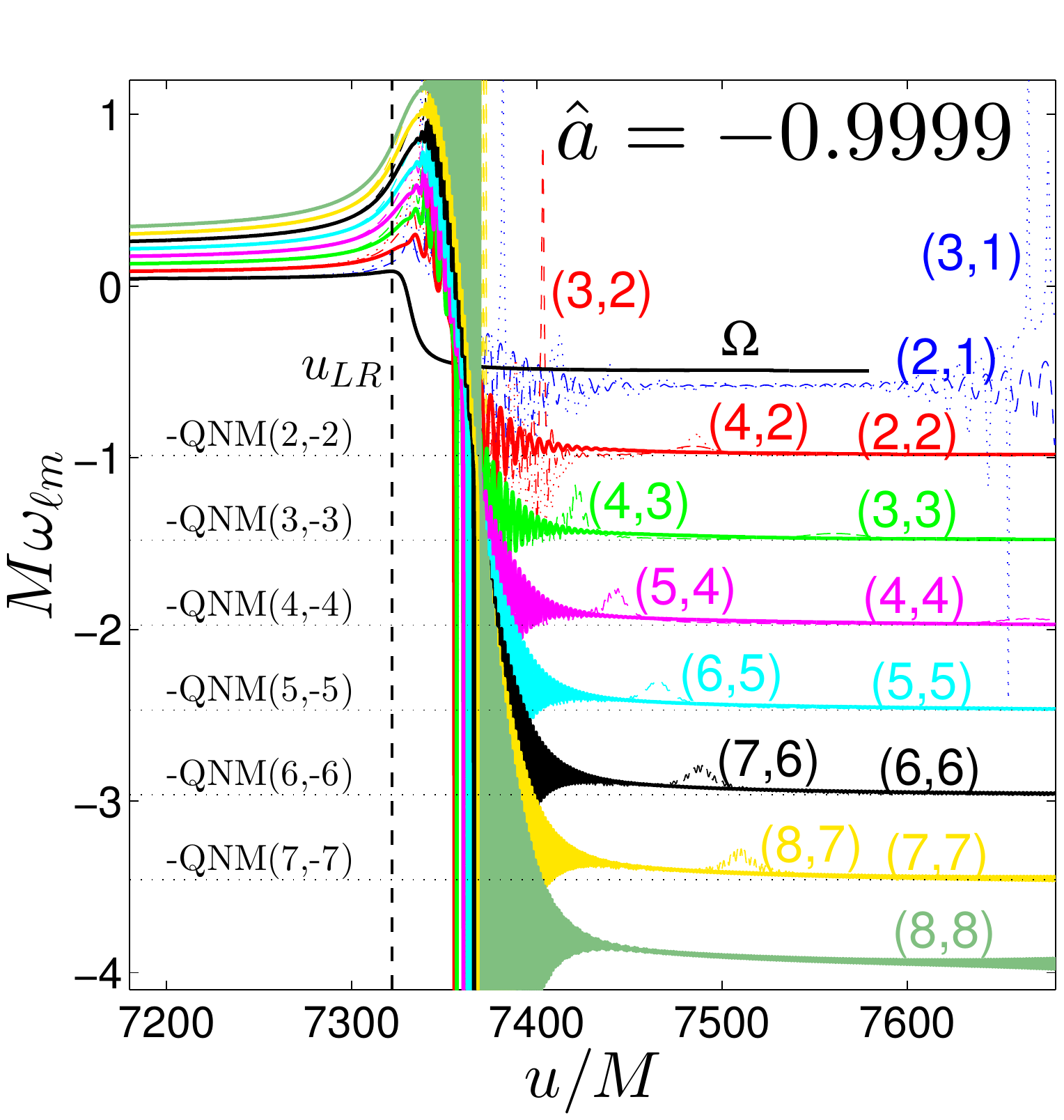}
  \includegraphics[width=0.32\textwidth]{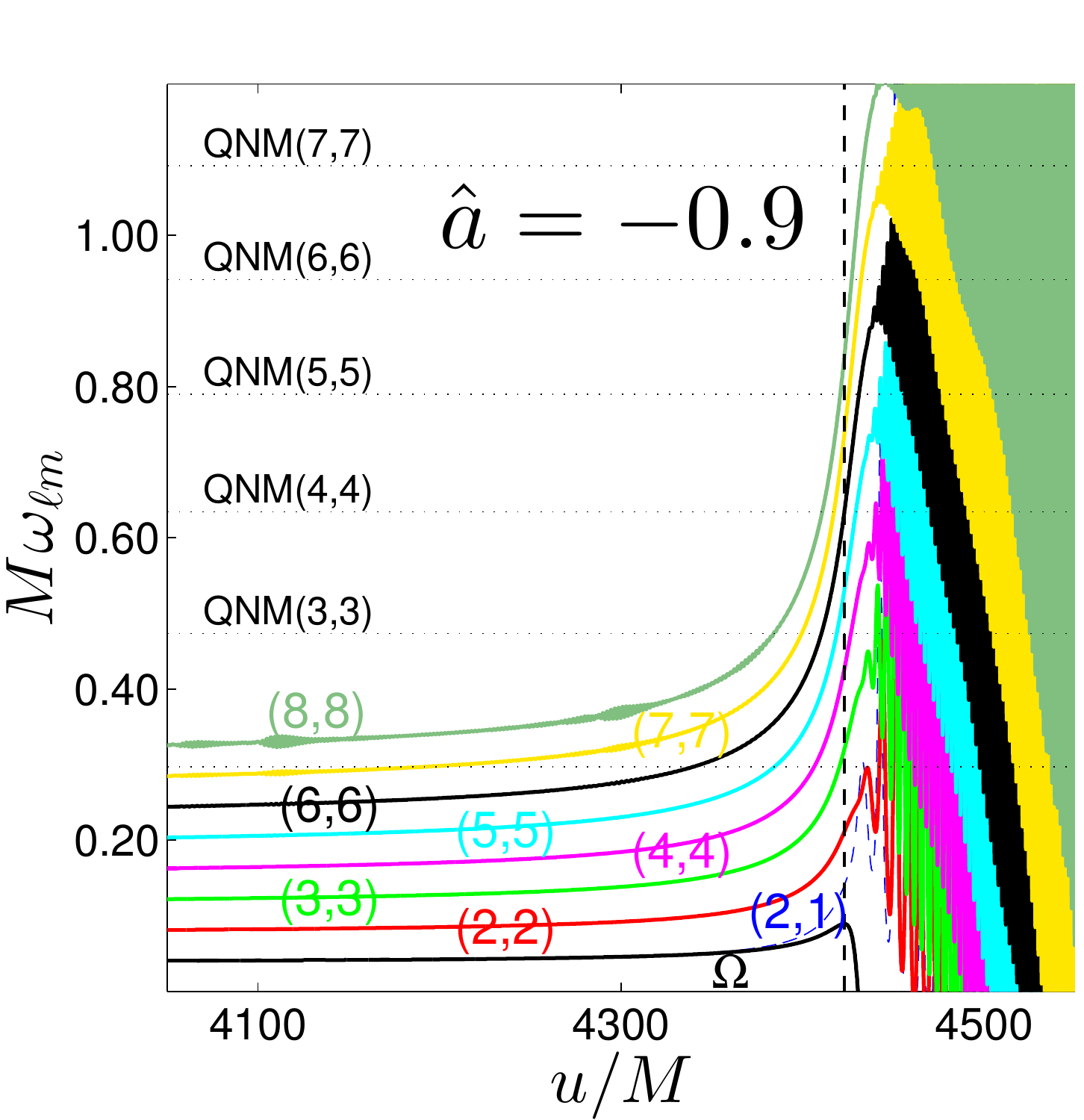}  
  \includegraphics[width=0.32\textwidth]{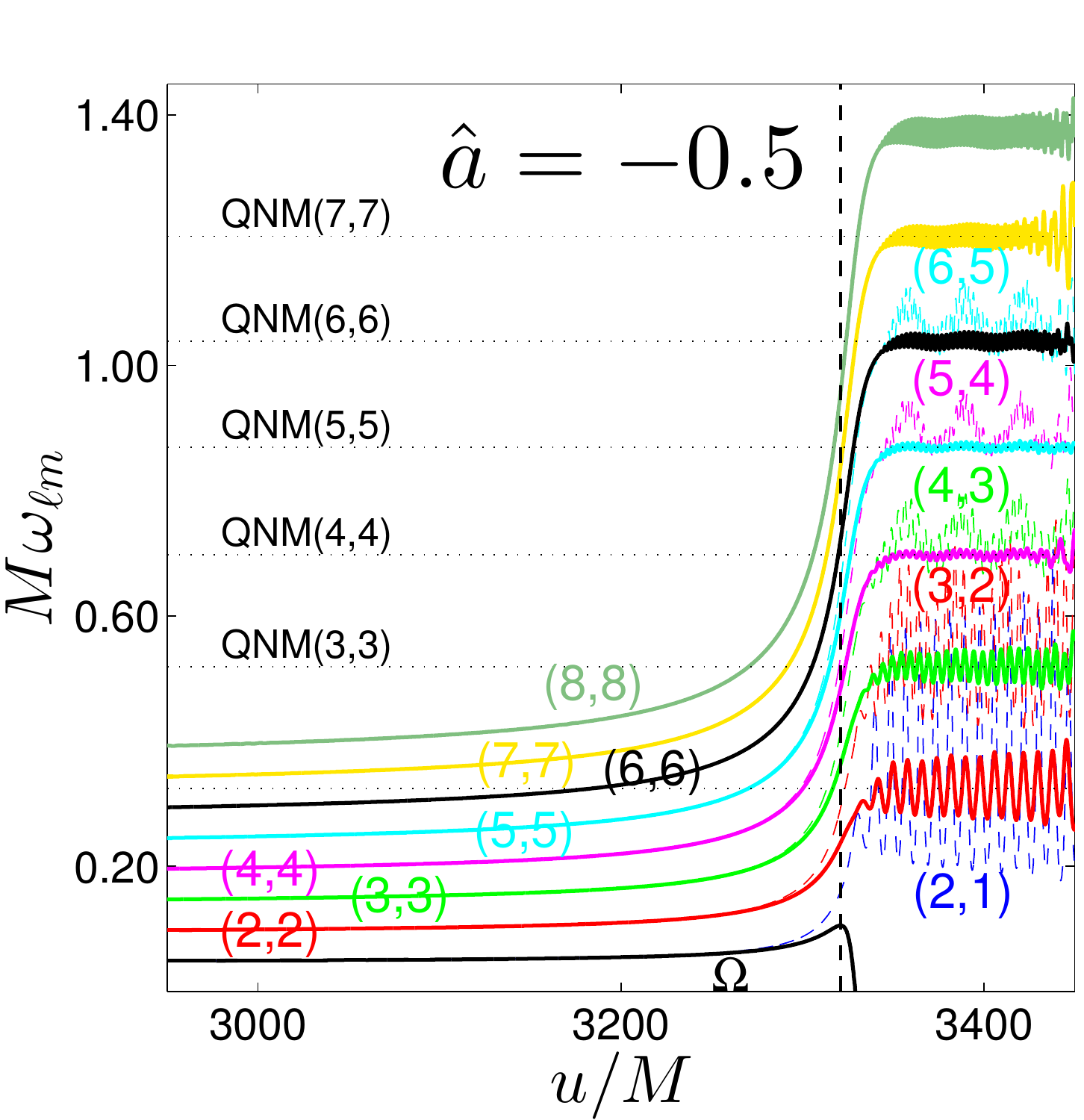}  \\
  \includegraphics[width=0.32\textwidth]{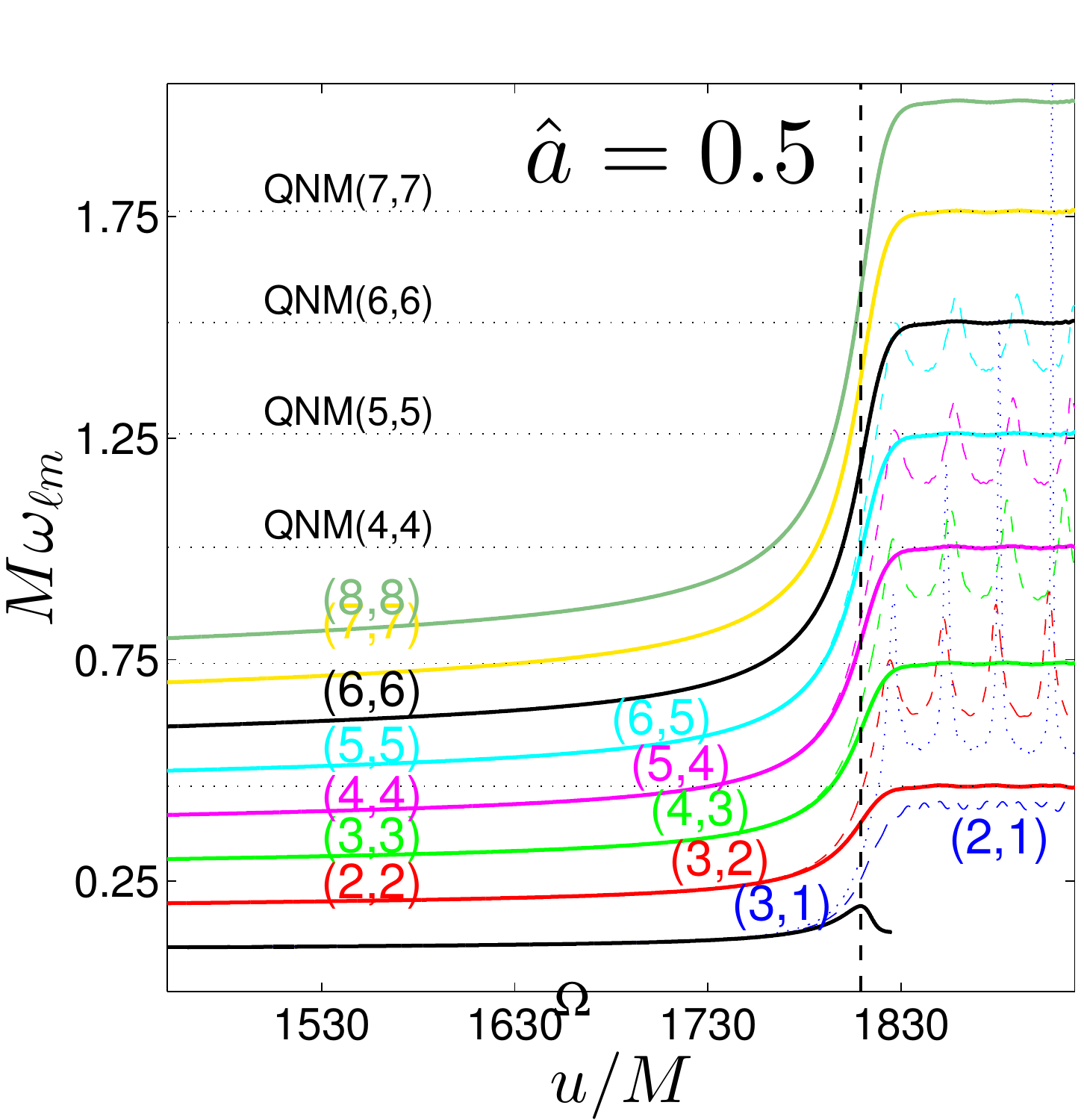}   
  \includegraphics[width=0.32\textwidth]{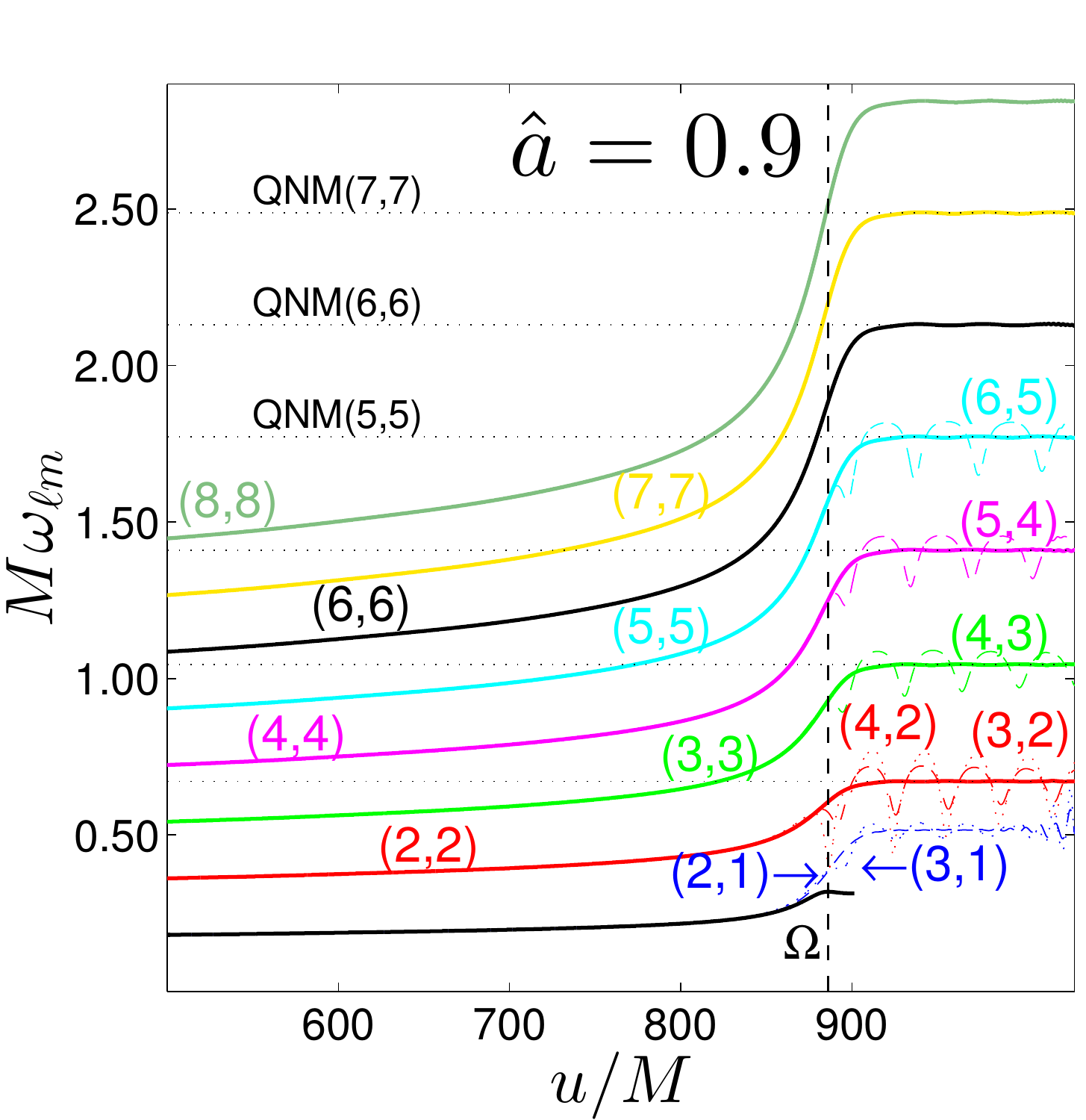}
  \includegraphics[width=0.32\textwidth]{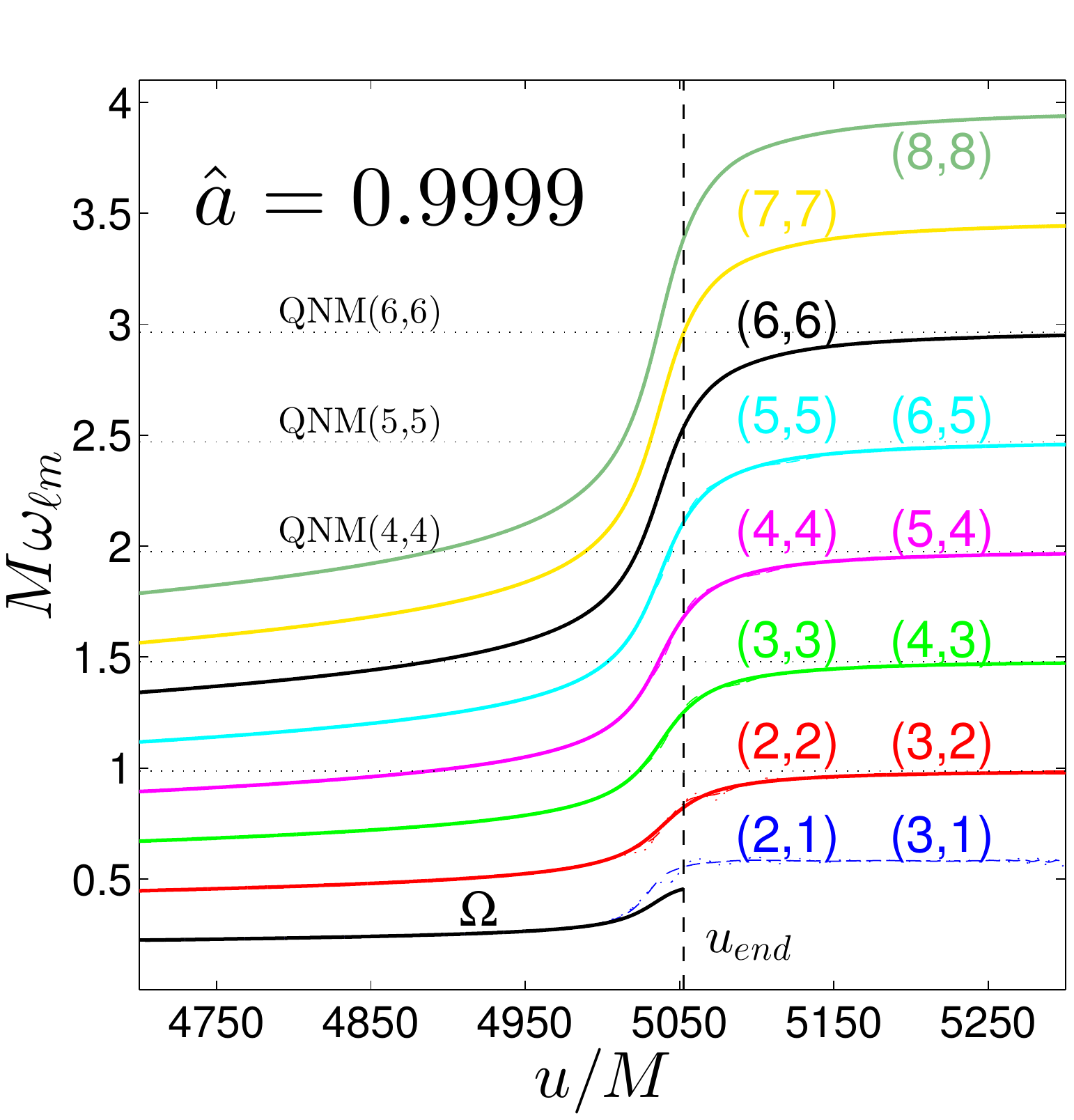}      
  \caption{Multipolar GW frequencies of the $\mu/M=10^{-3}$ insplunge waveforms
    for certain values of $\ha$.
    The vertical line on each plot marks the 
    crossing of the LR, except for $+0.9999$, where it marks the end 
    of the trajectory (in this case our trajectory stops slightly before the
    horizon, see discussion in text).
    Horizontal lines are the QNM
    frequencies of \cite{Berti:2009kk}.
    For nearly-extremal negative $\ha$ we find negative 
    frequencies due to the inversion of the trajectory after the LSO-crossing. }
  \label{fig:EOB_multipolar_structure_freq}
\end{figure}



Let us focus first on the dependence of the waveform amplitude on $\ha$.
Figure~\ref{fig:EOB_multipolar_structure_amp} gathers the time evolutions
of the ($\mu$-normalized) amplitudes $\A_{\lm}\equiv A_{\lm}/\mu\equiv|\Psi_\lm|/ \mu$ 
of several multipoles for representative
values of $\ha$. This figure is complemented by Fig.~\ref{fig:multipolar_peaks}
that highlights certain properties of the maxima of the $\A_{\lm}$'s.
The left panel of Fig.~\ref{fig:multipolar_peaks} shows the global 
variation of $\A_\lm^{\max}=\max(\A_\lm)$ versus $(\ell,m,\ha)$. The right panel of the figure
is the vertical plane projection of the left panel restricted to modes up to $\ell=m=4$
for clarity. We make the following observations:
\begin{itemize} 

\item[(i)]{} In the quasi-circular inspiral, the hierarchy between the $(2,1)$ and $(3,3)$ 
modes is inverted as the spin 
goes from negative to positive values. The $(2,1)$ mode becomes progressively 
less important for $\ha\to 1$. More precisely, looking at the maxima (right 
panel of Fig.~\ref{fig:multipolar_peaks}), $\A_{21}^{\max}\geq \A_{33}^{\max}$ for 
$\ha \lesssim  0.6$ and $\A_{21}^{\max}< \A_{33}^{\max}$ for $\ha \gtrsim  0.6$. 
Similarly, $\A_{31}$ is close to $\A_{43}$ during the inspiral for $\ha = -0.9$,
but at least an order of magnitude smaller
for $\ha=+0.9$. This feature is even more prominent for higher spins 
with $\A_{43}$ and $\A_{31}$ practically coinciding for $\ha=-0.9999$
(see top left plot of Fig.~\ref{fig:EOB_multipolar_structure_amp}) .
This behavior is due to the analytical structure of the leading-order
spin-dependent term entering the $(2,1)$ and $(3,1)$ waveform (see for example
Eqs.~(38b) and~(38e) of Ref.~\cite{Pan:2010hz}). When taking the limit of a nonspinning
point particle, the leading, spin-dependent PN correction
to the Newtonian prefactor is of the form $(1-\frac{3}{2}\ha v)$ for 
$(2,1)$ and  $(1-2\ha v^3)$ for $(3,1)$, which explains why their amplitudes 
are larger for $\ha < 0$ and smaller for $\ha >0$. 

\item[(ii)]{} Concerning multipoles with $m\neq 0$, the peak of each 
waveform flattens for $\ha\to 1$, and becomes sharper and more pronounced 
for $\ha\to -1$. This effect was pointed out in Ref.~\cite{Taracchini:2014zpa} 
for the therein computed modes, though explicitly shown only for $\ell=m=2$. 
Here we complete the results of Ref.~\cite{Taracchini:2014zpa}, showing how this 
flattening is a general property of all multipoles, as it is related to the (absence of) 
next-to-quasi-circular (NQC) corrections in the waveform. Defining the function $B=\Sigma/\Delta$, 
it is meaningful to use for this analysis the radial momentum $p_{r_{*}}=\sqrt{A/B} p_{r}$, i.e., 
the momentum conjugate  to the definition of a tortoise-like coordinate as 
$dr_{*}/dr\equiv \sqrt{B/A}$ (note that this
$r_{*}$ is different from the usual tortoise coordinate introduced above). 
The important point is that using $p_{r*}$ the $p_{r}$-dependent part
of the Hamiltonian just reads $p_{r_{*}}^{2}$, analogously to the Schwarzschild case~\cite{DN:2014prep}.
Consistently with the discussion in~\cite{Taracchini:2014zpa}, the magnitude of $p_{r_*}$ 
becomes very small when $\ha\to 1$ and so does the corresponding effect on the waveform amplitude.
On the contrary, due to the larger values of $p_{r_{*}}$ around light-ring crossing when $\ha\to -1$, noncircular 
effects are nonnegligible and yield the sharpness of the peak of each
multipolar amplitude. The dependence of $p_{r_*}$ on $\ha$ is illustrated 
in Fig.~\ref{fig:prstar}. In~\ref{app:nocirc} we present an heuristic discussion 
based on the leading-order Newtonian waveform to practically illustrate how 
NQC corrections can shape the multipolar waveform peaks close to merger.  

\item[(iii)]{} The effect of $p_{r_{*}}$ is particularly evident when inspecting the $m=0$ modes, 
that only depend on the radial part of the motion. For example, for $\ha=0.9$, $\A_{20}^{\max}=0.0009$, 
which is negligible compared to the dominant mode's peak $\A_{22}^{\max}=0.3212$.
By contrast, for $\ha=-0.9$ we have $\A_{20}^{\max}=0.1788$, which is of the same 
order as $A_{22}^{\max}=0.2738$ and $\A_{33}^{\max}=0.1996$.
This kind of information is gathered in
Fig.~\ref{fig:multipolar_peaks}, where the right panel of the  
figure clearly shows the ``growth'' of the subdominant modes over $1>\ha>-1$. 
This is particularly striking for the $m=0$ modes, that for $\ha\to -1$ attain values 
that are comparable to the $\ell=m$ ones. The consequence of this growth is that 
the well-known hierarchy of the modes at merger in the $\ha=0$ case (discussed in 
Appendix~A of~\cite{Bernuzzi:2010xj}) does no longer hold when $\ha\neq 0$.
\end{itemize}

Let us now turn to discuss the dependence of the multipolar
frequencies $M\omega_{\lm}$ on $\ha$, which is illustrated in Fig.~\ref{fig:EOB_multipolar_structure_freq}
for the same $\ha$'s as above. The plots show that
the QNM interference phenomenon mentioned in 
Sec.~\ref{sec:test:inspl} is greatly enhanced for $\ha<0$,
i.e.~retrograde plunges. This is due to the progressively larger
excitation of $m<0$ QNMs as $\ha\to -1$. 
These results are consistent with the findings of 
Refs.~\cite{Barausse:2011kb,Taracchini:2014zpa}.

Finally, though we do not discuss EOB waveform calibration in this work, we
list in Table~\ref{tab:peaks_a} useful information
extracted from the TE waveforms for a few values of $\ha$.
The numerical relativity completion of the EOB waveform requires the calibration of
next-to-quasicircular (NQC) corrections to data and the attachment
of the QNM waveform to the merger one. Both NQC corrections and QNM
attachment are usually performed after a careful analysis of the
properties of the merger waveform, i.e.~around the peak of $\hat{A}_{22}$.
The table lists the values of $\A_{\lm}^{\max}$ and the corresponding
GW frequencies together with the time lag between the peak of the
orbital frequency, $t_{\Omega^{\rm max}}$, and the peak of each  
multipole $u_{A^{\max}_{\lm}}$,
\be
\Delta t_{\lm}= u_{A^{\max}_{\lm}}-t_{\Omega^{\max}}.
\ee
We confirm the finding of~\cite{Barausse:2011kb}  that $\Delta t_{22}$ 
(as well as all $\Delta t_{\lm}$'s) strongly depends 
on the spin for $\ha \to 1$. The comparison of our $\Delta t_{\lm}$ with the 
values stated in Tab.\MakeUppercase{\romannumeral 3} of \cite{Barausse:2011kb} 
shows excellent agreement between the two codes, considered that they 
are completely independent and use different prescriptions for the radiation reaction.
Besides $\Delta t_{\lm}$ also the peak of $\hat{\Omega}$ 
is spin-dependent, because the peak of $\hat{\Omega}$ becomes progressively
less visible for $\ha\to 1$, so that $\hat{\Omega}$ becomes effectively 
monotonic in time for large spins. As advocated already 
in~\cite{Barausse:2011kb}, $\hat{\Omega}$ is not a good quantity to
identify an ``anchor'' point (like its maximum, that is a robust choice in
the nonspinning case) for EOB modeling purposes, namely
for determining effective next-to-quasi-circular corrections and for
matching the insplunge EOB waveform to the ringdown part. 
Reference~\cite{Barausse:2011kb} suggested to use the time when
the EOB insplunge waveform peaks. An alternative 
approach currently under investigation~\cite{DN:2014prep} 
is to use the peak of the (formally) ``pure orbital'' frequency. 
This frequency is defined as  
\be
\hat{\Omega}_{\rm orb}=\p_\phi\hat{H}_{\rm orb},
\ee 
that, differently from $\hOmg$, 
always has a neat isolated peak (before the particle gets to the horizon) 
due to the absence of spin-orbit effects. Table~\ref{tab:OmgOrb} in \ref{app:DataTables} displays 
the time lag $u_{A_{22}^{\rm max}}-\hat{t}_{\Omega_{\rm orb}^{\rm max}}$ between the peak 
of $\A_{22}$ and the peak of $\hOmg_{\rm orb}$. Interestingly, this time difference 
is small (order unity) and has a very mild dependence on spin up to $\ha=0.8$.

\subsection{Waveforms for nearly-extremal configurations}
\label{sec:inspl:09999}

The analytical and numerical setup of this work allows us to explore
nearly extremal configurations. In this section we discuss the
multipolar hierarchies at merger for the cases
$|\ha|=0.99,0.999,0.9999$. To our knowledge these are the first
results available of this kind (compare with \cite{Taracchini:2014zpa}.)

Since these simulations are technically more challenging than those for $|\ha|\leq0.99$, 
we needed artificial dissipation operators for stability (using the same resolution 
as before). Also, for such high positive
spins, the analytical radiation reaction (in both prescriptions) is
not only inaccurate but grows very rapidly around the LR and corrupts 
the numerical calculation of the trajectory. In order to perform the
simulations one can either
(i)~stop the particle dynamics before it reaches
the horizon or the LR and advect the source off the domain
(for example, for $\ha=0.999$ ($0.9999$) the LR is at
$\hr_{\rm LR}\sim1.05$ ($1.02$) while our trajectory stops at
$\hr_{end}=1.09$); or
(ii)~smoothly ``switch off'' the fluxes after the LSO (similarly to the
procedure of Sec.~\ref{sec:sf_flux}). Both procedures lead to qualitatively the same
results, and, although they might introduce a systematic effect in the
QNM waveforms, we do not observe any obvious unphysical features in using (i). 


Figures~\ref{fig:EOB_multipolar_structure_amp}
and~\ref{fig:EOB_multipolar_structure_freq} include the amplitudes
$\A_\lm$ and frequencies $M\omega_\lm$ for $|\ha|=0.9999$ for 
various dominant multipoles. (The analogous plots for $|\ha|=0.99,0.999$ resemble
that one and do not convey more information). 

The features discussed for lower spins remain valid, i.e. flattening of
the waveform amplitude around merger for $\ha\rightarrow1$ and sharpening as
 $\ha\rightarrow-1$. 
%
The most interesting features are observed during the QNM ringing. First, the
top left plot of Fig.~\ref{fig:EOB_multipolar_structure_freq} shows
that for nearly-extremal negative $\ha$ the QNM waveforms are 
characterized by negative frequencies. The frequencies correspond to
the QNM ringing modes with $m<0$~\cite{Berti:2009kk}.
As noted in~\cite{Taracchini:2014zpa}, they are excited by the change
of sense of rotation of the particle ($\hOmg$ has a zero) during the plunge. 
Our analysis extends the previous one by considering higher spins and 
including all the multipoles up to $\ell=m=8$. 
Second, for nearly-extremal cases the trapping of modes
leads to very weakly damped QNM's in the case of free QNM ringing (no
particle), see e.g.~\cite{Andersson:1999wj,Glampedakis:2001js,Berti:2009kk,Harms:2013ib,Yang:2013uba}.
We qualitatively confirm this behavior also for the QNM's excited by a particle.
Damping times can be measured by fitting the exponential decays of
Fig.~\ref{fig:EOB_multipolar_structure_amp}. 
Note that in the nearly-extremal regime the exponential decay holds for much longer times
than shown in the plots, e.g.~for $\ha=0.9999$ we fit over $[5400M,6800M]$ for the $\ell=m=2$
mode. The damping exponents of the dominant overtones computed for $\ha=0.9999$ are 
$\{-3.52, -39.11, -3.61\}\times10^{-3}$ for $(\ell,m) = \{(2,2),(2,1),(3,3)\}$
respectively. These numbers match at $\lesssim 4\%$ with the corresponding values for free QNM 
ringing~\cite{Berti:2009kk}.

\subsection{Finite extraction errors in waveforms}
\label{sec:inspl:finiter}

\begin{figure}[t]
  \centering
  \includegraphics[width=0.48\textwidth]{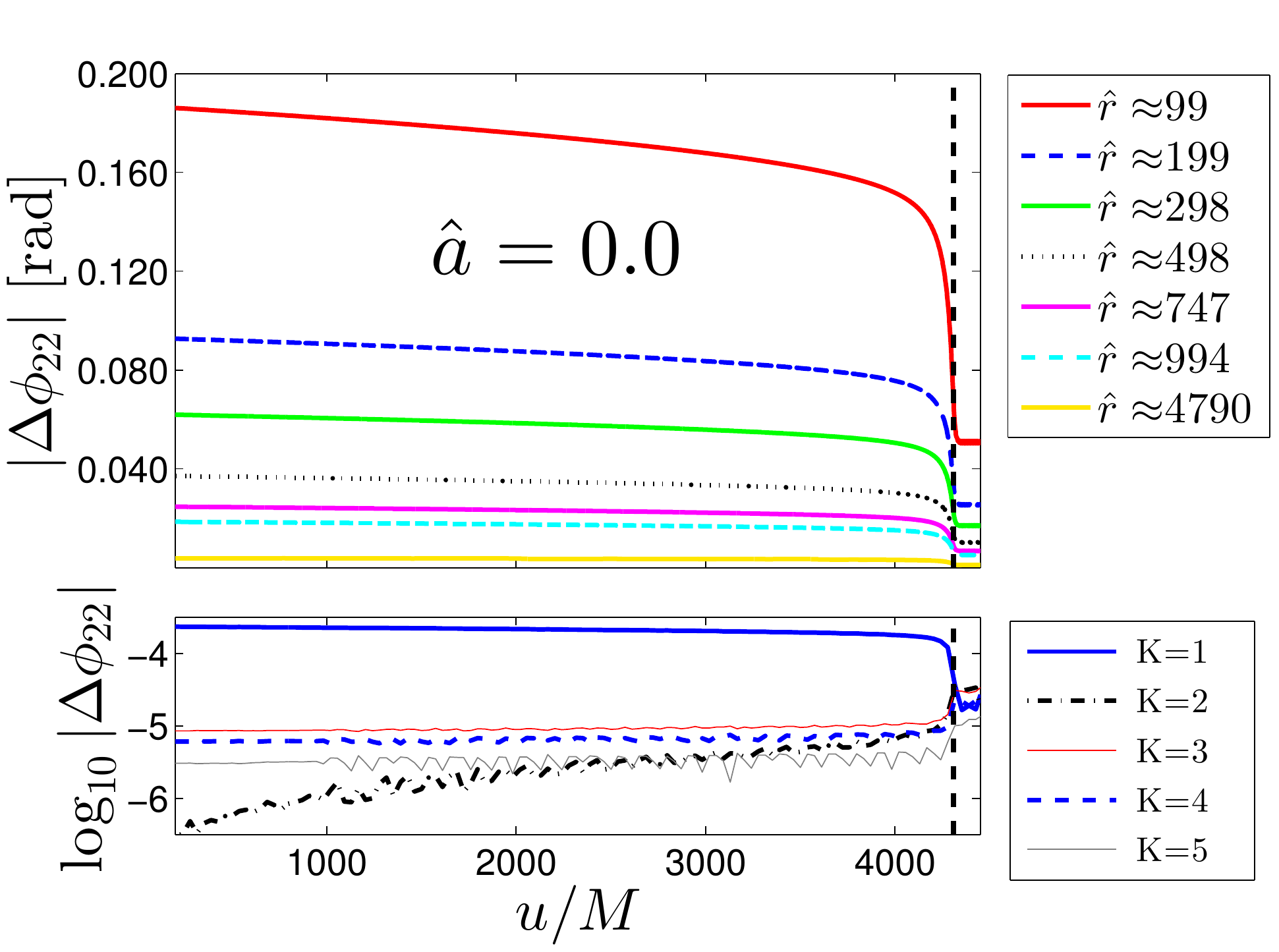} 
  \includegraphics[width=0.48\textwidth]{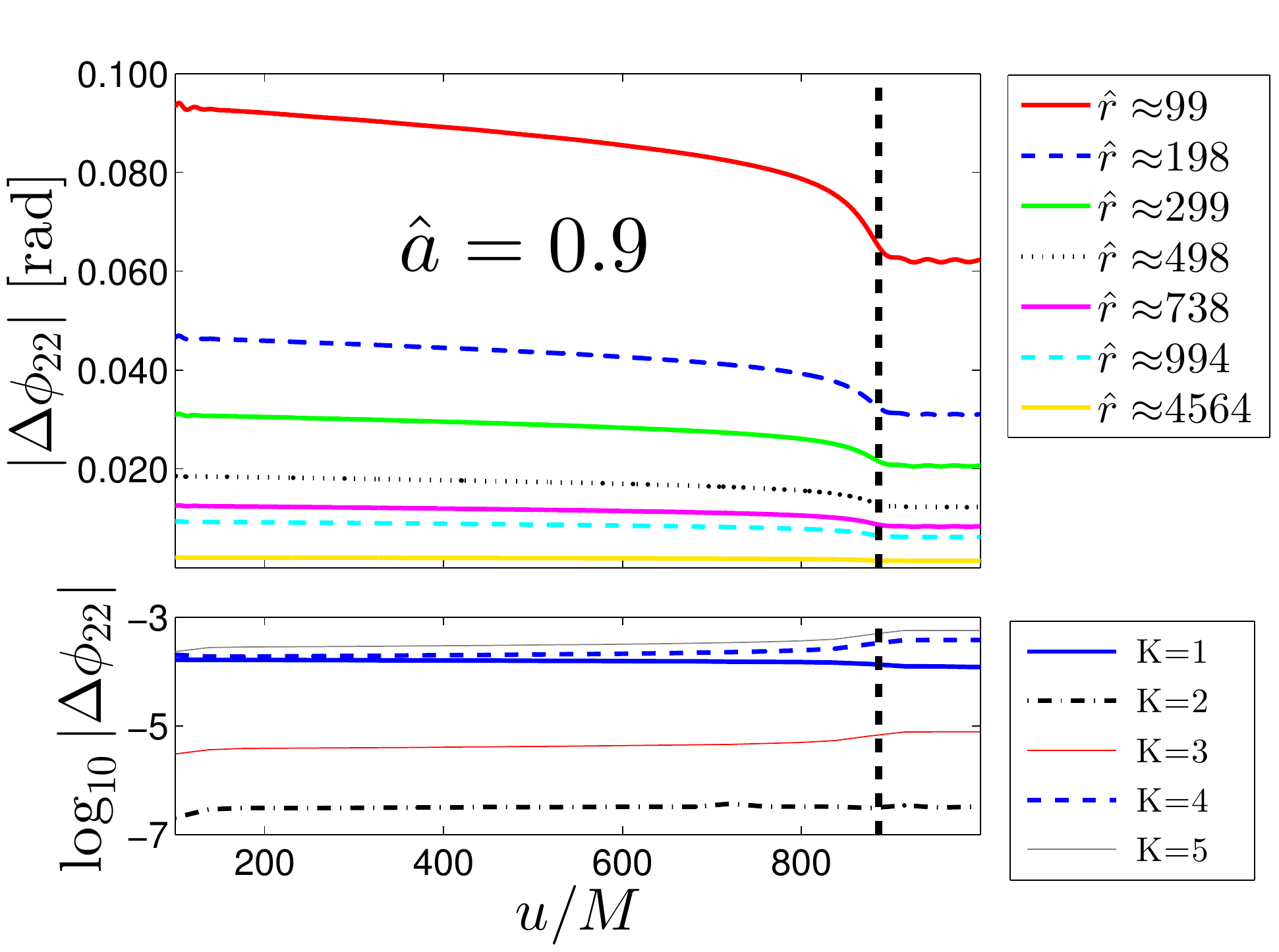} 
  \caption{Finite-extraction-radius effects on the phase of
    the $\psi_{4\, 22}$ waveforms from insplunges with $\ha=0$ (left) and $\ha=0.9$ (right).
    Top panels refer to absolute differences between scri and finite radii
    $\hr\sim(100,200,500,750,1000,4800)$. Bottom panels refer to
    absolute differences between the scri values and waveforms extrapolated
    according to Eq.~\eqref{eq:extrap_formulas} with various
    polynomial orders $K=1,2,3,4,5$. Note the $\log_{10}$ in the $y$-axis. 
    The vertical line represents the
    crossing of the LR.}
  \label{fig:EOB_r_extrapolation}
\end{figure}

To estimate finite extraction effects on the waveforms, we have
investigated systematic phase and amplitude differences between 
scri and finite radius waveforms, as well as the performance of an
extrapolation procedure commonly used in numerical relativity.

Figure~\ref{fig:EOB_r_extrapolation} (top panels) shows phasing results for two case
studies: the $\ha=0$ insplunge $\ell=m=2$ waveform 
of Sec.~\ref{sec:test:inspl} (left) and the $\ha=0.9$ insplunge
$\ell=m=2$ waveform (right). In the $\ha=0$ plot finite extraction errors
at $\hr\sim100$ amount to $\Delta\phi_{22}\sim0.16~rad$. This value is
not only significantly larger than truncation errors, but has a similar 
size as the dephasing due to horizon absorption
fluxes~\cite{Bernuzzi:2012ku}. Hence, extraction at these radii can
affect relevant physics. Similarly, phase errors in the $\ha=0.9$
plot are at the order of $\Delta\phi_{22}\sim0.08~rad$
($\sim0.03~rad)$ for $\hr\sim100$ ($\sim300$) accumulated to merger over
$25$ orbits. In all the cases analyzed, finite extraction errors are
typically larger at early times. The quantitative difference between
finite radius and scri waveforms is relevant when comparing and
calibrating the EOB waveform. For example, Ref.~\cite{Bernuzzi:2010xj}
has found that no time-phase alignment is needed if waveforms are
extracted at scri and  that the errors at early times are those
expected by the order of the PN approximation of the EOB waveform.

Many numerical relativity (either nonlinear or perturbative)
simulations calculate approximate 
GWs at finite radii and use extrapolation to estimate those at null
infinity. Our method allows us to test such a procedure.
In order to extrapolate waveforms extracted at finite
radius, we assume that the phase (and the amplitude) of the finite-radii 
waveform behaves as a $K$th order polynomial in $1/r$, 
\be
\label{eq:extrap_formulas}
f(u,r) = f^{(0)}(u) + \sum^K_{k=1} r^{-k} f^{(k)}(u) \,.
\ee
We fit this model for every radius and some choices of the
polynomial order $K$. The first term approximates the scri
waveform. The extrapolation procedure is applied here on the $\psi_4$
multipolar waveform considering radii
$\hr\sim(100,200,500,750,1000,4800)$. We find that the leading finite
radius effect on $\phi(u,r)$ is the $1/r$ behavior, i.e.~the term
$K=1$. Using larger $K$ can reduce the phase errors, but we do not
find a single prescription for $K>1$ that robustly improves the results among
different datasets. 
Figure~\ref{fig:EOB_r_extrapolation} (bottom panels) shows
these findings for the case study. Using linear extrapolation we
obtain typical phase differences of $\Delta\phi\lesssim10^{-3}$ with
respect to scri. Different results are obtained using extrapolations
with $K>1$. For example, for $\ha=0$ the choice $K=5$ gives the smallest
errors, but for $\ha=0.9$ that choice leads to a large error and $K=2$
is the optimal one. As suggested by the figure, in many cases
$K=2$ improves the $K=1$ result. Similar results are found for the
amplitude (not shown in the figure). The only important difference in
that case is that $A(u,r)$ clearly shows a parabolic behavior, thus
$K=1$ extrapolation cannot be used. We suggest $K=2$ as a
``safe option'' also in nonlinear numerical relativity
simulations~\footnote{Note in the nonlinear case there are more complications 
  due to the dynamical gauge (no fixed background), dynamical null structure, 
  nonlinear effects, and outer boundary errors.}.

\section{Horizon absorption during insplunge}
\label{sec:horabs}

\begin{figure}[t]
  \centering
   \includegraphics[width=0.49\textwidth]{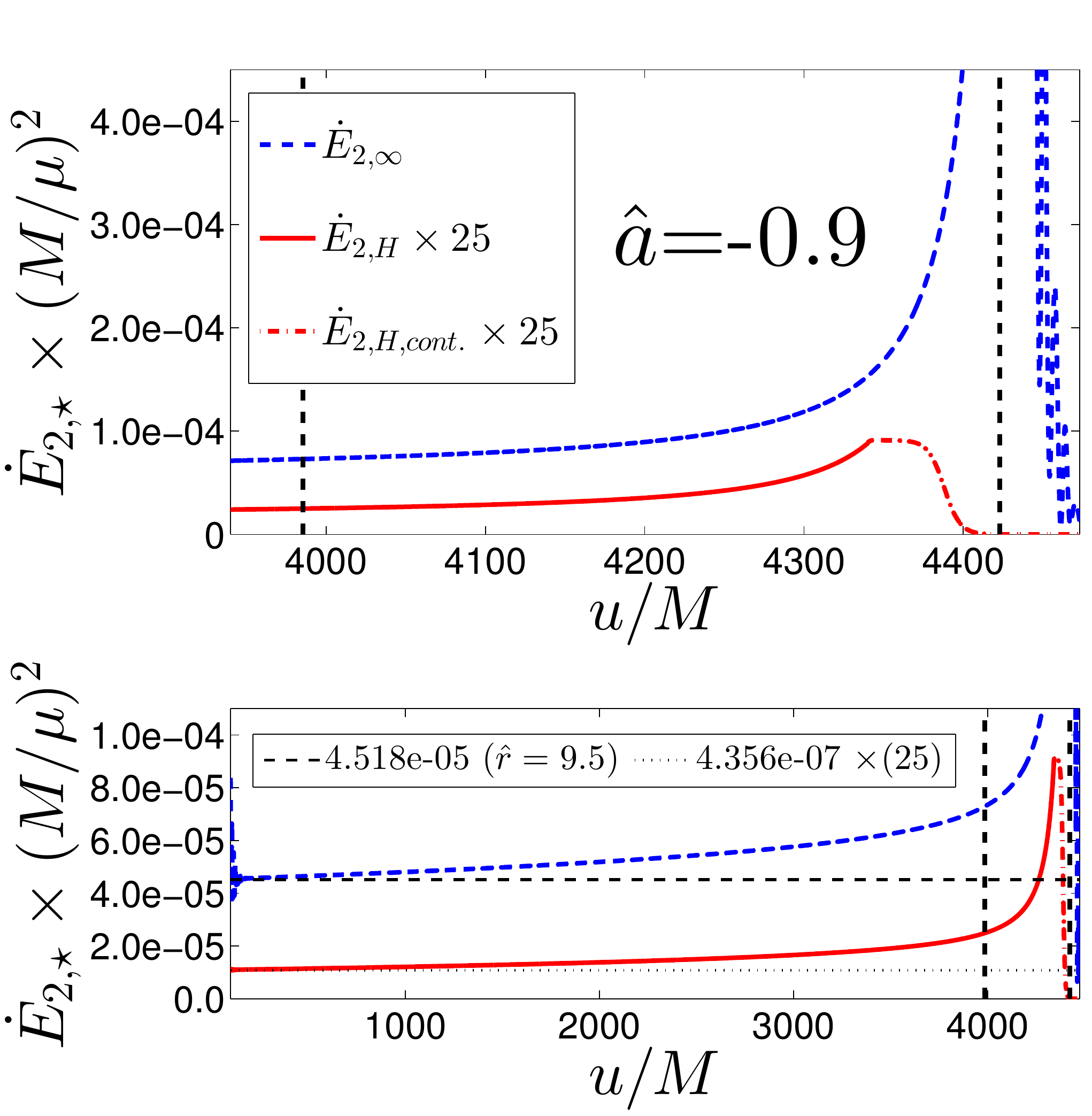} 
   \includegraphics[width=0.49\textwidth]{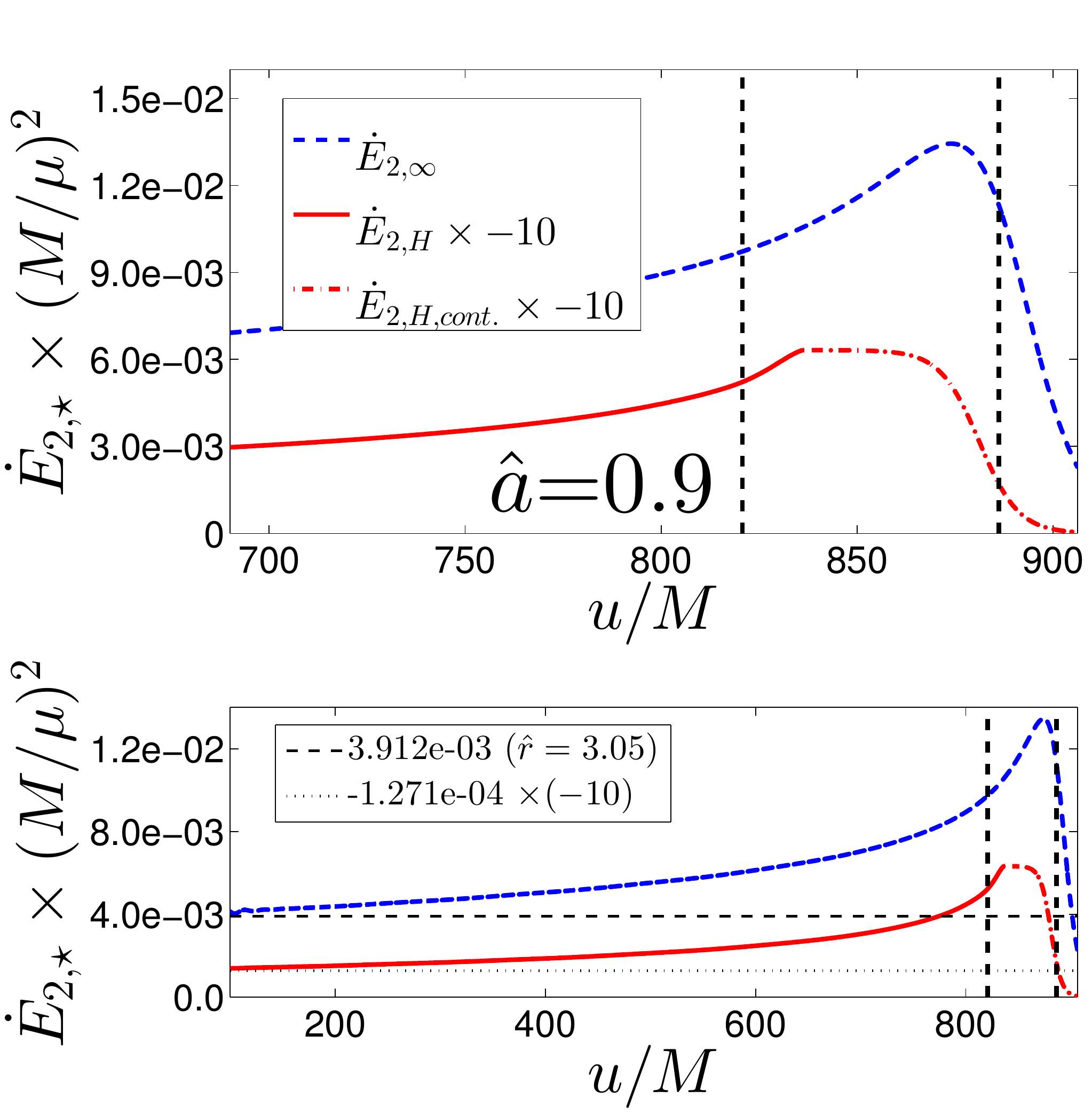} 
   \caption{Comparison of $m=2$ energy fluxes at the horizon (red) and scri (blue dashed)
     for $\ha=-0.9$ (left panel) and $\ha=0.9$ (right panel). 
     The time coordinate $u$ refers to the retarded time for scri
     and the advanced time $v$ for the horizon. The first vertical 
     lines mark the LSO and the second the LR. The horizontal 
     lines show the circular orbits' fluxes at the initial
     separation. Note the absorbed fluxes are scaled by 
     multiplicative factors, and, in the right panel, the factor is
     negative, i.e. the two fluxes have different signs. 
     The relative importance of 
     horizon fluxes to infinity fluxes remains around $\sim1\%$ ($\sim3\%$)
     for $\ha=-0.9$ ($0.9$).
     The waveforms at the horizon are contaminated by the source
     at late times. As a consequence, the horizon fluxes 
     are reliable until $v\approx v_{r_+}-100M$ as marked by the
     solid line (see discussion in the text).      
     The dashed lines illustrate how we analytically continue the 
     horizon fluxes in order to build a self-consistent radiation reaction
     for insplunge trajectories (see Sec.~\ref{sec:sf_flux}). }
   \label{fig:Flux_comparison_HRZSCR}
\end{figure}

Here, we discuss the fluxes absorbed by the black
hole. The theoretical tools to compute time-domain horizon fluxes are
provided by~\cite{Poisson:2004cw} and described in
Sec.~\ref{sec:TE:gw}. The accuracy of our
implementation was tested against frequency domain circular orbit
data as presented in Sec.~\ref{sec:test:ciro}.
In this Section, we compute, for the first time, horizon fluxes
for insplunge trajectories around spinning black holes. 

Horizon absorption is nonnegligible during the inspiral and
can be important at merger already in the nonspinning case~\cite{Bernuzzi:2012ku}.  
In the spinning case, it is expected to be more relevant because
absorption terms enter at 2.5 PN order (4PN for $\ha=0$),
e.g.~\cite{Taracchini:2013wfa}. So far, no numerical calculation of 
horizon-absorbed fluxes during the insplunge into a rotating black
hole has been performed. The formalism of
Poisson~\cite{Poisson:2004cw} allows us to perform this calculation beyond
the LSO, but not up to the LR, as we shall see. 

In Fig.~\ref{fig:Flux_comparison_HRZSCR} we compare the $m=2$
horizon absorbed energy flux with the infinity flux for $\ha=-0.9$ (left) and $\ha=0.9$ (right).
The horizontal lines indicate the fluxes for a circular orbit at the
initial separation of the insplunge, $\hr_0=9.5$ ($3.05$) for $\ha=-0.9$ ($0.9$). The agreement 
of our fluxes at initial times with the circular fluxes at $\hr_0$ is
about $\sim0.2\%$. During the 25 orbits up to the LSO the ratio of horizon-absorbed fluxes 
to null infinity fluxes amounts to $\sim 1\%$ ($\sim3\%$) for $\ha=-0.9$ ($0.9$).
These values are consistent with circular orbits' fluxes of Sec.~\ref{sec:test:ciro}. 

Between the LSO and the LR we observe a rapid variation of the horizon
fluxes. Unfortunately, our calculation becomes inaccurate and we stop it
some time before reaching the LR.  The reason of the failure is twofold: 
(i)~the wave-extraction at the horizon is corrupted by the source term
when the particle reaches the horizon (cf.~also Fig.~3 of \cite{Bernuzzi:2012ku}); 
(ii)~the time-domain formalism by Poisson relies on the
calculation of advanced-time integrals (see Sec.~\ref{sec:TE:gw}); 
the resulting absorbed fluxes are corrupted {\it earlier} than the
waveforms. In order to exclude the waveform corruption, the flux
integrals are considered up to $v\sim v_{r_+} - 100M$, where $v_{r_+}$
is the advanced time corresponding to the particle reaching the
horizon (see also Sec.~\ref{sec:num}).
For this reason a 2+1 formalism that is ``local'' in time, i.e.~only 
relies on data on hypersurfaces, seems desirable. We leave
the development of such a formalism, that will allow us to compute horizon 
fluxes up to the LR, to the future.

Finally, we stress that the absorbed fluxes calculated in this Section are
inconsistent since they are not taken into account in the trajectory. 
Analytical results for absorbed fluxes are available, but
the development of {\it resummed} expressions valid in the
strong-field--fast-motion regime and for generic spins has just
started~\cite{Nagar:2011aa,Taracchini:2013wfa}. 
In the next Section we develop a numerical method to calculate  
consistent horizon absorptions and evaluate their impact on the
waveform phasing for a test case.

\section{A numerical method to compute $\cal{O}(\nu)$ self-consistent GW fluxes}
\label{sec:sf_flux}

\begin{figure}[t]
  \centering
   \includegraphics[width=0.49\textwidth]{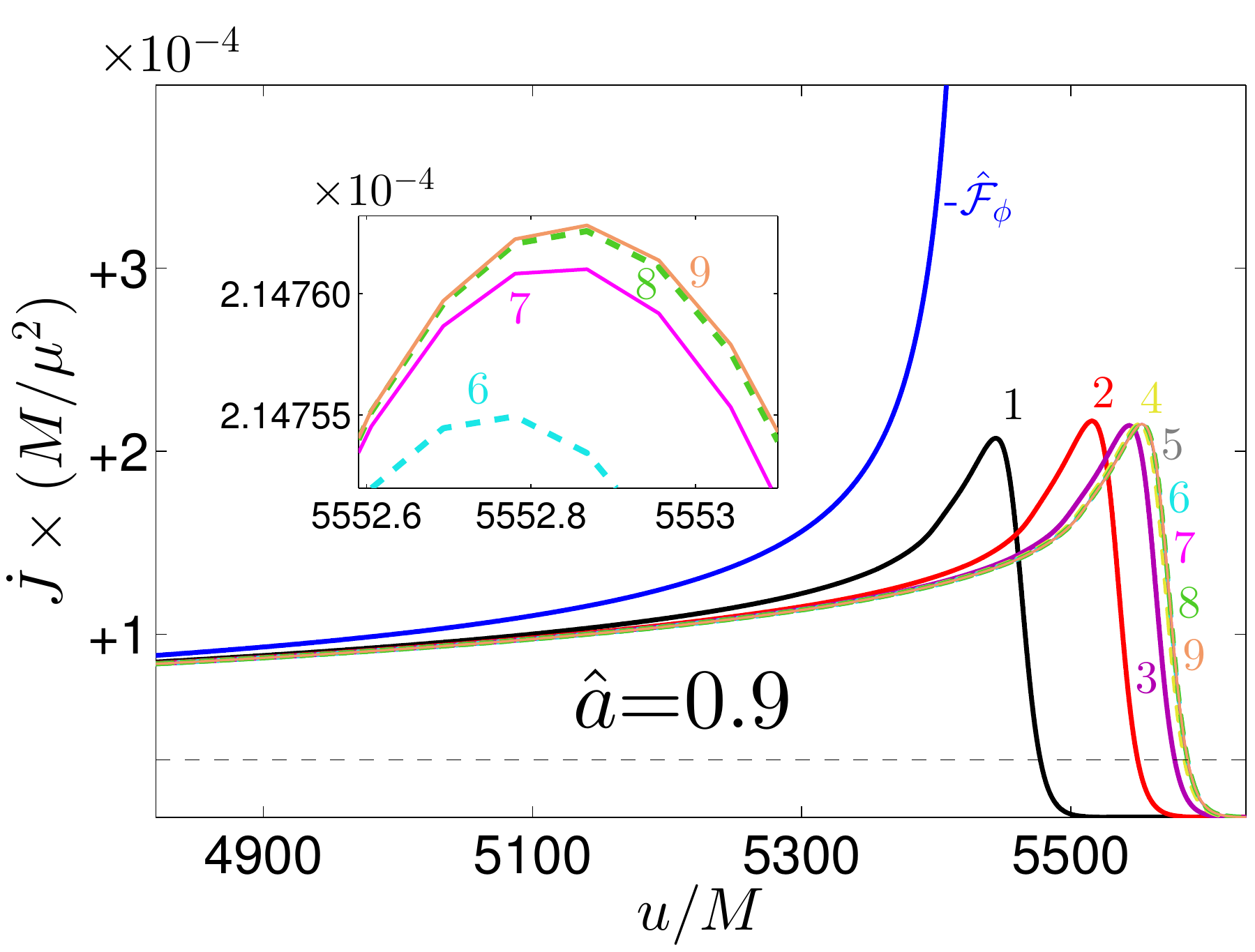} 
   \includegraphics[width=0.49\textwidth]{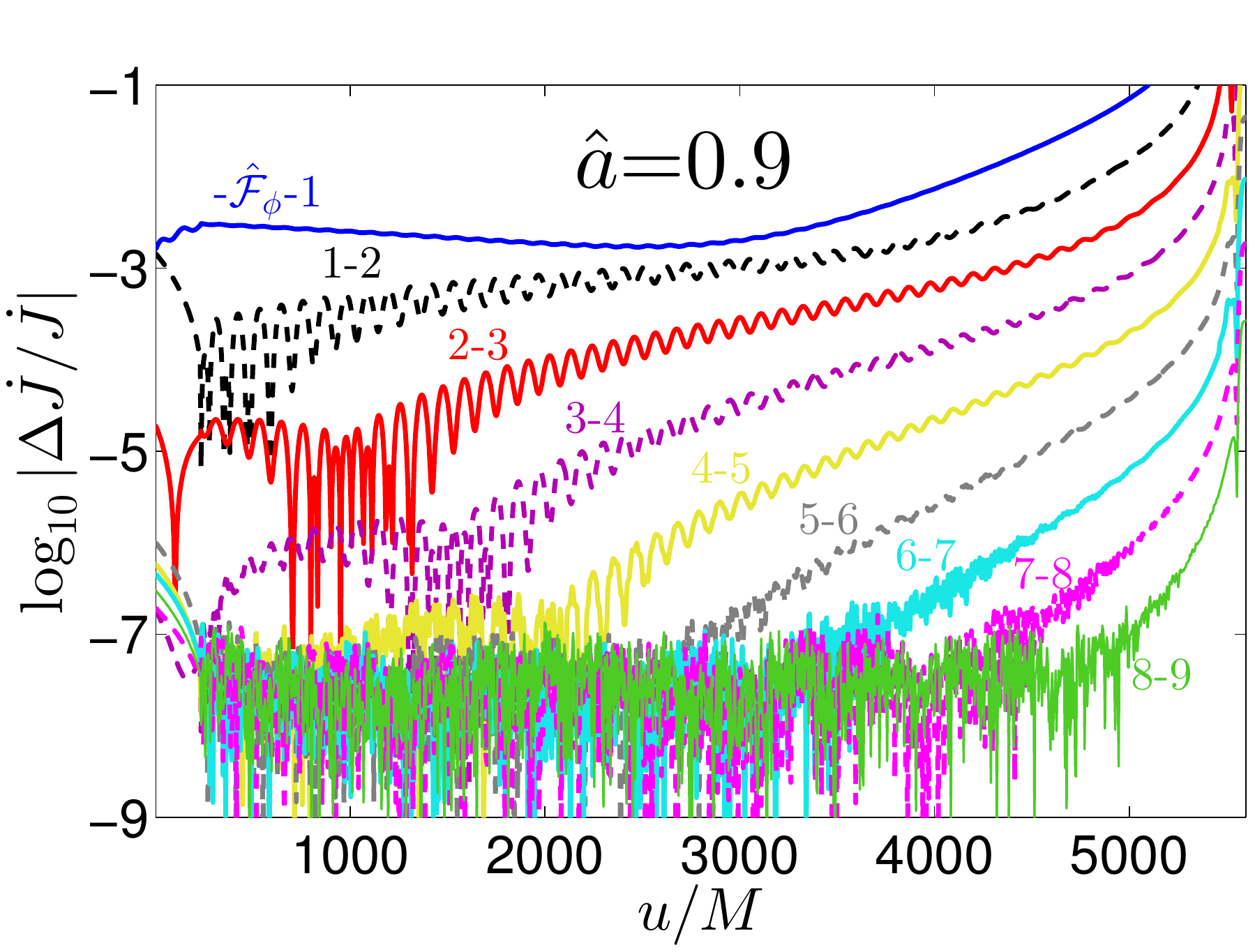} 
  \caption{Convergence of the iterative self-consistent procedure
    for the radiation reaction. Insplunge with $\ha=0.9$ and $r_0=4M$. 
    Left: GW fluxes after each iteration. Note the large differences
    with respect to the EOB radiation reaction (iteration 0) even 
    though we used the refining fit given in~\cite{Taracchini:2013wfa}.
    Right: Convergence of the fluxes after each iteration. 
    After $9$ iterations we obtain a self-consistent radiation reaction including both
    infinity and horizon fluxes.}
  \label{fig:SelfConsistency_Flux}
\end{figure}

In this Section we present a method to compute {\it exactly} the GW
fluxes at linear order in the symmetric mass ratio $\cal{O}(\nu)$. The
method is a simple and self-consistent procedure that iteratively
employs the GW flux extracted from the TE simulations in the particle
dynamics. We have tested it for various spin values; here we discuss
the case $\ha=0.9$. Differently from the $\ha=0.9$ simulation 
of previous Sections (see also Tab.~\ref{tab:test_flux}), 
the simulations of this Section start at $\hr_0=4$ corresponding to 
$\sim120$ orbits. We discuss, in particular, the differences with the EOB 
analytical radiation reaction and the effect of horizon absorbed fluxes.

At the first iteration, the fluxes are calculated by the analytical EOB formula, as
described in Sec.~\ref{sec:dyn_RR}. Then, we run a series of 16 simulations for
$m=1,...,8$ and $s=\pm2$ and compute the numerical fluxes at scri and 
at the horizon. These numerical fluxes are then employed for a new
calculation of the trajectory. At the next iteration, new fluxes are
computed by another series of 16 simulations. If a fixed point exists, 
and/or the first step is accurate enough, convergence should be
observed. This was the case for all the tests we performed,
i.e.~provided with a ``close-enough'' guess, the method converges and the
differences between the fluxes at each iteration consistently converge
to zero. The most difficult cases are those with positive high
spins, where EOB fluxes are less accurate. For example, for $\ha=+0.9$ we
were able to obtain iterative convergence only when the 
initial EOB trajectory was computed incorporating the refined $\rho_{\lm}$'s
obtained in Ref.~\cite{Taracchini:2013wfa} by fitting to numerical data
into the radiation reaction. 

The result of the iterative procedure is summarized in
Fig.~\ref{fig:SelfConsistency_Flux}. 
The left panel shows the complete (scri and horizon) GW flux at each
iteration. The peak position of the flux significantly
changes for the first iterations and in total from
$u_{peak}\approx5450M$ for the analytical flux to
$u_{peak}\approx5553M$ for the 9th iteration. 
On the contrary the peak amplitude remains approximately the
same after the first iteration (see inset), consistently
with the intuition that it is not a radiation-reaction-driven effect,
but rather its structure depends on the plunge phase.
The right panel shows that the relative differences
in the fluxes between the previous and the next iteration converge
very rapidly to zero. At iteration 9 the relative flux differences saturate
around $10^{-7}$ during most of the evolution, and the radiation emitted during the
insplunge is consistent with the one used for the particle to this level.

The impact of the consistent flux on the GW modeling is 
quantified by considering the difference in the number of gravitational wave cycles
$\Delta N_{\rm gw}$ between the final waveform (after 9
iterations) and the starting one (iteration 0): $\Delta N_{\rm gw}\sim6.7$. Note, however,
that the 0th iteration waveform is already very different from the one
computed in Sec.~\ref{sec:inspl:09} without the use of the fit
of~\cite{Taracchini:2013wfa}. The latter difference amounts to $\Delta
N_{\rm gw}\sim28.8$; so, overall, the self-consistent simulation
differs from the corresponding EOB one by about $\Delta N_{\rm gw}\sim35.5$.

We tested the importance of horizon absorbed fluxes by performing another
self-consistent calculation that neglects these contributions. We found that
the final trajectory is {\it shorter} and the particle reaches the
horizon $\sim117M$ earlier compared to the final trajectory that 
includes horizon absorption.
The fact that in presence of horizon absorption effects the insplunge 
is longer is explained by superradiance, i.e.~the net effect of 
horizon-absorption for $\ha=0.9$ is essentially an {\it emission} of energy/angular momentum. 
The differences in the dynamics correspond to $\Delta N_{\rm gw}\sim5.4$,
i.e.~$\sim127.1$ instead of $\sim124.4$ orbits before merger. This result
highlights the importance of horizon-absorbed fluxes during the insplunge. 

We mention two technical details. First, at early
times $\lesssim200M$ our 
numerical fluxes are corrupted by ``junk radiation'' due to the
imperfect initial data. For their use in the
trajectory calculation, we linearly extrapolate the flux from
$u\in[250,500]$ to $u=0$. This procedure does not introduce a
significant error. As a check we compared the extrapolated value at
$u=0$ with the circular flux at the relative radius and find an agreement of $\sim0.2\%$. 
Second, as mentioned in the previous Section, our horizon fluxes are
only reliable until $v\approx v_{r_+} - 100M$. For the numerical fluxes
we therefore switch off the absorbed fluxes at late times by multiplying them with 
a $\tanh()$ function that goes rapidly to zero. The
result of this procedure is shown by the dotted red lines of
Fig.~\ref{fig:Flux_comparison_HRZSCR}.

\section{Summary and Outlook}
\label{sec:conc}

We summarize the main findings of this paper and give an outlook for future developments.

\paragraph{New approach for time-domain solution of the 2+1 Teukolsky equation.}
We presented a new approach to compute
time domain solutions of the 2+1 Teukolsky equation (TE) based on
horizon-penetrating, hyperboloidal foliations of Kerr spacetime and spatial compactification. 
The coordinates are constructed from the ingoing-Kerr coordinates,
introducing a simple compactification and demanding the invariance of
the coordinate expression for outgoing radial characteristics (see Sec.~\ref{sec:coords}). 
The advantages of these coordinates in numerical applications are pointed out in 
Sec.~\ref{sec:num:coords}. 

In Sec.~\ref{sec:TE} we discussed the derivation of a regular
inhomogeneous 2+1 TE as well as the computation of the point-particle
source term in the new coordinates. A natural way to derive a regular TE in
horizon-penetrating coordinates is to use the Hartle-Hawking tetrad~\cite{Teukolsky:1973ha} or the one
in~\cite{Campanelli:2000nc} (we use the latter) with a subsequent rescaling for regularity at scri.

\paragraph{New wave generation algorithm for point-particle perturbations.}  
We implemented and tested a gravitational wave generation algorithm based on
the new coordinates. The code {\it Teukode}~\cite{Harms:2013ib},
employed in this work, uses a standard method-of-lines approach and
finite differencing operators of fourth and sixth order.
The Dirac $\delta$ functions involved in the
particle description are handled either using a narrow Gaussian or a
discrete
representation~\cite{Tornberg:2004JCoPh.200..462T,Engquist:2005JCoPh.207...28E,Sundararajan:2007jg}.  
We found that  the ``best'' representation, i.e.~the most accurate and
computationally most efficient, depends on the trajectory (see Sec.~\ref{sec:num:delta}). 

We have tested the code for the simple, but physically relevant, cases
of circular orbits (various rotating backgrounds) and a radial geodetic
plunge (Schwarzschild). In the computation of circular GW
fluxes at scri and at the horizon, our 2+1 simulations obtain a good
accuracy as verified in a comparison against frequency domain results. 

In Sec.~\ref{sec:test:inspl} the results of the 2+1 {\it Teukode} for a
nonrotating background are directly compared with those of the RWZE 
1+1 code \cite{Bernuzzi:2010xj,Bernuzzi:2011aj,Bernuzzi:2012ku}. At
the resolution employed, small systematic differences in the
multipolar phases and amplitudes (i.e.~larger then those
expected from self-convergence tests) are found between the two
codes. These differences are negligible for many practical purposes;
notably they do not influence, but confirm, the precise analysis
of~\cite{Bernuzzi:2010xj,Damour:2012ky}.

\paragraph{Modeling the late-inspiral-merger waveforms from
  large-mass-ratio spinning black hole binaries.}   
We applied our method to study GWs from large-mass-ratio and spinning black-hole binaries. 
The plunging particle's trajectory is calculated by an analytic and
nonadiabatic effective-one-body approach~\cite{Nagar:2006xv,Damour:2007xr}.

\paragraph{Consistency of analytical EOB fluxes.} 
In Sec.~\ref{sec:inspl:fluxes} we have compared and contrasted, 
for the first time, the numerical GW flux at scri (as computed from the TE) 
with two analytical prescriptions (``$\hat{\cal{F}}^{v_\phi}_{\phi}$'' and ``$\hat{\cal{F}}^{v_\Omega}_{\phi}$'') 
for the EOB radiation reaction.
For spins anti-aligned with the orbital angular momentum, the current, resummed, 
PN knowledge suffices to make the analytical fluxes agree
well with the numerical data. This is in-line with the nonspinning
case~\cite{Damour:2007xr,Bernuzzi:2010ty,Bernuzzi:2011aj}. 
For highly negative spins, the ${v_\phi}$-prescription is qualitatively and
quantitatively closer to the numerical data in the late plunge and merger
phases than the ${v_\Omega}$-prescription. The reason is the intrinsic
``less Keplerian'' character of $\hat{\cal{F}}_\phi^{v_\phi}$, represented 
by the uniform use of $v_\phi=r_\Omega \Omega$ instead of $v_\Omega=\Omega^{1/3}$ 
as the argument in the Newtonian prefactors of the subdominant $(2,1)$ and $(4,4)$ multipoles. 
As such, the uniform use of $v_\phi$ in the EOB resummed (circular) multipolar 
waveform (and flux) seems the most natural way to incorporate the 
violation of Kepler's law during the plunge; being qualitatively consistent
with the TE data, it offers a more suitable starting point for further improvements 
of the EOB insplunge waveform that are
needed in the late plunge and merger phase (such as the addition of NQC corrections
and the matching to the ringdown).
In the case of spins aligned with the orbital angular momentum, we found that the dominant
source of uncertainty in both EOB fluxes is the lack of high-order PN resummed
knowledge. A crucial step for the development of next generation EOB models for 
spinning binaries will therefore be the update of the EOB resummed multipolar waveforms 
with higher order PN corrections. Similar conclusions have been drawn recently
in~\cite{Taracchini:2014zpa} in an independent analysis based on the
waveform phasing. Note that the necessary PN information to update the
$\rho_\lm$'s of the EOB waveform is implicitly available in the work of
Shah~\cite{Shah:2014tka} (a result that complements the 22PN-accurate 
result for a particle orbiting a Schwarzschild black hole by Fujita~\cite{Fujita:2012cm}). 
Once the various PN multipoles are available and resummed 
to obtain the higher order PN corrections to the $\rho_\lm$'s, it will
be interesting to revise our study of the span of accuracy for the 
radiation reaction.

\paragraph{Multipolar insplunge waveforms at scri.} 
We computed multipolar waveforms spanning almost the entire black hole
rotation parameter range, i.e.~corotating and counterrotating
backgrounds up to nearly-extremal spins of $|\ha|=0.9999$. Results are
reported in Sec.~\ref{sec:inspl:09} and Sec.~\ref{sec:inspl:09999}.
The multipolar hierarchy at merger has been investigated in detail 
as a function of the spin, considering all the multipoles up to
$\ell=m=8$. Close to merger, the subdominant multipolar amplitudes
(notably the $m=0$ ones) are enhanced for retrograde orbits with respect to
prograde ones. We argue that this effect mirrors nonnegligible deviations from
circularity of the dynamics during the late-plunge and merger phase
(see in particular Fig.~\ref{fig:multipolar_peaks} and \ref{app:nocirc}).

The waveforms presented here are the first test-mass waveforms 
extracted at scri for inspirals on spinning backgrounds driven by 
a nonadiabatic EOB radiation reaction~\footnote{
  The work of~\cite{Taracchini:2014zpa} also computed waveforms at
  scri. Differently from here, however, the Teukolsky waveforms therein
  are generated by a radiation reaction computed numerically from
  frequency domain circular orbits' fluxes, and extracted with the algorithm
  presented in~\cite{Zenginoglu:2011zz}. See also Sec.~\ref{sec:inspl:09}.}.
We believe these results improve quantitatively over previous
calculations~\cite{Barausse:2011kb}, due to the extraction at scri and
the accuracy of the employed high-order finite differencing scheme. 
In particular, systematic uncertainties due to finite radius extraction and
extrapolation are discussed in Sec.~\ref{sec:inspl:finiter}.
Future work will be devoted to the analysis and the improvement
of the relative EOB analytical multipolar waveform following the lines
of~\cite{Bernuzzi:2010xj,Barausse:2011kb,Damour:2012ky}.

\paragraph{Horizon-absorbed fluxes in the time-domain.}
For the first time to our knowledge, we have applied the time-domain
formalism of Poisson~\cite{Poisson:2004cw} to calculate the GW fluxes
absorbed by the horizon during the insplunge. Results are collected in
Sec.~\ref{sec:horabs}. We calculated the 
horizon fluxes up to (and slightly beyond) the LSO. 
The TE solution indicates that during the last
25 orbits and up to the LSO the horizon-absorbed flux contributes to the total GW flux 
at least $3\%$ ($1\%$) for $\ha=0.9$ ($\ha=-0.9$). 
These values are consistent with those from adiabatic circular orbits' fluxes. 
Notably, it was not possible to obtain reliable results up to the LR due to
inaccuracies related to the calculation of the advanced-time integrals
and the presence of the particle source term. We argued that, in
our setup, a local formalism using data only from the
hypersurfaces would be preferable. 

In future work we will develop an improved analytical description within the EOB 
framework extending the work in~\cite{Nagar:2011aa,Taracchini:2013wfa} to
evaluate the relevance of absorbed fluxes for GW astronomy with
intermediate-mass-ratio astrophysical binaries~\cite{Bernuzzi:2012ku}.

\paragraph{A numerical method for a $\cal{O}(\nu)$-consistent radiation reaction.} 
We proposed a numerical, iterative method to calculate
self-consistently the GW fluxes to first order in the symmetric mass
ratio, $\cal{O}(\nu)$. As a proof of principle, a calculation of the $\cal{O}(\nu)$ consistent 
radiation reaction for $\ha=0.9$, including horizon absorption is
presented in Sec.~\ref{sec:sf_flux}. The self-consistent
simulation produces a waveform that differs by $\Delta N_{\rm gw}\sim35.5$ gravitational wave
cycles from the one using the EOB radiation reaction.
Two self-consistent simulations, one with and the other without horizon absorption, 
differ by $\Delta N_{\rm gw}\sim5.4$. 
This result further highlights the importance of horizon-absorbed fluxes
during insplunge.

The method can be used alternatively to the EOB analytical radiation
reaction when the analytical information is poor or not
sufficient. For example, it could be employed for rapidly spinning and/or
precessing binaries, for the horizon absorbed fluxes, and for generic
orbits, including eccentric/scattering configurations.

\ack

We thank Scott Hughes for making available his data on circular
orbits.
This work was supported in part by  DFG grant SFB/Transregio~7
``Gravitational Wave Astronomy''. 
E.H. thanks IHES for hospitality during the development of
part of this work. 
A.Z.~was supported in part by NSF grant PHY-1068881 and by a 
Sherman Fairchild Foundation grant to Caltech.

\appendix

\newpage
\section{Data and Tables}
\label{app:DataTables}
Tab.~\ref{tab:test_flux}
shows parameters and important times (crossing of LR etc.) 
and orbital frequencies of the dynamics.
Tab.~\ref{tab:peaks_a} lists important features 
of the waveforms at merger, which will be useful for waveform
modeling and calibration studies. Tab.~\ref{tab:OmgOrb}
shows the mild dependence of 
$u_{A_{22}^{\rm max}}-\hat{t}_{\Omega_{\rm orb}^{\rm max}}$
discussed in Sec.~\ref{sec:inspl:09}. \vspace{-6mm}

\begin{table}[h]
  \center
  \caption{Key numbers for the simulations discussed in this work 
    (see Figs.~\ref{fig:EOB_fluxes_TKvsDNBB0}, \ref{fig:EOB_fluxes}, 
    \ref{fig:EOB_multipolar_structure_amp}, \ref{fig:multipolar_peaks}, 
    \ref{fig:EOB_multipolar_structure_freq} 
    ).
    From left to right: $\hr_0$ is the initial separation, $M\Omega_0= \left( \hr^{3/2} + \ha\right)^{-1}$ 
    the initial (circular) orbital frequency; $\hOmg_{\rm LSO}\equiv \hOmg\left(\hat{t}_{\rm LSO}\right)$ 
    and $\hOmg_{\rm LR}\equiv \hOmg\left(\hat{t}_{\rm LR}\right)$ refer to the orbital frequency 
    of the particle at the LSO and LR crossing respectively,
    (at $\hat{t}_{LR}$). The last column of the table lists the time corresponding
    to $\max(M\Omega)$.
    For $\ha\rightarrow 1$ the trajectories stop slightly outside the LR 
    (see discussion in Sec.~\ref{sec:inspl:09999}). Note, how  
    (i)~for positive spins the inspiral starts already in the strong-field regime
    and (ii)~for $\ha\neq 0$, $\hat{t}_{\Omega^{\rm max}}\neq \hat{t}_{\rm LR}$ with progressively
    larger differences as $\ha\to 1$.}
  \label{tab:test_flux}
    \begin{tabular}[t]{c||cc|ccc|ccc|c}
      \hline
      \hline
        $\ha $   &  $\hr_0$  & $M\Omega_0$ & $\hr_{\rm LSO}$ & $M\Omega_{\rm LSO}$ & $\hat{t}_{\rm LSO}$ & $\hr_{\rm LR }$ & $M\Omega_{\rm LR }$ & $\hat{t}_{\rm LR }$ & $\hat{t}_{\Omega^{\rm max}}$ \\
        \hline 
        -0.9999 &  10.00 & 0.03266 & 9.000 & 0.0385 & 6858.3 & 4.000 & 0.03846 & 7321.7 & 7321.3  \\ 
        -0.9995 &  9.90 & 0.03317 & 8.999 & 0.0385 & 5541.0 & 4.000 & 0.03846 & 6004.4 & 6004.0  \\ 
        -0.9990 &  9.80 & 0.03369 & 8.997 & 0.0385 & 4382.9 & 3.999 & 0.03847 & 4846.3 & 4845.8  \\ 
        -0.9950 &  9.75 & 0.03396 & 8.986 & 0.0385 & 3963.1 & 3.996 & 0.03854 & 4425.2 & 4424.8  \\ 
        -0.9900 &  9.50 & 0.03535 & 8.972 & 0.0386 & 1931.5 & 3.991 & 0.03863 & 2392.6 & 2392.2  \\ 
        -0.9700 &  9.40 & 0.03591 & 8.916 & 0.0390 & 1629.7 & 3.973 & 0.03898 & 2085.5 & 2085.1  \\ 
        -0.9500 &  9.50 & 0.03530 & 8.859 & 0.0393 & 2747.8 & 3.955 & 0.03934 & 3198.1 & 3197.7  \\ 
        -0.9000 &  9.50 & 0.03523 & 8.717 & 0.0403 & 3985.5 & 3.910 & 0.04025 & 4423.4 & 4423.0  \\ 
        -0.8000 &  9.20 & 0.03689 & 8.432 & 0.0422 & 3668.1 & 3.819 & 0.04222 & 4080.8 & 4080.4  \\ 
        -0.7000 &  8.90 & 0.03868 & 8.143 & 0.0444 & 3397.0 & 3.725 & 0.04436 & 3785.2 & 3784.8  \\ 
        -0.6000 &  8.60 & 0.04062 & 7.851 & 0.0467 & 3168.7 & 3.630 & 0.04673 & 3533.0 & 3532.7  \\ 
        -0.5000 &  8.30 & 0.04271 & 7.555 & 0.0493 & 2980.4 & 3.532 & 0.04934 & 3321.3 & 3321.0  \\ 
        -0.4000 &  8.00 & 0.04499 & 7.254 & 0.0522 & 2829.6 & 3.432 & 0.05224 & 3147.7 & 3147.5  \\ 
        -0.3000 &  7.70 & 0.04747 & 6.949 & 0.0555 & 2714.6 & 3.329 & 0.05548 & 3010.4 & 3010.2  \\ 
        -0.2000 &  7.40 & 0.05018 & 6.639 & 0.0591 & 2634.3 & 3.223 & 0.05913 & 2908.4 & 2908.3  \\ 
        -0.1000 &  7.10 & 0.05314 & 6.323 & 0.0633 & 2588.9 & 3.113 & 0.06328 & 2841.8 & 2841.8  \\ 
        0.0000 &  7.00 & 0.05399 & 6.000 & 0.0680 & 4076.1 & 3.000 & 0.06802 & 4308.4 & 4308.4  \\ 
        0.1000 &  6.40 & 0.06138 & 5.669 & 0.0735 & 2012.0 & 2.882 & 0.07352 & 2224.2 & 2224.3  \\ 
        0.2000 &  6.10 & 0.06551 & 5.329 & 0.0800 & 2088.2 & 2.759 & 0.07995 & 2281.0 & 2281.1  \\ 
        0.3000 &  5.80 & 0.07009 & 4.979 & 0.0876 & 2207.2 & 2.630 & 0.08762 & 2381.0 & 2381.2  \\ 
        0.4000 &  5.40 & 0.07723 & 4.614 & 0.0969 & 1862.8 & 2.493 & 0.09694 & 2018.3 & 2018.6  \\ 
        0.5000 &  5.01 & 0.08537 & 4.233 & 0.1085 & 1671.1 & 2.347 & 0.10854 & 1808.8 & 1809.2  \\ 
        0.6000 &  4.70 & 0.09268 & 3.829 & 0.1235 & 1914.2 & 2.189 & 0.12351 & 2034.5 & 2035.0  \\ 
        0.7000 &  4.10 & 0.11109 & 3.393 & 0.1438 & 1126.9 & 2.013 & 0.14379 & 1230.1 & 1230.9  \\ 
        0.8000 &  3.80 & 0.12184 & 2.907 & 0.1736 & 1571.6 & 1.811 & 0.17360 & 1657.3 & 1658.5  \\ 
        0.9000 &  3.05 & 0.16060 & 2.321 & 0.2251 & 820.7 & 1.558 & 0.22514 & 883.6 & 886.2  \\ 
        0.9500 &  3.02 & 0.16134 & 1.937 & 0.2732 & 1432.9 & 1.386 & 0.27316 & 1472.5 & 1491.6  \\ 
        0.9700 &  3.30 & 0.14358 & 1.738 & 0.3037 & 2813.7 & 1.296 & 0.30368 & 2841.9 & 2862.4  \\ 
        0.9900 &  3.01 & 0.16097 & 1.454 & 0.3510 & 2010.0 & 1.168 & 0.35101 & 2032.6 & 2058.5  \\ 
        0.9950 &  3.60 & 0.12779 & 1.341 & 0.3722 & 4914.9 & 1.118 & 0.37215 & 4941.1 & 4945.2  \\ 
        0.9990 &  3.60 & 0.12772 & 1.182 & 0.4137 & 5018.1  & 1.052 & 0.45258 & $\times$ & 5032.5 \\ 
        0.9995 &  3.60 & 0.12771 & 1.140 & 0.4308 & 5034.2  & 1.037 & 0.45309 & $\times$ & 5043.6 \\ 
        0.9999 &  3.60 & 0.12771 & 1.079 & 0.4537 & $\times$ & 1.016 & 0.45368 & $\times$  & 5052.5 \\ 
      \hline
      \hline
    \end{tabular}    
\end{table}

\begin{table}[h]
    %
    \caption{Properties of multipolar waveforms at merger for representative
      values of $\ha$. See Table~\ref{tab:peaks_0} for definitions. 
      The values in brackets are the numbers found in~\cite{Barausse:2011kb}. }  
    \label{tab:peaks_a}
    \begin{tabular}[t]{cc|lllc|clcc}
      \hline
      \hline
      $\ell$ & $m$ & $\ha$ &$\Delta t_\lm$ & $\hat{A}_\lm^{\max} $ &
       $M\omega_\lm^{A_\lm^{\max}} $  & $\ha$ & $\Delta t_\lm$ & $\hat{A}_\lm^{\max}$ &
      $M\omega_\lm^{A_\lm^{\max}} $ \\
      \hline
      2 & 2 & 0.5  &$-7.23$ ($-7.22$)    &  0.3147 &  0.3396    &   $-0.5$ &$-0.03$ ($-0.08$)  &  0.2820  &  0.2378  \\
      2 & 1 & 0.5  &\;\;\;3.83  &  0.0666  &  0.2912            &   $-0.5$ &\;\;\;12.79  &  0.1508  &  0.2391  \\
      2 & 0 & 0.5  &\;\;\;8.45  &  0.0155  &  $\times$          &   $-0.5$ &\;\;\;12.81  &  0.1108  &  $\times$  \\  
      3 & 3 & 0.5  &$-1.99$  &  0.0576  &  0.5678               &   $-0.5$ &\;\;\;2.76  &  0.0480  &  0.3916  \\ 
      3 & 2 & 0.5  &$-1.61$  &  0.0146  &  0.4262               &   $-0.5$ &\;\;\;12.15  &  0.0220  &  0.4422  \\ 
      3 & 1 & 0.5  &\;\;\;3.12  &  0.0025  &  0.3514            &   $-0.5$ &\;\;\;12.10  &  0.0115  &  0.3162  \\  
      3 & 0 & 0.5  &\;\;\;8.94  &  0.0004  &  $\times$          &   $-0.5$ &\;\;\;17.93  &  0.0091  &  $\times$  \\  
      4 & 4 & 0.5  &\;\;\;0.32  &  0.0169  &  0.7954            &   $-0.5$ &\;\;\;4.40  &  0.0132  &  0.5458  \\ 
      4 & 3 & 0.5  &\;\;\;0.54  &  0.0046  &  0.6535            &   $-0.5$ &\;\;\;10.10  &  0.0056  &  0.5934  \\ 
      4 & 2 & 0.5  &\;\;\;1.06  &  0.0009  &  0.5306            &   $-0.5$ &\;\;\;12.44  &  0.0027  &  0.5719  \\ 
      4 & 1 & 0.5  &\;\;\;4.81  &  0.0001  &  0.4958            &   $-0.5$ &\;\;\;16.38  &  0.0015  &  0.4168  \\ 
      4 & 0 & 0.5  &\;\;\;8.69  &  2.16e-05  &  $\times$        &   $-0.5$ &\;\;\;16.22  &  0.0011  &  $\times$  \\  
      \hline
      \hline
      2 & 2 & 0.7  &$-12.74$ ($-12.77$)  &  0.3228    &0.3886 & $-0.7$ &  0.76 ($\times$) &  0.2776  &  0.2279  \\ 
      2 & 1 & 0.7 &$-0.02$  &  0.0472  &  0.2950            & $-0.7$ & 13.62   &0.1728   &  0.2095  \\  
      2 & 0 & 0.7 &\;\;\;6.50  &  0.0061  &  $\times$           & $-0.7$ & 13.43   &0.1418   &  $\times$  \\  
      3 & 3 & 0.7 &$-5.10$  &  0.0611  &  0.6505            & $-0.7$ & 3.28    &0.0468   &  0.3729  \\  
      3 & 2 & 0.7 &$-9.57$  &  0.0143  &  0.4236            & $-0.7$ & 13.24   & 0.0241  &  0.4015  \\  
      3 & 1 & 0.7 &$-1.12$  &  0.0018  &  0.3196            & $-0.7$ & 18.65   & 0.0151  &  0.2230  \\  
      3 & 0 & 0.7 &\;\;\;8.30  &  0.0001  &  $\times$           & $-0.7$ & 18.38   & 0.0138  &  $\times$  \\ 
      4 & 4 & 0.7 &$-2.12$  &  0.0183  &  0.9117            & $-0.7$ & 4.90    &0.0128   &  0.5192  \\ 
      4 & 3 & 0.7 &$-4.66$  &  0.0047  &  0.6832            & $-0.7$ & 12.92   & 0.0058  &  0.5907  \\  
      4 & 2 & 0.7 &$-9.29$  &  0.0007  &  0.4515            & $-0.7$ & 17.17   & 0.0032  &  0.5043  \\  
      4 & 1 & 0.7 &$-0.39$  &  5.97e-05  &  0.3849          & $-0.7$ & 16.79   &0.0022   &  0.3207  \\  
      4 & 0 & 0.7 &\;\;\;27.91  &  4.68e-06  &  $\times$        & $-0.7$ & 16.61   &0.0018   &  $\times$  \\  
      \hline
      \hline
      2 & 2 & 0.9  &$-39.16$ ($-39.09$)  &  0.3212  &0.4771       &$-0.9$  & 1.54 (1.60)  &  0.2738  &  0.2198  \\ 
      2 & 1 & 0.9 &$-35.01$            &  0.0249  &  0.2509     &$-0.9$  & 14.36  &  0.1996  &  0.1738  \\ 
      2 & 0 & 0.9 & \;\;\;0.09             &  0.0009  &  $\times$   &$-0.9$  & 14.03  &  0.1788  &  $\times$  \\  
      3 & 3 & 0.9 &$-18.03$            &  0.0645  &  0.8013     &$-0.9$  & 3.75  &  0.0459  &  0.3567  \\  
      3 & 2 & 0.9 &$-35.46$            &  0.0148  &  0.4860     &$-0.9$  & 13.94  &  0.0267  &  0.3442  \\  
      3 & 1 & 0.9 &$-26.01$            &  0.0010  &  0.2711     &$-0.9$  & 19.13  &  0.0209  &  0.1295  \\  
      3 & 0 & 0.9 & \;\;\;5.18             &  1.75e-05  &  $\times$ &$-0.9$  & 18.82  &  0.0203  &  $\times$  \\  
      4 & 4 & 0.9 &$-12.11$            &  0.0201  &  1.1223     &$-0.9$  & 5.37  &  0.0125  &  0.4960  \\ 
      4 & 3 & 0.9 &$-19.96$            &  0.0052  &  0.7959     &$-0.9$  & 13.93  &  0.0062  &  0.5318  \\ 
      4 & 2 & 0.9 &$-51.22$            &  0.0007  &  0.4624     &$-0.9$  & 17.65  &  0.0042  &  0.3645  \\ 
      4 & 1 & 0.9 &$-13.43$            &  3.11e-05  &  0.3218   &$-0.9$  &  17.18  &  0.0032  &  0.2089  \\ 
      4 & 0 & 0.9 & \;\;\;21.41            &  4.17e-07  &  $\times$ &$-0.9$  &  16.98  &  0.0029  &  $\times$  \\ 
      \hline
      \hline
    \end{tabular}          
\end{table}

\begin{table}[t]
\centering
    \caption{Time lag between $\hOmg^{\rm max}$, $\hOmg^{\rm max}_{\rm orb}$ and $\hat{A}_{22}^{\rm max}$ varying $\ha$. The time difference
     $u_{A_{22}^{\rm max}}-\hat{t}_{\Omega_{\rm orb}^{\rm max}}$
     exhibits a mild dependence on $\ha$ up to $\ha=0.8$.  }   
   \label{tab:OmgOrb}
   \begin{tabular}[t]{cccc}
     \hline
     \hline
     $\ha$  & $\Delta t_{22}$ & $\hat{t}_{\Omega^{\rm max}}-\hat{t}_{\Omega^{\rm max}_{\rm orb}} $ & $u_{A_{22}^{\rm max}}-\hat{t}_{\Omega_{\rm orb}^{\rm max}}$ \\
     \hline
           $-0.9999$ & $\;\;\;1.94$ & $-4.09$ & $-2.15$ \\
           $-0.9990$ & $\;\;\;1.93$ & $-4.09$ & $-2.16$ \\
           $-0.9900$ & $\;\;\;1.90$ & $-4.07$ & $-2.17$ \\
           $-0.9500$ & $\;\;\;1.74$ & $-3.97$ & $-2.23$ \\
           $-0.9000$ & $\;\;\;1.54$ & $-3.84$ & $-2.30$ \\
           $-0.8000$ & $\;\;\;1.15$ & $-3.56$ & $-2.41$ \\
           $-0.7000$ & $\;\;\;0.76$ & $-3.26$ & $-2.50$ \\
           $-0.6000$ & $\;\;\;0.37$ & $-2.93$ & $-2.56$ \\
           $-0.5000$ & $-0.03$      & $-2.56$ & $-2.59$ \\
           $-0.4000$ & $-0.43$      & $-2.16$ & $-2.59$ \\
           $-0.3000$ & $-0.86$      & $-1.71$ & $-2.57$ \\
           $-0.2000$ & $-1.33$      & $-1.21$ & $-2.54$ \\
           $-0.1000$ & $-1.83$      & $-0.65$ & $-2.48$ \\
      $\;\;\;0.0000$ & $-2.38$      & $\;\;\;0.00$ & $-2.38$\\
      $\;\;\;0.1000$ & $-3.02$      & $\;\;\;0.75$ & $-2.27$ \\
      $\;\;\;0.2000$ & $-3.76$      & $\;\;\;1.64$ & $-2.12$ \\
      $\;\;\;0.3000$ & $-4.64$      & $\;\;\;2.73$ & $-1.91$ \\
      $\;\;\;0.4000$ & $-5.76$      & $\;\;\;4.12$ & $-1.64$ \\
      $\;\;\;0.5000$ & $-7.24$      & $\;\;\;6.00$ & $-1.24$ \\
      $\;\;\;0.6000$ & $-9.35$      & $\;\;\;8.80$ & $-0.55$ \\
      $\;\;\;0.7000$ & $-12.74$     & $\;\;\;13.63$ &$\;\;\; 0.89$ \\
      $\;\;\;0.8000$ & $-19.36$     & $\;\;\;24.86$ &$\;\;\; 5.50$ \\
      $\;\;\;0.9000$ & $-39.16$     & $\;\;\;88.26$ &$\;\;\; 49.10$ \\
      $\;\;\;0.9500$ & $-85.79$     & $\;\;\;323.13$ & $\;\;\;237.34$ \\
      $\;\;\;0.9900$ & $-156.26$    & $\;\;\;670.37$ & $\;\;\;514.11$ \\
      $\;\;\;0.9990$ & $-154.70$    & $\;\;\;862.53$ & $\;\;\;707.83$ \\
      $\;\;\;0.9999$ & $-154.96$    & $\;\;\;1122.53$ &$\;\;\;967.57$ \\
     \hline
     \hline
    \end{tabular}          
\end{table}    

\newpage

\section{Hamiltonian dynamics}
\label{app:ham}

The geodesic motion of a particle on a fixed background metric can be
expressed in Hamiltonian form using coordinates $q^{\alpha}$ (e.g.~$q^\alpha=(t,r,\theta,\phi)$)
and the conjugate momenta $p_{\alpha}$ (e.g.~$P_{\alpha}=(P_t,P_r,P_\theta,P_\phi)$).
The affine parameter $\lambda$ disappears.  One starts from the
four dimensional Hamiltonian $\cal{H}$ written as 
\be
\label{superH} 
{\cal H} = \dfrac{1}{2}g^{\alpha \beta}P_\alpha P_\beta
= -\dfrac{1}{2}\mu^2\ , 
\ee 
where the second equation comes from using the length of the geodesic as
the affine parameter itself. The reduced momenta are defined as 
\be
\frac{dq_\alpha}{d\lambda} = \dfrac{P_\alpha}{\mu} \equiv p_\alpha \ ,
\ee
which leads to the expression
\be
\label{eq:start} g^{\alpha \beta}p_\alpha p_\beta = -1\ .
\ee 
Since the dynamics does not depend explicitly on
the affine parameter $\lambda$ (,i.e.~the proper time) one can use the
time coordinate as the integration parameter and thus reduce the geodesic
equations generated by the superhamiltonian, Eq.~\eqref{superH},
to just six Hamilton equations. 
The Hamiltonian is defined from Eq.~\eqref{eq:start} as
\be \hat{H} \equiv \dfrac{H}{\mu}= -p_0 =
\dfrac{g^{0i}p_i}{g^{00}}+\left(
\dfrac{g^{0i}p_i}{g^{00}}-\dfrac{g^{ij}p_i p_j+1}{g^{00}} 
\right)^{1/2} \ . \ee 
Note that $\hat{H}$ is the energy that is conserved for geodesic
orbits, i.e. one of the first integrals of motion that one gets 
from the usual  Lagrangian description of the relativistic geodesic motion.
The equations of motion follow, 
\be
\dfrac{d q^i}{dt} = \dfrac{\p \H}{\p p_i} \ ; \quad \
\dfrac{d p_i}{dt} = -\dfrac{\p \H}{\p q^i} + \hat{\cal F}_i \ . 
\ee
We introduced explicitly at the r.h.s. of the second set of equations the dissipative 
terms $\hat{\cal F}_i$ that represent the radiation reaction. For example $\hat{\cal F}_\phi$
is prescribed in the EOB model as a resummed analytical expression for the 
radiation-reaction. 
The Hamiltonians relevant for this work are given below.
From the Schwarzschild metric
\be 
ds^2 = -A_S(\hr) dt^2 + A^{-1}_S(\hr) d\hr^2 + \hr^2(d\theta^2 + \sin^2\theta
d\phi^2)  \ ,
\ee 
one gets the well-known Schwarzschild Hamiltonian 
\be 
\hat{H}_{\rm Schw} = \sqrt{
A_S(\hr)\left(1+\dfrac{p_\phi^2}{\hr^2}\right) + A_S(\hr)^2 p_\hr^2} 
\ee 
where $A_S(\hr) = (1-2/\hr)$. Note that we use
normalized coordinates, $\hr\equiv r/M$. 
Moving to Kerr spacetime, the line element can be written as
(we adopt a common notation, see e.g.~\cite{Schnittman:2003tm})
\be 
ds^2 = -A dt^2 + \varpi^2(d\phi-\omega dt)^2 +
\dfrac{\Sigma}{\Delta} d\hr^2 + \Sigma d\theta^2 \ . 
\ee 
The inverse metric that enters the definition of the Hamiltonian reads
\be
\label{eq:ginv}
g^{\mu \nu} = \left(\begin{array}{cccc} -1/A &
0 & 0 & -\omega/A \\ 0 & \Delta/\Sigma & 0 & 0 \\ 0 & 0 &
1/\Sigma & 0 \\ -\omega/\alpha^2 & 0 & 0 &
1/\varpi^2-\omega^2/A \end{array}\right),
\ee
where one has defined
\begin{eqnarray} 
\label{eq:Sigma}
\Sigma &\equiv& \hr^2 +\ha^2 \cos^2\theta \\
\label{eq:Delta}
\Delta &\equiv& \hr^2 -2\hr + \ha^2 \\ 
A &\equiv& \frac{\Sigma \Delta}{\Sigma \Delta+2 \hr(\ha^2+\hr^2)} \\
\label{eq:omega}
\omega &\equiv& \frac{2 \hr \ha}{\Sigma\Delta + 2 \hr(\ha^2+\hr^2)} \\
\label{eq:varpi}
\varpi^2 &\equiv& \left[\frac{\Sigma\Delta + 2
\hr(\ha^2+\hr^2)}{\Sigma}\right] \sin^2\theta \ .
\end{eqnarray}
The Hamiltonian of a nonspinning particle on Kerr spacetime is finally written as
\be
\label{eq:H_general}
\hat{H}_{\rm Kerr} = \omega p_\phi + \sqrt{A\left(1+\dfrac{p_\phi^2}{\varpi^2}\right)
                 +\dfrac{A}{\Sigma}p_\theta^2 +
                 A\dfrac{\Delta}{\Sigma}p_\hr^2}  \ .
\ee

\section{Next-to-quasi-circular effects in the multipolar waveform amplitude}
\label{app:nocirc}

In Sec.~\ref{sec:inspl} we pointed out that (consistently with the 
discussion in Ref.~\cite{Taracchini:2014zpa}) next-to-quasi-circular (NQC) 
effects related to the growth of $p_{r_*}$ in the late-plunge and merger phase 
when $\ha\to -1$ are responsible for the corresponding  ``sharpening'' 
of the peaks of the multipolar amplitudes $\A_\lm$. On the contrary, 
when $\ha\to 1$ and $p_{r_*}$ is almost negligible, the peak is rather 
flat and barely distinguishable. In other words, when $\ha\to +1$ one has
a long, persistent, quasi-adiabatic inspiral until the particle locks 
to the black hole horizon. In this Appendix we add some details, showing 
how NQC effects can practically act to shape the waveform amplitude around
merger. For pedagogical purposes, we focus on the $\ell=2$ modes only and 
use Newtonian-like waveforms to illustrate the effect. 
The same structural behavior is valid also for the other multipoles.

\begin{figure}[t]
  \centering
  \includegraphics[width=0.49\textwidth]{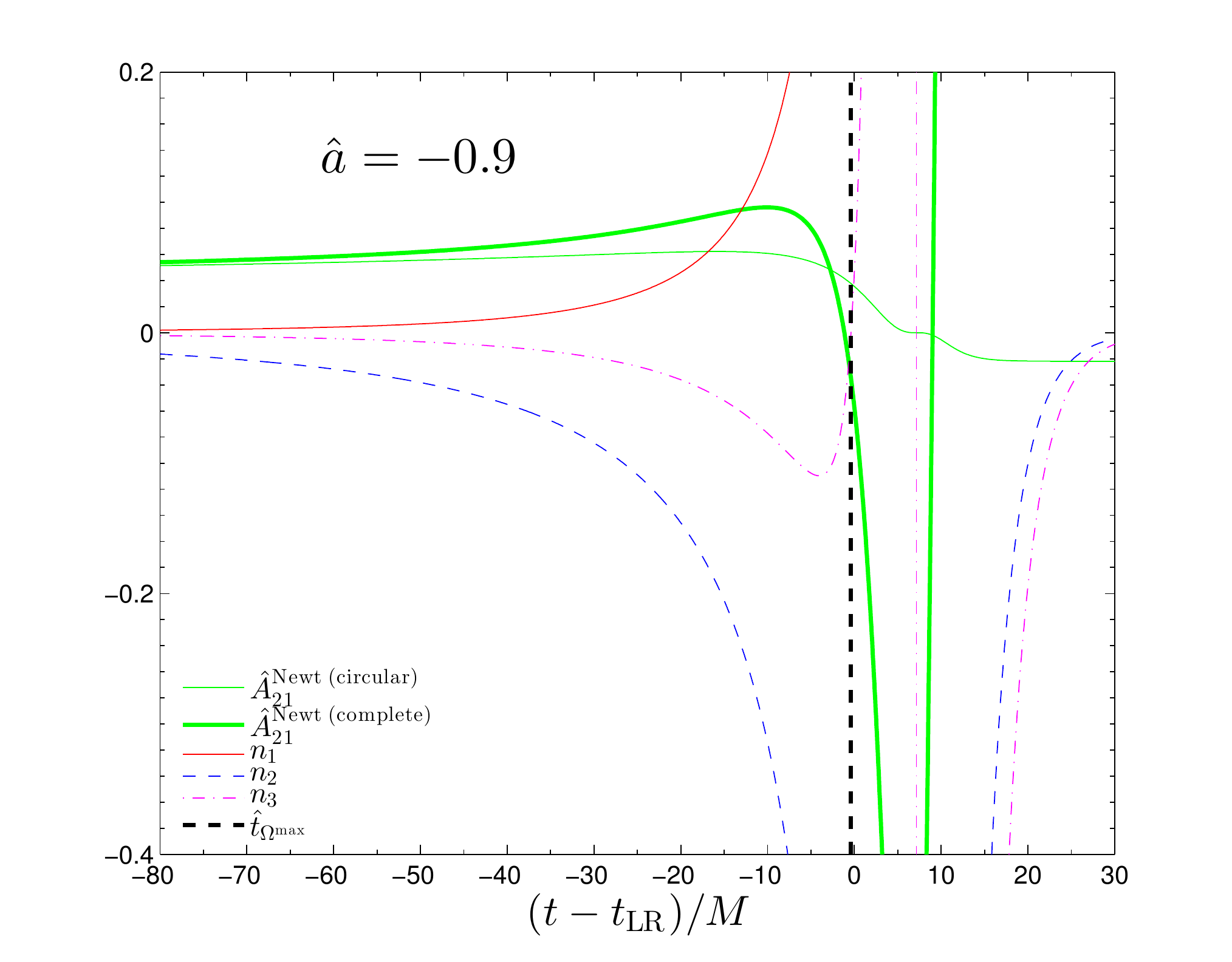}
  \includegraphics[width=0.49\textwidth]{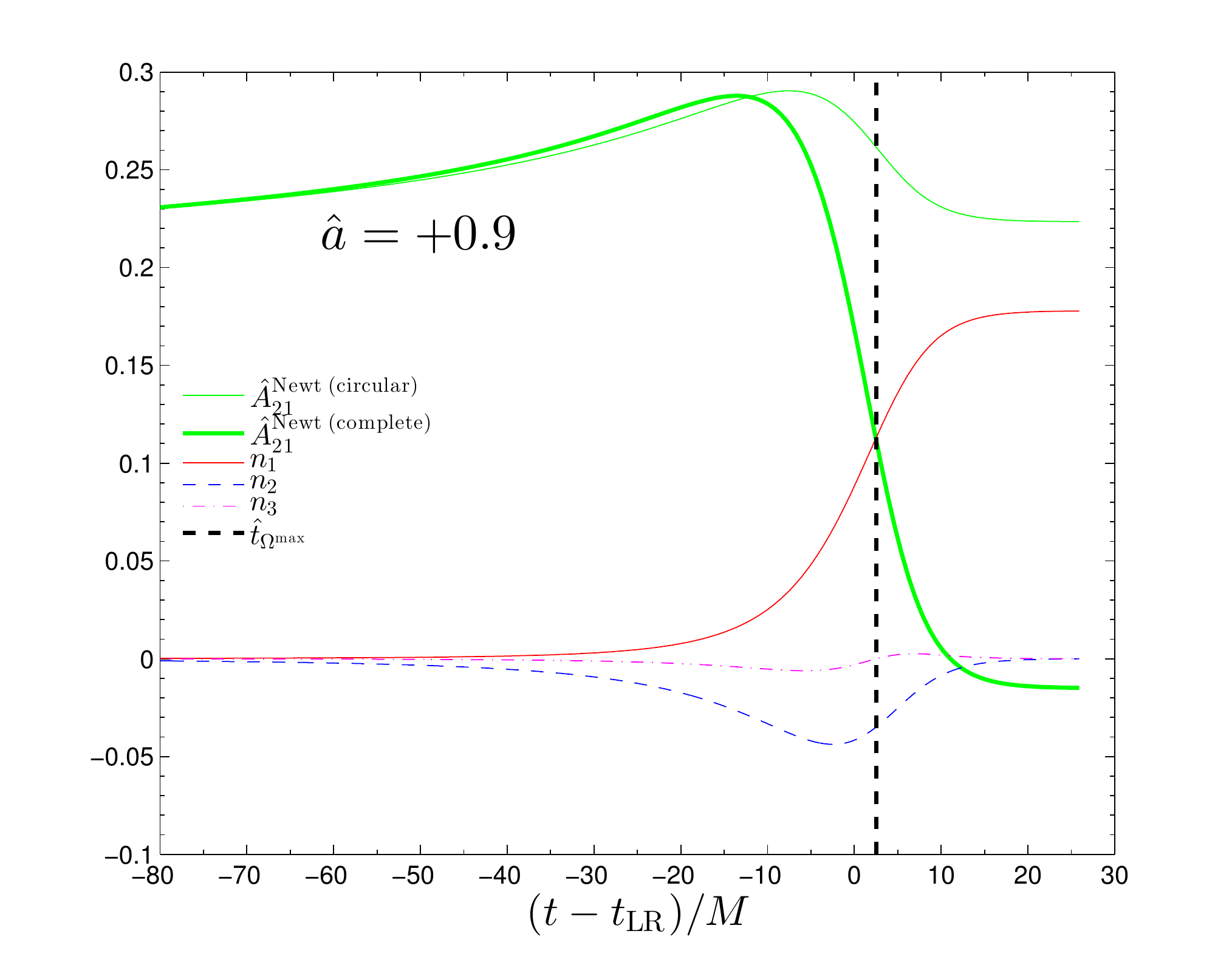}
  \caption{Effect of next-to-quasi-circular correction factors to the Newtonian 
   (circular) waveform for $\ha=\pm 0.9$. NQC terms are responsible for the 
    sharpening of the amplitude's peak when $\ha<0$. }  
  \label{fig:nqc21} 
\end{figure}

At the leading-order, Newtonian level the ($\mu$-normalized) 
quadrupolar mass moments $(\hat{I}_{20},\hat{I}_{22})$ and the
mass current $\hat{S}_{21}$ moment of a particle moving
along a trajectory expressed in terms of $(r,\phi)$ read
\begin{align}
\hat{I}_{22}  & =  \sqrt{\dfrac{\pi}{30}}\hr^{2}e^{{-2 \ii \phi}} ,\\
\hat{S}_{21}  & =  \sqrt{\dfrac{16\pi}{45}}\hr^{3}\hOmg e^{-\ii\phi},\\
\hat{I}_{20}  & = -\sqrt{\dfrac{\pi}{15}} \hr^{2}.
\end{align}
The leading-order waveform is computed taking two time derivatives of
these multipoles. The numerical constants here are chosen consistently
with our RWZ normalizations. Replacing  $dr/dt\to p_{r_*}$ (as they coincide 
in the Newtonian limit) in the two time derivatives of  
the multipolar moments, one finds 
\begin{align}
\label{A22}
\hat{A}_{22}^{\rm Newt}&=\hat{A}^0_{22}\hat{A}^{\rm NQC}_{22}=\sqrt{\dfrac{16\pi}{30}}v_\phi^2\left(1-\dfrac{n_1}{2} - \dfrac{n_2}{2}\right), \\
\label{A21}
\hat{A}_{21}^{\rm Newt}&=\hat{A}^0_{21}\hat{A}^{\rm NQC}_{21}=\sqrt{\dfrac{16\pi}{45}} v_\phi^3\left(1 - 6 n_1 - 3 n_2 - 6 n_3\right),\\
\hat{A}_{20}^{\rm Newt} &= \sqrt{\dfrac{4\pi}{15}}\left(\dot{p}_{r_*}^2 + r \dot{p}_{r_*}\right).
\end{align}
Here, the $m\neq 0$ amplitudes were factorized in a circular prefactor, $\hat{A}^0_{\lm}$, where we 
replaced $r \hOmg\to r_\Omega \hOmg=v_\phi$, and a NQC correction factor, $\hat{A}^{\rm NQC}_{\lm}$, 
that depends on $p_{r_*}$ through the factors
\begin{align}
n_{1} &= \dfrac{p_{r_{*}}^{2}}{(\hr\hOmg)^{2}}, \\
n_{2} &= \dfrac{\dot{p}_{r_*}}{\hr\hOmg^2},\\
n_{3} &= \dfrac{p_{r_*}}{\hr\hOmg^2}\dfrac{\dot{\hOmg}}{\hOmg}.
\end{align}
Useful conclusions can be driven from this simple Newtonian formulas. Let us first focus on the $(2,1)$ 
mode end explore its dependence on $\ha$. Figure~\ref{fig:nqc21} illustrates the effect of 
the NQC corrections $(n_{1},n_{2},n_{3})$ on the circular waveform (represented by a thin green line) 
when $\ha=-0.9$ (left panel) and $\ha=+0.9$ (right panel). Note that the vertical dashed line in the figure marks
the location of the peak of $\hOmg$ and that the plot is done versus $t-t_{\rm LR}$, so to 
easily identify the merger. When $\ha=-0.9$ one sees that the large effect of the  
noncircular factors $(n_{1},n_{2},n_{3})$ produces a sharpening of the 
waveform amplitude, while the  time location of the peak moves to the right, closer to the light-ring crossing. 
On the contrary, for $\ha=+0.9$ the much smaller amplitude of the NQC factor is unable to further sharpen 
the waveform peaks, but just results in moving it farther to the left of the light-ring crossing.
This simple Newtonian-based example is helpful in understanding heuristically the role of the NQC amplitude
effects incorporated in the TE waveforms displayed in 
Fig.~\ref{fig:EOB_multipolar_structure_amp}. 
A similar heuristic understanding can be driven also for 
the other multipoles. Evidently, in the actual TE waveform the numerical coefficients in Eq.~\eqref{A21} 
are replaced by functions of $1/r$ so that the actual behavior is more complicated than what is discussed
here, though the behavior of the NQC terms is the key element behind the sharpening of the peak of 
the $\A_{\lm}$ as $\ha\to -1$.

\begin{figure}[t]
  \centering
  \includegraphics[width=0.49\textwidth]{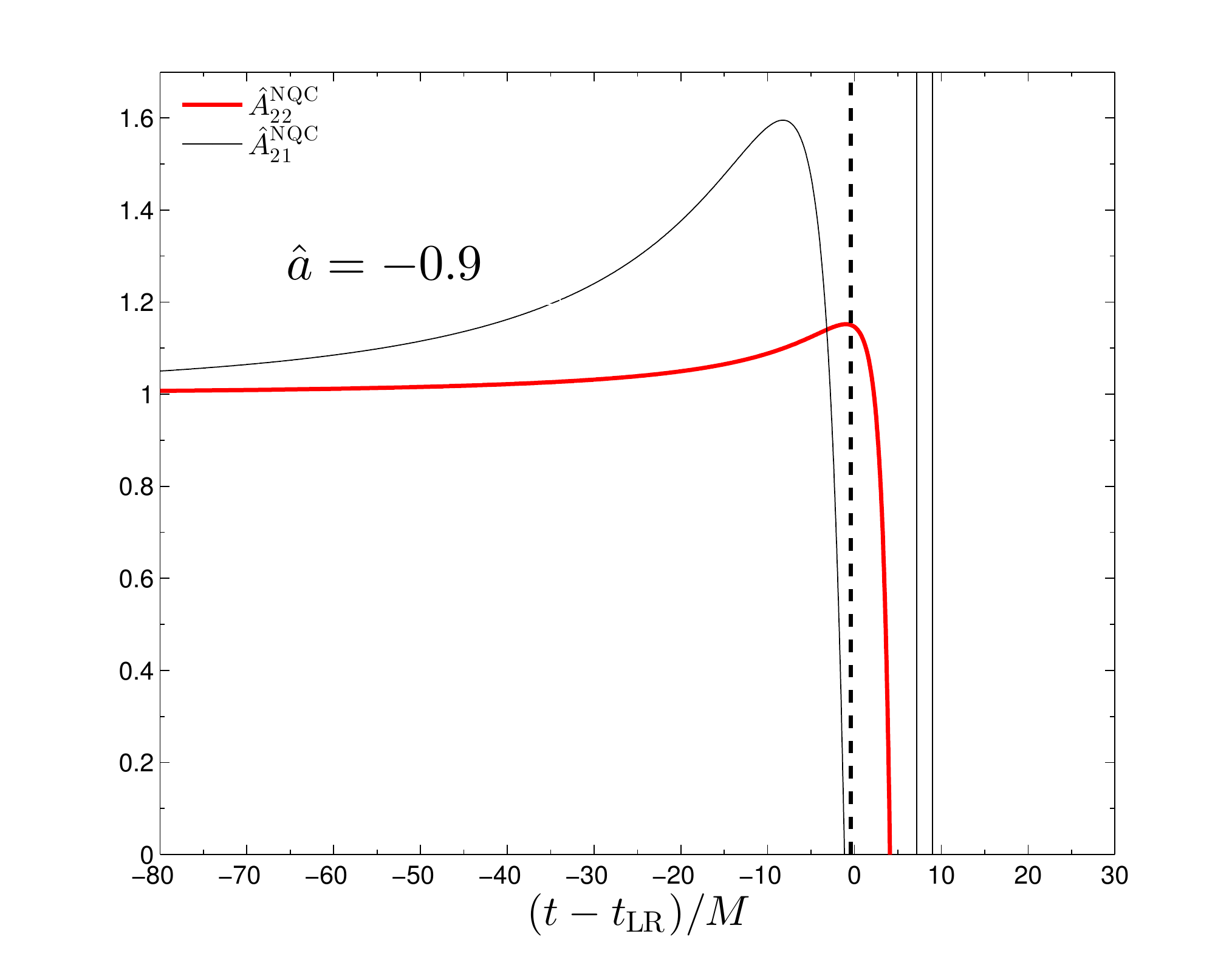}
  \caption{Comparing the NQC correction factors $\hat{A}^{\rm NQC}_{22}$  and $\hat{A}^{\rm NQC}_{21}$ for $\ha=-0.9$.}  
  \label{fig:22vs21} 
\end{figure}

The inspection of Figs.~\ref{fig:EOB_multipolar_structure_amp} also shows that, for a given value of
$\ha$ (,i.e. a given behavior of $p_{r_*}$), and a given value of $\ell$ the peaks of amplitudes with
$\ell=m$ are always less pronounced (with respect to the inspiral part) than the corresponding $0\leq m < \ell$ 
subdominant modes. This holds true for any value of $\ha$ and is explained as due to differences between the 
various $\hat{A}_\lm^{\rm NQC}$'s. Again, the Newtonian analysis helps us understanding the key physical elements.
For example, when one looks at the Newtonian NQC corrections, Eqs.~\eqref{A22} and~\eqref{A21}, one sees that,
for a given value of $p_{r_*}$, the absolute values of the coefficients of the $n_i$ factors 
are {\it smaller} than 1 in $\A^{\rm NQC}_{21}$, while they are {\it larger} than 1 in $\A^{\rm NQC}_{22}$. 
This implies that the NQC correction factor for $(2,1)$ is larger than the one for $(2,2)$, 
as illustrated in Fig.~\ref{fig:22vs21}. Note that $n_3$ is approximately degenerate with 
$n_1$ when using the Kepler's constraint for circular orbits: at leading order one has 
$\dot{\hOmg}/\hOmg = -2/3\, (p_{r_*}/r)$ that is $n_3\approx -2/3\, n_1$, which 
yields $\A_{21}\propto (1-2 n_1 - 3 n_2)$.
An analogous behavior is found for any other subdominant mode, in the sense that, 
for any $\ell$, the NQC correction for $0\leq m < \ell$ is always larger than for $\ell=m$, 
and progressively increasing as $m\to 0$. 
It is remarkable that the Newtonian analysis suffices in capturing the essential elements
of the hierarchical behavior we find in the complete waveform. We postpone to further studies a
detailed analysis of the effect on each multipole of the other PN-resummed corrections factors 
$(\hat{S}^{(\epsilon)},T_{\lm},\rho_\lm^\ell)$ entering the EOB resummed waveform, Eq.~\eqref{eq:h_lm},
as well as of the corresponding NQC corrections.

\section*{References}

\bibliographystyle{unsrt} 


\end{document}